\documentclass[letterpaper,onecolumn]{IEEEtran}
\usepackage[T1]{fontenc}
\usepackage[latin9]{inputenc}
\usepackage{amsmath}
\usepackage{amsthm}
\usepackage{amssymb}
\usepackage{mathtools}
\usepackage{bm}
\usepackage{graphicx}

\makeatletter
\theoremstyle{plain}
\newtheorem{thm}{\protect\theoremname}
\theoremstyle{plain}
\newtheorem{lem}[thm]{\protect\lemmaname}
 \theoremstyle{remark}
\newtheorem{rem}[thm]{\protect\remarkname}
\theoremstyle{definition}
\newtheorem{defn}[thm]{Definition}
\newtheorem{exam}[thm]{Example}
\theoremstyle{plain}
\newtheorem{cor}[thm]{\protect\corollaryname}

\usepackage{hyperref}

\makeatother

\usepackage{babel}
\providecommand{\corollaryname}{Corollary}
\providecommand{\lemmaname}{Lemma}
\providecommand{\remarkname}{Remark}
\providecommand{\theoremname}{Theorem}

\newcommand{\cA}{\mathcal{A}}
\newcommand{\cB}{\mathcal{B}}
\newcommand{\cC}{\mathcal{C}}

\newcommand{\cE}{\mathcal{E}}
\newcommand{\cF}{\mathcal{F}}
\newcommand{\cG}{\mathcal{G}}

\newcommand{\cK}{\mathcal{K}}
\newcommand{\cL}{\mathcal{L}}

\newcommand{\cO}{\mathcal{O}}
\newcommand{\cP}{\mathcal{P}}

\newcommand{\cX}{\mathcal{X}}
\newcommand{\cY}{\mathcal{Y}}

\newcommand{\bM}{\bm{M}}

\newcommand{\bX}{\bm{X}}
\newcommand{\bY}{\bm{Y}}
\newcommand{\bZ}{\bm{Z}}

\newcommand{\ba}{\bm{a}}
\newcommand{\bb}{\bm{b}}
\newcommand{\bc}{\bm{c}}

\newcommand{\bi}{\bm{i}}
\newcommand{\bj}{\bm{j}}

\newcommand{\bt}{\bm{t}}
\newcommand{\bu}{\bm{u}}
\newcommand{\bv}{\bm{v}}

\newcommand{\bx}{\bm{x}}
\newcommand{\by}{\bm{y}}
\newcommand{\bz}{\bm{z}}

\renewcommand{\mid}{\,|\,}
\renewcommand{\triangleq}{\coloneqq}
\newcommand{\qbinom}[2]{\genfrac{[}{]}{0pt}{}{#1}{#2}}

\renewcommand{\Pr}{\ensuremath{\mathbb{P}}}
\newcommand{\ex}[1]{\ensuremath{\mathbb{E}\left[ #1\right]}}
\newcommand{\pr}[1]{\ensuremath{\mathbb{P}\left[ #1\right]}}
\newcommand{\var}[1]{\ensuremath{\operatorname{Var}\left( #1\right)}}

\DeclareMathOperator{\mmse}{\sf mmse}

\DeclareMathOperator{\Span}{span} 

\global\long\def\expt{\mathbb{E}}

\global\long\def\Pr{\mathbb{P}}

\global\long\def\ind{\boldsymbol{1}}

\newcommand{\RM}{\ensuremath{\textnormal{RM}}}
\newcommand{\PRM}{\ensuremath{\textnormal{PRM}}} \newcommand{\dH}{\ensuremath{d_{\mathrm{H}}}} 

   \newcommand{\oca}{&}
  \newcommand{\tca}{}
  \newcommand{\tcb}{}
  \newcommand{\tc}[1]{}
 
\allowdisplaybreaks

\begin{document}

\title{From Symmetry to Capacity: Nested Codes on Binary Memoryless Symmetric Channels}
\author{Henry D.\ Pfister and Galen Reeves\thanks{The work of G.~Reeves and H.\,D.\,Pfister was supported in part by the National Science Foundation (NSF) under Grant Number~2308445.
Any opinions, findings, recommendations, and conclusions expressed in this material are those of the authors and do not necessarily reflect the views of these sponsors.
This research was completed in part while the authors were visiting the Simons Institute for the Theory of Computing.
H.\,D.\,Pfister is a member of the Department of Electrical and Computer Engineering, the Department of Mathematics, and the Duke Quantum Center, Duke University (email: henry.pfister@duke.edu).
G.\,Reeves is a member of the Department of Electrical and Computer Engineering and the Department of Statistical Science, Duke University (email: galen.reeves@duke.edu).
}
}

\maketitle

\begin{abstract}
The past decade has seen notable advances in our understanding of structured error-correcting codes, particularly binary Reed--Muller (RM) codes. While initial breakthroughs were for erasure channels based on symmetry, extending these results to the binary symmetric channel (BSC) and other binary memoryless symmetric (BMS) channels required new tools and conditions.
Recent work uses nesting to obtain multiple weakly correlated looks at each code bit to establish capacity-achieving performance under bit-MAP and block-MAP decoding.
This paper revisits and extends past approaches, aiming to simplify proofs, unify insights, remove unnecessary conditions, and provide new results.
By leveraging powerful results from the analysis of boolean functions, we derive recursive bounds using two or three looks at each stage.
This gives bounds on the bit-error probability that decay exponentially in the number of stages.
For the BSC, we incorporate level-k inequalities and hypercontractive techniques to achieve the faster decay rate required for vanishing block-error probability.
The same ideas also extend to product codes with RM component codes, which are transitive but not doubly-transitive in general, and yield vanishing bit-error and block-error probability at rates arbitrarily close to capacity.
The results are presented in a semi-tutorial style, providing both theoretical insights and practical implications for future research on structured codes.
\end{abstract}

\section{Introduction}

\subsection{Overview}

Over the past 10 years, researchers have significantly advanced our understanding of sequences of structured error-correcting codes and their ability to achieve capacity.
Much of this work has focused on binary Reed--Muller (RM) codes~\cite{Muller-ire54,Reed-ire54} and was driven by the goal of proving that they achieve capacity~\cite{Abbe-it15,Kudekar-stoc16,Kudekar-it17,Abbe-it20,hkazla-stoc21,Reeves-it23,Abbe-focs23}.
A detailed discussion of relevant prior work until 2017 can be found in~\cite{Kudekar-it17}.
For a tutorial that covers RM codes and relevant prior work until 2020, we suggest~\cite{Abbe-now23}.

For erasure channels, in 2016 it was shown that sequences of error-correcting codes achieve capacity under symmetry conditions that are relatively mild and easy to verify~\cite{Kudekar-stoc16,Kumar-itw16,Kudekar-it17}.
The main arguments are based on connecting the decoding problem to the analysis of boolean functions~\cite{ODonnell-2014} and using well-known sharp threshold results~\cite{Kahn-focs88,Talagrand-ap94,Friedgut-procams96,Bourgain-gafa97}.
When combined with a certain hypercontractive inequality~\cite{Samorodnitsky-it19}, these results also imply that RM codes achieve vanishing error rates on the binary symmetric channel (BSC) albeit at code rates bounded away from capacity~\cite{hkazla-stoc21}.

However, extending the erasure results to establish capacity-achieving performance on the BSC (and other symmetric channels) requires additional tools and stronger conditions.
For example, \cite{Reeves-it23,Reeves-isit23} focus on sequences of doubly-transitive codes that satisfy a two-look property.
In particular, this requires that each code has two different puncturing patterns with small overlap that give codes of roughly the same rate as the original.
RM codes naturally satisfy this property because longer RM codes can always be punctured down to shorter RM codes of the same order.
This condition is sufficient to show the code sequence achieves capacity under bit-MAP decoding and it provided the first proof of this fact for RM codes on binary memoryless symmetric (BMS) channels~\cite{Reeves-it23}.
Building on this, \cite{Abbe-focs23} showed that RM codes actually allow many weakly correlated looks.
This establishes a fast decay for the bit-error probability, which can be combined with a weight-enumerator argument (improving an earlier result in~\cite{Mondelli-izsc16}), to show that RM codes achieve capacity under block-MAP decoding.

In this paper, we revisit previous approaches introduced for non-erasure channels and provide some extensions and simplifications.
The style is somewhat tutorial and serves to provide a unified picture of past results while simplifying the proofs and removing some unnecessary conditions.
Our approach combines a few powerful results from the analysis of boolean functions~\cite{ODonnell-2014}. 
Historically, the idea of using boolean functions to analyze error-correcting codes was introduced by Z\'{e}mor~\cite{Zemor-wac94} as a generalization of earlier work by Margulis~\cite{Margulis-ppi74} and developed in~\cite{Tillich-cpc00,Tillich-it04}.
More recently, techniques from boolean functions have been used to analyze sequences of codes in a variety of papers~\cite{Kudekar-it17,Sberlo-soda20,hkazla-stoc21,Reeves-it23,Reeves-isit23,Abbe-focs23}.

At a high level, this paper considers how some ideas from~\cite{Reeves-it23} and~\cite{Abbe-focs23} can be combined to provide more general arguments.
Instead of combining many weakly correlated looks together all at once (as in~\cite{Abbe-focs23}), we derive recursive bounds based on combining two looks (as identified in~\cite{Reeves-it23}) at each stage.
For the soft-decoding functions analyzed in~\cite{Reeves-it23} for BMS channels, we show that two looks at each stage are sufficient to achieve an exponential decay rate in the number of recursive steps.
The following informal theorem is a simple consequence of~Lemma~\ref{lem:nested_sym_mmse_bound} (see Theorem~\ref{thm:bms} for its application to RM codes).
We note that its statement involves some terms defined in Section~\ref{sec:bg}.

\begin{thm} [Informal Bit Error]
Consider a BMS channel with capacity $C$ and a strongly nested sequence of doubly-transitive codes  $\cC_k$ with normalized overlap $\rho_k \leq \rho < 1$ for all $k\in \{0,1,\ldots\}$.
If the rate of $\cC_0$ is less than $C$, then there is a $\delta >0$ such that the bit-error probability $P_b (\cC_k)$ of code $\cC_k$ satisfies
\[ P_b (\cC_k) \leq \left( \frac{1+\rho}{2} \right)^k \left(\frac{1-\delta}{\delta} \right). \]
\end{thm}

In the above theorem, the assumed conditions for the code sequence imply that the code rates must be decreasing.
While the error probability is guaranteed to decay as stated, this is only useful if we can choose $k$ to be relatively large without decreasing the code rate too much.
Fortunately, for the sequence of RM codes that we consider in this paper, the rate difference vanishes with increasing $m$ as long as we choose $\cC_0 = \RM(r_m,m)$ for any $r_m \in \mathbb{N}$ and $k_m =o(\sqrt{m})$.

This rate of decay is reasonably fast but existing techniques to establish vanishing block-error probability from bit-error bounds require an even faster decay rate.
To obtain such a bound, we specialize our analysis for the BSC and use an argument that combines three-looks at each stage.
For codes with additional symmetry, we get an improved bound based on hypercontractivity via the  ``level-$k$'' inequality.
For the BSC, one part of the analysis  focuses on a boolean function that maps the received sequence to the extrinsic bit-MAP estimate of a single bit.
 This type of decoding function was also used~\cite{Abbe-focs23}.
The following informal theorem is essentially a restatement of Theorem~\ref{thm:fast_rm_err_bound_bsc}.

\begin{thm}[Informal Block Error]
Consider a BSC with capacity $C\in (0,1)$.
For all $s \in \mathbb{N}$, there exist infinitely many pairs $(r,m)\in \mathbb {N}^2$ with $m\to \infty$ such that $\cC=\RM(r,m)$ has rate $R\geq C - 2/s$ and satisfies
\[ P_b (\cC)
\leq \exp\left(-\frac{1}{9s} \sqrt{m} \ln(m) \right). \]
In addition, the block-error probability satisfies
\[ P_B (\cC) \leq \sqrt{P_b (\cC)} + O\left(2^{-2^{m/3}}\right). \]
Thus, one can achieve a block-error probability decaying exponentially in $\sqrt{m} \ln m$ for a code whose rate is within $2/s$ of channel capacity.
\end{thm}

The vanishing block-error probability is based on the list-decoding argument in~\cite{Abbe-focs23}.
To explain this, we encapsulate a key RM weight-enumerator estimate in Lemma~\ref{lem:as_weight_enum} which can be extracted from the proof of~\cite[Lemma~9]{Abbe-focs23}.

A key innovation in this work is the use of three-look recursions in Section~\ref{sec:bsc} to achieve faster decay rates for codes with additional symmetry.
The new analysis in Lemma~\ref{lem:gl2_boolean_bound} is based on splitting the Fourier coefficients of a decoding function into two groups: coefficients indexed by sets of size $\lesssim \log N$ and coefficients indexed by sets of size $\gtrsim \log N$.
The first group is bounded using hypercontractivity via the biased ``level-$k$'' inequality (e.g., Lemma~\ref{lem:biased_level_k}) whereas the second group is bounded using a correlation inequality which takes advantage of the additional symmetry (e.g., Lemma~\ref{lem:f_bool_restrict_var} and Lemma~\ref{lem:gl_sym}).

 \subsection{Outline}

\begin{itemize}
\item Section~\ref{sec:bg}, Background and Review, contains background material such as standard definitions, overviews of past work, and a few small observations that follow from past results.
It also outlines the list-decoding argument used to convert a fast bit-error decay into a vanishing block-error probability, including the weight-enumerator estimate extracted from~\cite{Abbe-focs23}.
\item Section~\ref{sec:bec}, Two-Look Analysis for the BEC, applies the basic idea of the analysis to doubly transitive codes on the BEC.
It is pedagogical in that it provides a simpler introduction to the method even though the resulting bound is weaker than the one in~\cite{Kudekar-it17}.
\item Section~\ref{sec:bms}, Two-Look Analysis for BMS Channels, applies the same idea to any BMS channel and provides a relatively strong bound on the bit-error probability for nested sequences of doubly-transitive codes with two looks per stage.
It is novel in the sense that it extends results for RM codes in~\cite{Abbe-focs23} to a wider class of codes. 
\item Section~\ref{sec:aobf}, Analysis of Boolean Functions, provides an overview of the analysis of boolean functions and generalizes the symmetry result from~\cite{Abbe-focs23} that forms the backbone of our analysis.
It also extends some known results (e.g., the ``level-$k$'' inequality) to our setting and shows how this can be used to achieve faster decay rates for codes with additional symmetry.
\item Section~\ref{sec:bsc}, Block Recovery for RM Codes on the BSC, provides novel results that show how our basic approach can be applied to the BSC to achieve faster decay rates via the  ``level-$k$'' inequality and hypercontractivity.
\item Section~\ref{sec:product_rm}, RM Product Codes Achieve Capacity, extends the recursive analysis to the product codes with \RM\ component codes.
These codes are transitive but not doubly transitive in general; nevertheless, the same two-look and three-look recursions apply and give a bit-error probability decaying exponentially in $\sqrt{\ln N}\ln\ln N$ on the BSC at rates arbitrarily close to capacity.
A two-pass row--column list-decoding argument establishes vanishing block-error probability.
\item Section~\ref{sec:tzs}, From BSC Block Error to BMS Block Error, demonstrates how results from~\cite{Tillich-cpc00} and~\cite{Sasoglu-phd11} can be combined to show that achieving capacity on the BSC with respect to block-error probability is sufficient to achieve capacity on any BMS channel under mild growth conditions on the minimum distance. 
\item Some proofs and definitions are relegated to the Appendix including the glossary of notation in Appendix~\ref{app:gon}.
\end{itemize}

\begin{figure*}[!t]
\centering
\includegraphics[scale=0.95]{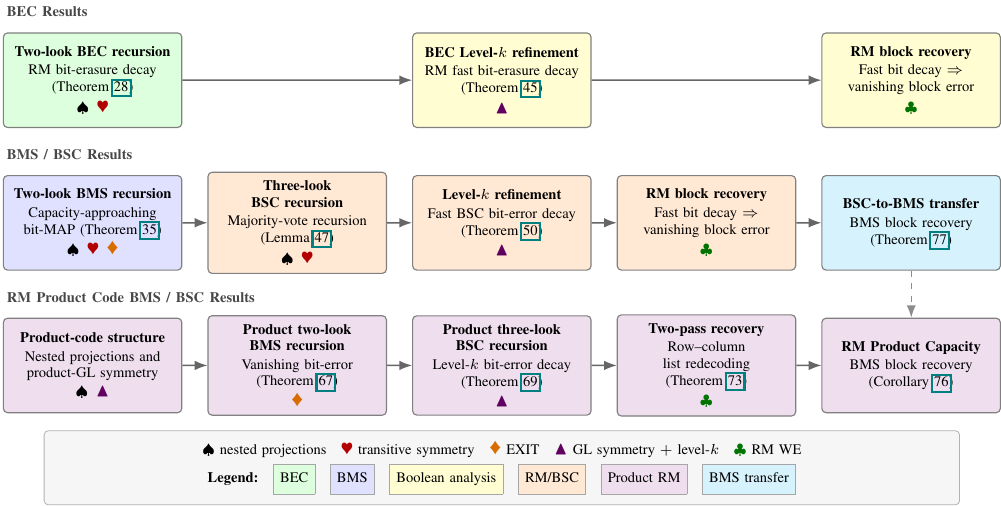}
\caption{Proof map for the principal results. Solid arrows follow the proof progressions for the BEC, BMS/BSC, and product case.
The dashed arrow indicates that the general BSC-to-BMS transfer is used for the RM product code.
Box colors identify the manuscript section containing each displayed result, while the markers indicate connections to key lemmas.}
\label{fig:proof-map}
\end{figure*}
 
\section{Background and Review} \label{sec:bg}

 \subsection{Standard Notation} \label{sec:std_notation}

The real numbers are denoted by $\mathbb{R}$.
The natural numbers are denoted by $\mathbb{N} \coloneqq \{1,2,\ldots\}$ and $\mathbb{N}_0 \coloneqq  \mathbb{N} \cup \{0\}$.
For $N \in \mathbb{N}_0$, a range of natural numbers is denoted by $[N] \coloneqq \{0,1,\ldots,N-1\}$.
The finite field with $q$ elements, where $q$ is a prime power, is denoted by $\mathbb{F}_q$. 
The set of permutations on $N$ elements (i.e., bijective functions mapping $[N]$ to $[N]$) is denoted by $\mathbb{S}_N$.
 For a set $S$, let $2^S$ denote the set of all subsets (i.e., the power set) of $S$.

For a set $\cX$, the $N$-element vector $\bx \in \cX^N$ is denoted by boldface and is indexed from 0 so that $\bx=(x_0,\ldots,x_{N-1})$.
For an $M$-element index set $A=\{a_0,a_1,\ldots,a_{M-1}\}\subseteq [N]$ with $a_0<a_1<\cdots<a_{M-1}$, we define the subvector $x_A = (x_{a_0},x_{a_1},\ldots,x_{a_{M-1}}) \in \cX^M$ without using boldface.
Likewise, $x_{\sim i}$ is shorthand for $x_{[N]\setminus\{i\}}$ and $x_{\sim A}$ is shorthand for $x_{[N]\setminus A}$.
The reordering of a vector $\bx \in \cX^N$ by a permutation $\pi \in \mathbb{S}_N$ is denoted by $\pi \bx$ and defined\footnote{Our convention is based on converting $\pi$ to a permutation matrix $C_\pi$ whose $i$-th column is zero except for  a one in row $\pi(i)$ and applying this matrix to the column vector $\bm{x}=(x_0,\ldots,x_{N-1})$ to get $\pi \bm{x} \coloneqq C_\pi \bm{x} = (x_{\pi^{-1}(0)},\ldots,x_{\pi^{-1}(N-1)})$.} by $(\pi \bx)_i = x_{\pi^{-1}(i)}$.
For a set of vectors $\cC \subseteq \cX^N$, we can apply $\pi$ to each element to get the new set of vectors $\pi \cC = \{ \bx \in \cX^N \,|\, \pi^{-1}\bx \in \cC \}$.
A single random variable is denoted by a capital letter (e.g., $X,Y,Z$). Vectors of random variables are denoted by boldface capital letters (e.g., $\bX,\bY,\bZ$). For a bounded random variable $X$, the $L^p$-norm ($p \ge 1$) is denoted by  $\|X\|_p \coloneqq ( \ex{|X|^p})^{1/p}$.

\subsection{Preliminaries}

To begin, we recall some mostly standard definitions.
 
\begin{defn} \label{def:blc}
A \emph{linear code} $\mathcal{C} \subseteq \mathbb{F}_q^N$ is a subspace of the vector space $\mathbb{F}_q^N$ over the scalar field $\mathbb{F}_q$.
The \emph{code rate} of $\cC$ is defined to be $R(\cC) = \dim (\cC) / N$, where $\dim{\cC}$ equals the dimension of the subspace $\cC$.
\end{defn}

 \begin{defn} \label{def:code_equivalence}
Two codes $\cC, \cC' \subseteq \mathbb{F}_q^N$ are called equal if they have exactly the same set of codewords.
They are called \emph{equivalent} if there is a permutation $\pi \in \mathbb{S}_N$ such that, $\cC' = \pi \cC$.
\end{defn}
 
\begin{defn} \label{def:code_auto}
For a code $\cC \subseteq \cX^N$ over a finite set $\cX$, the \emph{permutation automorphism group} $\cG$ is
\[\cG \triangleq \{ \pi\in \mathbb{S}_N \,|\, \forall \bc \in \cC, \pi \bc \in \cC \}. \]
We say a code is (doubly) transitive if its automorphism group is (doubly) transitive.
\end{defn}

\begin{defn} \label{def:code_proj_nest}
For a code $\cC \subseteq \cX^N$, we define its \emph{projection} onto a subset $A \subseteq [N]$ of coordinates by
\[ \cC |_A \coloneqq \{ \bc_A \in \cX^{|A|} \, | \, \bc \in \cC \}. \]
A code $\cC' \subseteq \cX^{|A|}$ is said to be \emph{nested} inside $\cC$ at $A$ if $\cC|_A = \cC'$.
\end{defn}

 \begin{defn} \label{def:binary_rm}
Let the set of $m$-variate polynomials with coefficients in $\mathbb{F}_{2}$ and degree at most $r$ be
\[ \cF (r,m) \coloneqq \left\{ \sum_{ \bi\in\{0,1\}^{m}:\left| \bi\right|\leq r} \!\!\! a_{\bi}\left(\prod_{j=0}^{m-1}x_{j}^{i_{j}}\right) \,\middle|\,a_{\bi}\in\mathbb{F}_{2}\right\}, \]
where $\left| \bi\right|\triangleq\sum_{j=0}^{m-1}i_{j}$.
We note that powers other than 0 and 1 are not included in $\cF(r,m)$ because $x^{2}=x$ for all $x\in\mathbb{F}_{2}$.

The \emph{binary Reed--Muller code} $\RM(r,m)$ of length $N=2^{m}$ is constructed by evaluating each polynomial in $\cF(r,m)$ at all points $\bx \in \mathbb{F}_2^m$ to get
\begin{align*}
\tca \RM(r,m) \tcb &\,\coloneqq \! \left\{ \bc\in\mathbb{F}_{2}^{N}\,\middle|\, \! f\in\mathcal{F}(r,m),\, \! c_{\ell}=f\big(\theta_{m}(\ell)\big) , \ell\in[N]  \right\},
\end{align*}
where we define $\theta_{m}\colon[N]\to\mathbb{F}_{2}^{m}$ to be the bijective function that maps the integer $\ell\in[N]$ to its binary expansion $ \bb\in\mathbb{F}_{2}^{m}$.
We assume that $b_{0}$ is the LSB and $b_{m-1}$ is the MSB and, for example, this implies that $\theta_{5}(19)=(b_0,\ldots,b_4)=(1,1,0,0,1)$. 
\end{defn}

\begin{defn} \label{def:bec}
The \emph{binary erasure channel} with erasure rate $p$ (or BEC($p$)) is a stochastic mapping from the input alphabet $\cX = \{0,1\}$ to the output alphabet $\cY = \{0,1,*\}$.
For this channel with input $x$, the output equals $x$ with probability $1-p$ and it equals $*$ otherwise (i.e., with probability $p$).
\end{defn}

\begin{defn} \label{def:bsc}
The \emph{binary symmetric channel} with error rate $p$ (or BSC($p$)) is a stochastic mapping from the input alphabet $\cX = \{0,1\}$ to the output alphabet $\cY = \{0,1\}$.
For this channel with input $x$, the output equals $x$ with probability $1-p$ and it equals $1-x$ otherwise (i.e., with probability $p$).
\end{defn}
 
\begin{defn} \label{def:bms}
A \emph{binary memoryless symmetric channel} (or BMS channel) $W$ is a stochastic mapping from the input alphabet $\cX = \{\pm 1\}$ to the output alphabet $\cY = \mathbb{R}$.
 Let $X\in \cX$ be the input random variable and $Y\in \cY$ be the output random variable with the transition probability given by $\Pr(Y\in A \,|\, X=x) \coloneqq W(A \mid x)$ for any Borel measurable set $A\subseteq \mathbb{R}$. 
Channel symmetry is enforced by the condition $W(A \mid -\!1)=W(-A \mid 1)$~\cite[p.~178]{RU-2008}.
From this, we also have the sign-flip decomposition $Y=XZ$ for BMS channels where $Z$ is independent of $X$ and distributed according to $W(A \mid 1)$~\cite[p.~182]{RU-2008}. 
This also implies that $Y=0$ represents an erasure because $W(\{0\}\mid 1)=W(\{0\} \mid -\!1)$.
 \end{defn}

\subsection{Decoding Functions}
\label{sec:dec_fun}

A common theme shared by the results in~\cite{Kudekar-it17,Kumar-itw16,Reeves-it23,Reeves-isit23,Abbe-focs23} is to analyze the decoding function for a single distinguished symbol of a codeword drawn from a code with transitive symmetry.
Since the code is transitive, the choice of the distinguished symbol is arbitrary and the first symbol is chosen for simplicity.
In particular, let $\bX \in \cX^N$ be a uniform random codeword from a transitive code $\cC \subseteq \cX^{N}$ and $\bY \in \cY^N$ be an observation of $\bX$ through a noisy channel.
Then, the focus is on an optimal decoding function for $X_0$ given the \emph{extrinsic} observation $Y_{\sim 0} = (Y_1,Y_2,\ldots,Y_{N-1})$ which does not reveal $Y_0$.

 In~\cite{Kudekar-it17,Kumar-itw16}, this is applied to linear codes over erasure channels.
Thus, the channel input alphabet satisfies $\cX = \mathbb{F}_q$ and the output alphabet $\cY = \cX \cup \{*\}$, where $*$ indicates erasure.
In this case, the posterior distribution of $X_0$ given $Y_{\sim 0}$ is either uniform (i.e., it contains no information about $X_0$) or it is concentrated on a single value (i.e., the correct value is recovered without error).
In addition, one can determine which of these two possibilities will occur by looking only at the locations of the erasures and not at the unerased values.
Let the \emph{erasure indicator} vector $\bZ \in \{0,1\}^N$ be defined by $Z_i = 1$ if $Y_i = *$ and $Z_i = 0$ otherwise.
Then, the analysis focuses on a boolean function $f\colon \{0,1\}^{N-1} \to \{0,1\}$ which maps $Z_{\sim 0}$ to 1 if $X_0$ cannot be recovered and 0 otherwise.
 
In~\cite{Reeves-it23}, the focus is on binary-input memoryless symmetric (BMS) channels with $\cX = \{\pm 1\}$ and $\cY = \mathbb{R}$.
To model BMS channels~\cite{RU-2008}, it suffices to define $\bZ \in \mathbb{R}^N$ to be a vector of i.i.d.\ random variables (where the distribution depends on the exact channel) and then let $Y_i = X_i Z_i$.
Thus, the output $\bY$ consists of i.i.d.\ random variables whose signs are modulated by $\bX$.
One natural decoding function of interest is the \emph{extrinsic conditional mean} $f\colon \mathbb{R}^{N-1} \to [-1,1]$ defined by
\begin{equation} \label{eq:cond_exp_dec_fun}
f(y_{\sim 0}) = \ex {X_0 \, | \, Y_{\sim 0} = y_{\sim 0}}.
\end{equation}
This function has many nice properties that facilitate the proof in~\cite{Reeves-it23}.
In~\cite{Reeves-isit23}, the approach from~\cite{Reeves-it23} is generalized to linear codes over $\mathbb{F}_q$ by focusing on decoding functions that return the posterior distribution of $X_0$ given $Y_{\sim 0}$.

In~\cite{Abbe-focs23}, the focus is on binary linear codes over the binary symmetric channel (BSC) with $\cX = \cY = \{0,1\}$.
The decoding function $f\colon \{0,1\}^{N-1} \to \{0,1\}$ is chosen to be a maximum-likelihood decoding function that maps $Y_{\sim 0}$ to the most likely value of $X_0$.
While ties are possible in this case, they can be broken arbitrarily without affecting the analysis.

A key element in these results is the idea that a symmetric code $\cC$ will give rise to a symmetric decoding function $f$.
The following definition is used to make this precise.
\begin{defn} \label{def:symf}
Let $\cY$ be a set and $f\colon \cY^n \to \mathbb{R}$ be a function.
The \emph{symmetry group} of $f$ is defined to be
\[ \mathrm{Sym}(f) \triangleq \{ \pi \in \mathbb{S}_n \, |\, \forall \by \in \cY^n, f(\pi \by) = f(\by) \}. \]
\end{defn}

\begin{rem}
The decoding functions defined above are chosen for their mathematical properties without regard for efficient evaluation.
A related line of work is to design and analyze practically implementable decoding functions for related families of codes~\cite{Ye-it20,Rameshwar-arxiv24,Siddheshwar-isit24}.
\end{rem}

\subsection{Erasure Channels}

Consider a sequence of doubly-transitive linear codes over $\mathbb{F}_q$ with increasing length and rate converging to $R \in (0,1)$.
The proof technique from~\cite{Kudekar-stoc16,Kudekar-it17} shows that such a sequence achieves capacity on the symbol erasure channel under bit-MAP decoding.

The proof technique utilizes the erasure indicator decoding function described in Section~\ref{sec:dec_fun} and consists of the following steps.
First, it is observed that the symbol erasure rate after decoding is the expectation of a symmetric monotone boolean function.
Using this fact, a celebrated result from the theory of boolean functions~\cite{Talagrand-ap94,Friedgut-procams96} is applied to show that, for any $\delta>0$, the symbol erasure rate changes from $\delta$ to $1-\delta$ as the channel erasure probability increases by $O(1/\ln N)$.
Finally, the extrinsic information transfer (EXIT) area theorem is applied to show that this transition point converges to the maximal erasure rate allowed by Shannon's channel coding theorem, which is $1-R$.

For code families such as $\RM$ codes and primitive narrow-sense BCH codes, a few more steps show that the block failure probability also vanishes below the same erasure threshold point.
A little later, the symmetry condition required to achieve capacity under bit-MAP decoding was relaxed to something only a bit stronger than transitivity~\cite{Kumar-itw16}.
This extension allowed its application to additional code families~\cite{Natarajan-it23}.
 
\subsection{Beyond Erasures: Multiple Looks via Nested Codes}
\label{sec:mult_looks}

All current techniques used to prove sequences of RM codes achieve capacity on non-erasure channels are based on code sequences satisfying some type of nesting property~\cite{Reeves-it23,Reeves-isit23,Abbe-focs23}.
In particular, to prove that a sequence of doubly-transitive linear codes $\cC_k \subseteq \smash{\mathbb{F}_q^{N_k}}$ achieves capacity on a BMS channel, the papers \cite{Reeves-it23,Reeves-isit23}~implicitly use an auxiliary sequence of transitive codes $\cB_k \subseteq \smash{\mathbb{F}_q^{N_k '}}$ satisfying  $N_k ' > N_k$ and $R(\cB_k) < R(\cC_k)$.
For these sequences, it is required that each $\cC_k$ can be found in two distinct ways as a nested code inside $\cB_k$.
\begin{defn}[Nesting]
We say that a code sequence $\cC_k \subseteq \smash{\mathbb{F}_q^{N_k}}$ satisfies the \emph{weak nesting} property if there is an auxiliary code sequence $\cB_k \subseteq \smash{\mathbb{F}_q^{N_k '}}$ with $N_k ' > N_k$ satisfying: for each $k\in \mathbb{N}_0$, there are distinct subsets $A,B \subseteq [N_{k}']$ with $|A|=|B|=N_k$ such that the projections $\cB_k |_A$ and $\cB_k |_B$ are both equivalent to $\cC_k$.
The \emph{normalized overlap} between these two nested copies is denoted by $\rho_k = |A \cap B|/|A|$.
If we also have $\cB_k = \cC_{k+1}$, then we say the sequence satisfies the \emph{strong nesting} property.
\end{defn}
These nesting conditions provide two estimates of any bit whose index is in $A\cap B$.
For doubly transitive codes, we can upper bound correlation between these estimates in terms of the normalized overlap $\rho_k$.

 While the proofs in~\cite{Reeves-it23,Reeves-isit23} rely on the weak nesting property, the approach taken in this work requires the strong nesting property.
The proof in~\cite{Abbe-focs23} essentially uses the strong nesting property as well but does so less explicitly.
The idea is that, if $\cC_k$ has bit-error probability strictly below $1/2$, then the two observations can be combined to show that the longer code $\cB_k$ has a much lower bit-error probability.

The proofs in~\cite{Reeves-it23,Reeves-isit23} also require that the difference in code rate (i.e., $\Delta_k = R(\cC_k) - R(\cB_k)$) vanishes.
From a high-level point of view, this is required by all of these approaches because one needs the rate of $\cC_k$ to be just below capacity in order to have the bit-error probability less than $1/2$ and, if the rate difference does not vanish, then the rate of $\cB_k$ will have a strict gap from capacity.
In~\cite{Pfister-arxiv25c}, we provide a tutorial introduction to the proofs in~\cite{Reeves-it23,Reeves-isit23} with some slight improvements. 
In that work, the normalized overlap between the two nested codes can approach 1 but not too quickly (e.g., $\Delta_k=o(1-\rho_k)$).
 
We now describe a sequence of $\RM$ codes that satisfies these conditions and the strong nesting condition.
In particular, we consider the code sequence $\cC_k = \RM(r_m,m+k)$ where $r_m = \lfloor m/2 + \gamma \sqrt{m} \rfloor$ and let $\cB_k = \cC_{k+1}$.
The central limit theorem implies that, for fixed $k$ with $m$ increasing, the code rate $R(\cC_0)$ converges to $\Phi(2\gamma)$ where $\Phi(z) \coloneqq (2\pi)^{-1/2} \int_{-\infty}^z \exp( - u^2/2) \, du$ is the cumulative distribution function of a standard Gaussian random variable~\cite[Remark~24]{Kudekar-it17}.
Lemma~\ref{lem:rate} below formalizes a few related bounds on code rates for this sequence~\cite[Lemma~8]{Reeves-it23}. 
We note that each code $\cC_k$ is doubly transitive because all $\RM$ codes are doubly transitive.
Next, we recall from~\cite[Lemma~9]{Reeves-it23} that two partially overlapping copies of $\RM(r,m+k)$ can be found in $\RM(r,m+k+1)$.
\begin{lem} [{\cite[Lemma~8]{Reeves-it23}}]  \label{lem:rate}
For $\cC = \RM(r,m)$, the rate is bounded by
\begin{equation} \label{lem:rate_vs_phi}
\left|R(\cC)-\Phi\left(\frac{2 r-m}{\sqrt{m}}\right) \right| \le \frac{1}{\sqrt{2 \pi  m}}.
\end{equation}
Moreover, if $r =\lfloor m/2+ \sqrt{m}\,\Phi^{-1} (R) /2 \rfloor$, then 
\begin{equation} \label{lem:rate_vs_floor}
R -\frac{3}{\sqrt{2\pi m}} \leq R(\cC) \leq R +  \frac{1}{\sqrt{2\pi m}}.
\end{equation}
Finally, for $\cC_k = \RM(r,m+k)$, one gets  $R(\cC_{k}) \geq R(\cC_0) - \frac{k}{2\sqrt{m}}$.
\end{lem}

 \begin{proof}
See Appendix~\ref{proof:lem:rate}.
\end{proof}

To describe the nesting property in detail, we index the bits of $\cB_k = \cC_{k+1}$ by $A = [2^{m+k+1}]$ and define $B = [2^{m+k}] \subset A$ and $C = \left([2^{m+k-1}] \cup ([2^{m+k-1}]+2^{m+k}) \right) \subset A$.
 From Definition~\ref{def:binary_rm}, we see that $\theta_{m+k+1}(B) = \mathbb{F}_2^{m+k} \times \{0\}$ corresponds to evaluating the polynomials in $\cF(r,m+k+1)$ on the subset of points where $x_{m+k} = 0$.
Likewise, we see that $\theta_{m+k+1}(C) = \mathbb{F}_2^{m+k-1} \times \{0\} \times \mathbb{F}_2$ corresponds to evaluating the polynomials in $\cF(r,m+k+1)$ on the subset of points where $x_{m+k-1} = 0$.
 The size of the normalized overlap between these two sets is $\rho_k = |B \cap C|/|B| = 1/2$.
Thus, these two nested codes can be used to provide weakly correlated estimates of bit-0.
We also note that this RM code sequence satisfies the strong nesting property $\cB_k = \cC_{k+1}$. 

\subsection{Multiple Looks, Subspace Codes, and Sunflowers}
\label{sec:multiple_looks}

 The projection of $\RM(r,m)$ onto a set of $2^{m-t}$ coordinates, whose binary expansions in $\mathbb{F}_2^m$ form an affine subspace of dimension $m-t$, is equivalent to $\RM(r,m-t)$.

For the analysis of RM codes on BMS channels, this property was used in~\cite[Lemma 9]{Reeves-it23} to construct two ``$\RM(r,m-t)$ looks'' at bit-0 of an $\RM(r,m)$ code based on two observation sets with fractional overlap $2^{-t}$.
In particular, the two subspaces of $\mathbb{F}_2^m$ were chosen to be
\begin{align*}
U_0 &= \mathbb{F}_2^{m-2t} \times \mathbb{F}_2^{t} \times \{0\}^t \\
U_1 &= \mathbb{F}_2^{m-2t} \times \{0\}^t \times \mathbb{F}_2^t .
\end{align*}
Since $U_0 \cap U_1 = \mathbb{F}_2^{m-2t} \times \{0\}^{2t}$, this implies that the fractional overlap between the subspaces is indeed
\[| U_0 \cap U_1 | / | U_0 | = 2^{m-2t}/2^{m-t} = 2^{-t} . \]
The construction above generalizes naturally to give multiple looks with small fractional overlap.
\begin{defn}[Multiple Looks]
\label{def:mult_look}
For all $s,t \in \mathbb{N}$ with $st\leq m$, one can get $s$ different $\RM(r,m-(s-1)t)$ looks with fractional overlap $2^{-t}$.
In particular, the $i$-th look uses the index subspace
\[U_i = \mathbb{F}_2^{m-st} \times \{0\}^{ti} \times \mathbb{F}_2^{t} \times \{0\}^{t(s-1-i)}. \]
In this case, for all $i,j\in [s]$ with $i\neq j$, the intersection is \[ U_i \cap U_j  = \mathbb{F}_2^{m-st} \times \{0\}^{st} = \cap_{i\in [s]} U_i. \]
Thus, the fractional overlap between the subspaces is $|U_i \cap U_j|/|U_i|= 2^{m-st} / 2^{m-(s-1)t} = 2^{-t}$.
For example, if we choose $s=3$ and $t=2$, then we get 3 $\RM (r,m-4)$ looks with fractional overlap $1/4$.
\end{defn}

\begin{rem}
In~\cite{Abbe-focs23}, the two-look approach was generalized to construct a much larger collection of subspaces called a \emph{subspace sunflower}.
That construction significantly improves the trade-off between the number of looks and their quality.
 In particular, for all $t \in \mathbb{N}$ such that $2t\leq m$, there are actually $2^t+1$ $\RM(r,m-t)$ looks with fractional overlap $2^{-t}$.
In \cite{Abbe-focs23}, a Gilbert-Varshamov type argument is used to construct subspace sunflowers with good parameters.
In Appendix~\ref{app:sunflower},  we also outline a close connection to subspace codes that provides deterministic optimal constructions.
 \end{rem}

The new proof of vanishing block-error probability provided in this paper uses only the simple multiple-look construction in Definition~\ref{def:mult_look} and does not rely on the sunflower construction.
For tensor products of two \RM\ codes, which are studied in Section~\ref{sec:product_rm}, the construction in Definition~\ref{def:mult_look} can be performed independently in two coordinate spaces to define subspaces that give multiple rectangular looks.

\begin{rem} \label{rem:floral_block}
At present, significantly improved bounds based on subspace sunflowers have yet to be realized.
For example, this work shows that roughly the same error rate is achieved without using subspace sunflowers.
However, if one could intrinsically improve the correlation bound between sunflower petal estimates, then one could combine these with the subspace sunflower to provide a direct proof of vanishing block-error probability without even relying on weight enumerator arguments.
\end{rem}

\subsection{Binary Memoryless Symmetric (BMS) Channels}

The first proof showing that binary $\RM$ codes achieve capacity on BMS channels under bit-MAP decoding can be found in~\cite{Reeves-it23,Reeves-arxiv21}.

The proof embeds an arbitrary BMS channel $W$ with capacity $C$ into a continuous family $W_t$ of BMS channels parameterized by $t\in[0,1]$ where $W_{1-C} = W$.
For $t \in [0,1]$, this embedding is such that the capacity of $W_t$ equals $1-t$ and $W_{t}$ is degraded with respect to $W_{t'}$ for $t' \in [0,t)$.
In this approach, one focuses on the decoding function in~\eqref{eq:cond_exp_dec_fun} for sequences of RM codes with increasing length and converging rate.
It is shown that, as a function of $t$,  the variance $\var{f(Y_{\sim 0})}$ converges to zero for almost all $t\in [0,1]$.

The variance bound is derived from the construction of two looks described in Section~\ref{sec:mult_looks}. 
From that, it is established that  $\mmse(X_0|Y_{\sim 0})$, as a function of the channel parameter, converges to an increasing step function that jumps from 0 to 1.
Finally, a modified version of the generalized EXIT area theorem is used to show that the jump must occur at the point where the channel capacity equals the code rate.

In~\cite{Reeves-isit23}, the original approach is simplified and generalized to $q$-ary channels using the same two-look property.
Two key differences are that the standard EXIT area theorem suffices and no channel embedding is required.
In~\cite{Abbe-focs23}, the original approach was expanded to establish an upper bound on the bit-error probability that decays more quickly with block length.
The argument focuses on the BSC and uses the maximum-likelihood decoding function $f$ (described in Section~\ref{sec:dec_fun}) and a bound on the variance of $\ex{f(Y_{\sim 0})|Y_A}$ for a convenient choice of $A$ (c.f., Lemma~\ref{lem:f_bool_restrict_var}).
It also introduces the concept of a subspace sunflower to construct a large number of weakly correlated estimates of bit-0.
In combination with a list-decoding argument, an even stronger bound was used to establish that the block-error probability must also vanish at the point where the channel capacity equals the code rate.

The results contained herein show that the same improvements can be achieved by combining the variance bound in~\cite{Abbe-focs23} with the standard ``level-$k$'' inequality for boolean functions without relying on any floral subspaces.
In~\cite{Pfister-arxiv25c}, a tutorial introduction is provided for the BEC using slightly improved versions of the proofs in~\cite{Reeves-it23,Reeves-isit23}.

\subsection{Correlation Bound for Boolean Functions}

Let $f\colon \{0,1\}^n \to \mathbb{R}$ be a real function of boolean variables and let $\bX \in \{0,1\}^n$ be a vector of independent Bernoulli random variables.
Let the sets $A,B \subseteq [n]$ form a partition of $[n]$.
We also treat $f$ as a function of two variables by writing $f(\bx) = f(x_A, x_B)$.
The \emph{random restriction} $f_A : \{0,1\}^{|A|} \to \mathbb{R}$ of $f$ to the subset $A\subseteq [n]$ is formed by averaging over the variables indexed by $[n]\setminus A$ and defined by
\begin{equation} \label{eq:fA}
f_A (x_A) \triangleq \ex{f(X_A,X_B)\,|\,X_A = x_A}.
\end{equation}
For more background on boolean functions, see Appendix~\ref{app:bfa} and~\cite{ODonnell-2014}.

\begin{lem} \label{lem:f_bool_corr} Let $f\colon \{0,1\}^n \to \mathbb{R}$ be a function where the sets $A,B \subseteq [n]$ form a partition of $[n]$.
Let $X,X' \in \{0,1\}^n$ be random vectors such that, given $X_A$, the vectors $X_B$ and $X_B '$ are conditionally independent and identically distributed.
If, for some $\rho \in [0,1]$, we have
\[\var{\ex{f(X_A,X_B)|X_A}} \leq \rho \, \var{f(X_A,X_B)},\]
then we find that
\begin{align*}
&\ex{f(X_A,X_B) \, f(X_A,X_B')} \leq \tcb \tca
\tc{\qquad} \rho \, \ex{f(X_A,X_B)^2} + (1-\rho) \ex{f(X_A,X_B)}^2.
\end{align*}
\end{lem} 

\begin{proof}
Note that $X_A '$ does not play any role in this proof and a concrete choice satisfying the conditions is when $X,X'$ are independent with i.i.d.\ entries.
From this setup, we have
\begin{align*}
\tca \ex{f(X_A,X_B) \, f(X_A,X_B')} \tcb
& \;= \ex{\ex{f(X_A,X_B) f(X_A,X_B')|X_A}\vphantom{\ex{a}^2}} \\
&\;= \ex{\ex{f(X_A,X_B)|X_A} \ex{f(X_A,X_B')|X_A}\vphantom{\ex{a}^2}} \\
&\;= \ex{\ex{f(X_A,X_B)|X_A}^2} \\
&\;= \underbrace{\left(\ex{\ex{f(X_A,X_B)|X_A}^2} - \ex{f(X_A,X_B)}^2 \right)}_{\var{\ex{f(X_A,X_B)|X_A}}} \tca \tcb \tca \tc{\qquad\qquad} + \, \ex{f(X_A,X_B)}^2  \\
&\;\leq \rho \underbrace{\left(\ex{f(X_A,X_B)^2} - \ex{f(X_A,X_B)}^2 \right)}_{\var{f(X_A,X_B)}} \tca \tcb \tca \tc{\qquad\qquad} + \, \ex{f(X_A,X_B)}^2  \\
&\;= \rho \, \ex{f(X_A,X_B)^2} + (1-\rho) \ex{f(X_A,X_B)}^2. \tag*{\qedhere}
\end{align*}
\end{proof}

\begin{rem}
This inequality provides a linear interpolation in $\rho$ between the case where $f(X_A,X_B)$ and $f(X_A,X_B')$ are independent (i.e., the expectation of the product is the product of the expectations) and the case where they are perfectly correlated.
To get the first case, $f(X_A,X_B)$ must essentially depend only on $X_B$ and one has $\rho=0$.
To get the second case, $f(X_A,X_B)$ must essentially depend only on $X_A$ and one has $\rho=1$.
\end{rem}

\begin{rem}
We will see in Section~\ref{sec:fourier}, that the symmetry of $f$ can be used to compute a $\rho$ such that
\[ \var{\ex{f(X_A,X_B)|X_A}} \leq \rho \, \var{f(X_A,X_B)}.\]
In particular, if the function $f$ has transitive symmetry, then this bound holds for $\rho = |A|/(|A|+|B|)$ (e.g., see Corollary~\ref{cor:symmetry_bounds}).
\end{rem}

\subsection{Low-Weight Competitors and List Decoding}
\label{sec:list_dec}

A vanishing bit-error probability for a code sequence implies a vanishing block-error probability if the former decays as $o(d_m/N_m)$, where $d_m$ and $N_m$ are the minimum distance and blocklength of the $m$-th code~\cite{Kudekar-it17}.
Known bounds for the bit-error probability of RM codes do not decay this quickly, so one instead uses bounds based on the weight enumerator~\cite{Kudekar-isit16,Abbe-focs23}.
One successful approach is to use the bit decisions to define a list of nearby codewords from which the most likely codeword is chosen using the original channel observations.
For a linear code, the resulting block-error probability is governed by the probability that some low-weight shift of the transmitted codeword is more likely than the transmitted codeword~\cite{Abbe-focs23}.

This subsection collects the facts about that event that are used in Sections~\ref{sec:aobf},~\ref{sec:bsc}, and~\ref{sec:product_rm}.

For the input bit $b\in\{0,1\}$, let $W(y\mid b)$ be the transition law (or density) for a BMS channel when the output is $y\in \cY$ and the BPSK input is $(-1)^b$. Let
\begin{equation} \label{eq:bhattacharyya}
\beta \triangleq \sum_{y \in \cY} \sqrt{W(y\,|\,0)\,W(y\,|\,1)}
\end{equation}
denote its Bhattacharyya parameter, where the sum is replaced by an integral for continuous output alphabets.
For a binary vector $\bc$ of length $N$, we write $W^N (\by \,|\, \bc) = \prod_{i \in [N]} W(y_i \,|\, c_i)$ for the probability of receiving $\by$ when $\bc$ is transmitted over $N$ uses of $W$, and $\dH(\cdot,\cdot)$ denotes Hamming distance.
For a binary linear code $\cC$, let $A_w (\cC)$ be the number of codewords of Hamming weight $w$ and, for $\beta \in [0,1]$ and  $h \geq 0$, define the \emph{truncated union bound}
\begin{equation} \label{eq:competition_function}
T_{\cC,\beta}(h) \triangleq \sum_{w=1}^{\lfloor h \rfloor} A_w (\cC)\, \beta^w .
\end{equation}

The following lemmas encapsulate the argument to bound the block-error probability of RM codes in~\cite{Abbe-focs23}.
\begin{lem} \label{lem:low_weight_competition}
Let $\cC$ be a length-$N$ binary linear code and let $\bY$ be the output of a BMS channel, with Bhattacharyya parameter $\beta$, when the random codeword $\bX \in \cC$ is transmitted.  For $h\geq0$, let $E_h$ be the event that a different codeword within distance $2h$ of $\bX$ is at least as likely as $\bX$. Then,
\[  \pr{E_h} \leq T_{\cC,\beta}(2h).\]
\end{lem}

\begin{proof}
Condition on $\bX = \bx$ and observe that translation invariance of a linear code and of a BMS channel imply that the resulting probability does not depend on $\bx$.
Thus, we may assume $\bX = \bm{0}$.
For a fixed nonzero $\bc$ of weight $w$, it is well-known~\cite[Chap.~5]{Gallager-1968} that
\begin{align*}
\tc{\phantom{a}} \tca \tc{\!\!\!\!\!\!} \pr{W^N(\bY\mid \bc)\geq W^N(\bY\mid\bm{0})} \tcb
&\leq \sum_{\by}\sqrt{W^N(\by\mid\bm{0})W^N(\by\mid \bc)} \\
&=\bigg(\sum_{y \in \cY} \sqrt{W(y\mid 0)W(y\mid 1)}\bigg)^{\! w}=\beta^w.
\end{align*}
Indeed, for the likelihood-comparison event $W^N(\by\mid\bm{0})\leq\sqrt{W^N(\by\mid\bm{0})W^N(\by\mid \bc)}$, one can enlarge the sum to all $\by$ to prove the first inequality.
The equality then follows because the two conditional distributions differ only in the $w$ coordinates where $\bc$ is nonzero.
A union bound over the codewords of weight at most $2h$ gives the stated result.
\end{proof}

The next lemma makes no assumptions about how the candidate vector is produced by the first-stage decoder.
\begin{lem} \label{lem:list_success}
In the setting of Lemma~\ref{lem:low_weight_competition}, let $\widehat \bX \in \{0,1\}^N$ be an arbitrary random vector on the same probability space and fix $h\geq 0$.
Consider the list decoder that returns
\[ \overline \bX \in \arg\max_{\bc \in \cC \,:\, \dH (\bc,\widehat \bX) \leq h} W^N (\bY \mid \bc), \]
where an empty list or an unresolved tie is declared a failure.
Then, we have
\[ \pr{\overline \bX \neq \bX} \leq \pr{\dH (\widehat \bX, \bX ) > h} + T_{\cC,\beta}(2h). \]
\end{lem}

\begin{proof}
Suppose $\dH (\widehat \bX, \bX) \leq h$ so that the transmitted codeword lies in the list.
If the decoder nevertheless fails, then there is a $\bc \neq \bX$ on the list satisfying $W^N(\bY \mid \bc) \geq W^N (\bY\mid \bX)$ and the triangle inequality gives
\[ \dH (\bc,\bX) \leq \dH (\bc,\widehat \bX) + \dH (\widehat \bX,\bX) \leq 2h. \]
Thus, the failure event is in the union of $\{\dH (\widehat \bX, \bX) > h\}$ and the competition event of Lemma~\ref{lem:low_weight_competition}.
\end{proof}

It remains to bound $T_{\cC,\beta}(h)$ for RM codes.
The following weight-enumerator estimate is established in the proof of~\cite[Lemma~9]{Abbe-focs23} and it is the only property of RM weight enumerators used in this paper.
We note that $i$ indexes an arbitrary RM code sequence and not the recursion stage $k$.

\begin{lem}[{\cite[Proof of Lemma~9]{Abbe-focs23}}] \label{lem:as_weight_enum}
Fix $\beta \in (0,1)$ and $c>0$.
Let $(r_i)$ and $(\tilde m_i)$ be sequences of positive integers with $\tilde m_i \to \infty$ and $|\tilde m_i - 2 r_i | = O(\sqrt{\tilde m_i})$.
Then, we have
\begin{equation} \label{eq:as_weight_enum}
T_{\RM(r_i,\tilde m_i),\beta} \Big( 2^{\tilde m_i - c\sqrt{\tilde m_i}\log_2 \tilde m_i} \Big) = O\Big( 2^{-2^{\tilde m_i/3}} \Big).
\end{equation}
\end{lem}

Since $T_{\cC,\beta}(h)$ is non-decreasing in $h$ and $c>0$ is arbitrary, the estimate in~\eqref{eq:as_weight_enum} also bounds $T_{\RM(r_i,\tilde m_i),\beta}(h_i)$ for any cutoff sequence satisfying $h_i \leq 2^{\tilde m_i - \nu \sqrt{\tilde m_i}\log_2 \tilde m_i }$ for fixed $\nu>0$. The applications in Sections~\ref{sec:aobf},~\ref{sec:bsc}, and~\ref{sec:product_rm} verify this condition explicitly.

\section{Two-Look Analysis for the BEC}
\label{sec:bec}

\subsection{Setup}

We start by focusing on the BEC because this turns the extrinsic decoder into a monotone boolean function and provides a clean setting to introduce the recursion before treating real-valued BMS decoding functions.

Consider a transitive binary linear code $\cC$ of length-$N$ with a distinguished bit.
Since $\cC$ is transitive, we assume without loss of generality that the distinguished bit has index 0.
Let the $N-1$ non-zero bit indices be partitioned into 4 disjoint sets $A,B,C,D \subseteq [N] \setminus \{0\}$ and define the code projections $\cC' = \cC|_{\{0\}\cup A\cup B}$ and $\cC'' = \cC|_{\{0\} \cup A \cup C}$.
Let $\bm{X} = (X_0,\ldots,X_{N-1}) \in \cX^N$ be a uniform random codeword from $\cC$ and let $\bm{Y} = (Y_0,\ldots,Y_{N-1}) \in \cY^N$ be an observation of $\bm{X}$ through a BEC with erasure probability $p$.
The extrinsic bit-erasure probability for code $\cC$ from BEC observations is denoted by $P_e (\cC) $ and equals the probability that bit-0 of $\cC$ is not recoverable via bit-MAP decoding from $Y_{\sim 0}$ (i.e., this is the \emph{erasure probability} of optimal extrinsic decoding).

Let $\bm{Z} = (Z_0,\ldots,Z_{N-1}) \in \{0,1\}^N$ be the erasure indicator vector for the received sequence $\bm{Y}$ where, for all $i\in [N]$, we let $Z_i = 1$ if $Y_i$ is an erasure and $Z_i = 0$ otherwise.
We define the boolean function $g: \{0,1\}^{|A|} \times \{0,1\}^{|B|} \times \{0,1\}^{|C|} \times \{0,1\}^{|D|} \to \{0,1\}$ so that $g(z_{A},z_B,z_C,z_D)$ is the indicator function of the event that bit-0 of $\cC$ is not recoverable via bit-MAP decoding from $Y_{\sim 0}$.
It is well known that this event only depends on the erasure pattern and not the transmitted codeword (e.g., see~\cite{Kudekar-stoc16}).
It follows that
\[ P_e (\cC) = \ex{g(Z_{A},Z_B,Z_C,Z_D)}. \]

Likewise, let $f\colon \{0,1\}^n \to \{0,1\}$ be the boolean function with $n=|A|+|B|$ that is the indicator function of the event that bit-0 of $\cC$ is not recoverable via bit-MAP decoding from $(Y_A,Y_B)$.
It follows that
\[ f(z_{A},z_B) = g(z_{A},z_B,\bm{1},\bm{1}) \]
and thus
\[ P_e (\cC') = \ex{f(Z_{A},Z_B)}. \]

\begin{figure}[t]
\begin{center}
\includegraphics[scale=0.9]{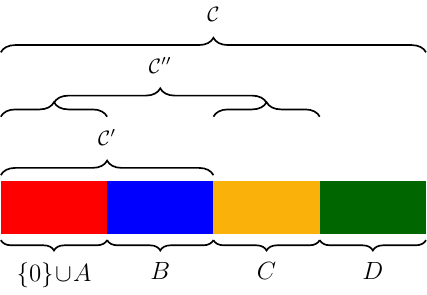}
\end{center}
\caption{Diagram showing the nesting structure of $\cC$, $\cC'$, and $\cC''$ with sizes shown for the RM code sequence.}
\end{figure}

\subsection{Recursion}

Let $G'$ be the automorphism group of $\cC'$ and $G_0' \subset G'$ be the subgroup that stabilizes bit 0 (i.e., $G_0' = \{\sigma \in G'\,|\, \sigma(0)=0\}$).
The symmetry group of the decoding function $f$ (e.g., see Definition~\ref{def:symf}) is inherited from the code $\cC'$ and each $\sigma \in G_0'$ defines a $\pi \in \mathrm{Sym}{(f)}$. One annoying issue is that the code $\cC'$ has length $N'=|A|+|B|+1$ and the decoding function $f$ has $n=N'-1 = |A|+|B|$ variables because the extrinsic decoder does not have access to $Y_0$.
Thus, mapping the code automorphism $\sigma \in G'$ to a functional symmetry $\pi \in \mathrm{Sym}(f)$ requires the awkward definition $\pi(i) = \sigma(i+1)-1$ for $i\in [n]$.

Let $F$ denote the subgroup of $\mathrm{Sym}(f)$ generated by $G_0'$.
In the sequel, $\Pi$ denotes a uniform random element of a subgroup $H \subseteq F$ that we are free to choose and, unless stated otherwise, $H=F$.
This freedom is justified because Lemma~\ref{lem:f_bool_restrict_var} holds for an arbitrary probability distribution $q$ on $\mathrm{Sym}(f)$ and the results below depend on the distribution of $\Pi$ only through the coefficient $\rho = \max_{S\subseteq [n]: S\neq \emptyset} \Pr(\Pi(S) \subseteq A )$.
For the product $\RM$ codes of Section~\ref{sec:product_rm}, choosing a convenient subgroup $H$ makes $\rho$ easy to evaluate.

\begin{lem} \label{lem:code_bound_rho}
Let $\rho = \max_{S\subseteq [n]: S\neq \emptyset} \Pr(\Pi(S) \subseteq A )$. If the two code projections are equal (i.e., $\cC'=\cC''$), then we find that
\begin{align*}
P_{e} (\cC) &\leq (1-\rho) P_e (\cC')^2 + \rho P_e (\cC') \\
\frac{P_{e}(\cC)}{1-P_{e} (\cC)}  &\leq \left(\frac{1}{2-\rho}\right)\frac{P_{e}(\cC')}{1-P_{e} (\cC')}.
\end{align*}
\end{lem} 

\begin{proof}
First, we note that combining the definition of $\rho$
with  Lemma~\ref{lem:f_bool_restrict_var} (with $q$ uniform over the chosen subgroup $H \subseteq F$) shows that \[\var{\ex{f(Z_A,Z_B)|Z_A}} \leq \rho \var{f(Z_A,Z_B)}.\]
Second, using the above setup, we have
\begin{align*}
P_e (\cC)
&= \ex{g(Z_A,Z_B,Z_C,Z_D)} \\
&\leq \ex{g(Z_A,Z_B,\bm{1},\bm{1}) g(Z_A,\bm{1},Z_C,\bm{1})} \\
&= \ex{f(Z_A,Z_B) f(Z_A,Z_C)} \\
&\leq \rho \underbrace{\ex{f(Z_A,Z_B)^2}}_{\ex{f(Z_A,Z_B)}} + (1-\rho) \ex{f(Z_A,Z_B)}^2  \\
&= \rho P_e (\cC') + (1-\rho) P_e (\cC')^2, \tag*{\qedhere}
\end{align*}
where the first inequality follows from the monotonicity of $g$ and the second inequality follows from Lemma~\ref{lem:f_bool_corr}. This gives the first stated result.
For the second stated result, we start by applying $\phi(x)=x/(1-x)$ to both sides of this expression.
Since $\phi$ is increasing on $[0,1)$, this gives the inequality
\[ \frac{P_{e}(\cC)}{1-P_{e} (\cC)} \leq \left( \frac{P_{e}(\cC')(1-\rho) +\rho}{P_{e}(\cC')(1-\rho)+1} \right) \frac{P_{e}(\cC')}{1-P_{e} (\cC')}. \]
Since the term in parentheses is increasing in $P_{e}(\cC')$, it achieves its upper bound of $1/(2-\rho)$ when $P_{e}(\cC')=1$.
This completes the proof.
\end{proof}

\begin{rem}
An interesting question is whether this can be extended to handle cases where $\cC' \neq \cC''$.
\end{rem}

\begin{exam} \label{exam:bec_rm_seq}
Now, we consider the code sequence $\cC_k = \RM(r,m+k)$ and describe the connection to Lemmas~\ref{lem:code_bound_rho} and~\ref{lem:nested_sym_err_bound}.
To start, we recall from~\cite[Lemma~9]{Reeves-it23} that two partially overlapping copies of $\cC_{k}$ can be found in $\cC_{k+1}$.
In particular, we index the bits in $\cC=\cC_{k+1}$ by $[2^{m+k+1}]$ and define $\cC'$ and $\cC''$ in Lemma~\ref{lem:code_bound_rho} using the construction above with $A=\{1,2,\ldots,2^{m+k-1}-1\}$, $B=\{2^{m+k-1},2^{m+k-1}+1,\ldots,2^{m+k}-1\}$, $C=\{2^{m+k},2^{m+k}+1,\ldots,2^{m+k}+2^{m+k-1}-1\}$, and $D = [2^{m+k+1}]\setminus (\{0\}\cup A \cup B \cup C)$.

Since $\RM$ codes have doubly transitive automorphism groups, their extrinsic decoding functions have transitive symmetry groups and $\mathrm{Sym}(f)$ (i.e., the symmetry group of the boolean extrinsic decoding function for $\cC'$) is transitive.
Thus, Corollary~\ref{cor:symmetry_bounds} shows that $\rho$ in Lemma~\ref{lem:code_bound_rho} satisfies $\rho=|A|/n=(2^{m+k-1}-1)/(2^{m+k}-1)\leq 1/2$ for all $k\in \mathbb{N}_0$.
\end{exam}

The following lemma shows that one can start with $P_e (\cC_0) = 1-\delta$ for any $\delta > 0$ and achieve exponential rate of decay in the number of stages.
It can be applied to any sequence of transitive codes satisfying the strong nesting property with a correlation bound of $\rho$.
But we are not aware of any meaningful consequences that do not involve $\RM$ codes.
Though not shown here, we note that the given exponential rate $(-\ln P_e (\cC_k))/k = \ln(2-\rho_0)$ can be asymptotically improved to $(-\ln P_e (\cC_k))/k \approx -\ln \rho$ by first using $\Omega(\ln (1/\delta))$ steps to bring the error rate from $1-\delta$ down to a small enough constant.

\begin{lem} \label{lem:nested_sym_err_bound}
Consider a sequence of codes $\cC_k$  where, for each $k\geq 0$, there are subsets $A,B,C,D$ satisfying Lemma~\ref{lem:code_bound_rho} with $\rho = \rho_k$.
Namely, we require that $\cC = \cC_{k+1}$ supports $\cC' = \cC_{k}$ on the indices $\{0\}\cup A \cup B$ and $\cC''=\cC_{k}$ on the indices $\{0\}\cup A \cup C$.
If $P_e (\cC_0) \leq 1-\delta$ and $\rho_k \leq \rho_0 < 1$ for all $k\geq 0$, then we have
\[ P_e (\cC_k) \leq \frac{P_e (\cC_k)}{1-P_e (\cC_k)} \leq \left(\frac{1}{2-\rho_0} \right)^k \left(\frac{1-\delta}{\delta}\right). \]
\end{lem}

 \begin{proof}
In this setting, Lemma~\ref{lem:code_bound_rho} shows that
\[ \frac{P_{e}(\cC_{k+1})}{1-P_{e} (\cC_{k+1})}  \leq \left(\frac{1}{2-\rho_k}\right)\frac{P_{e}(\cC_k)}{1-P_{e} (\cC_k)}. \]
Since $1/(2-\rho_k)\leq 1/(2-\rho_0)$, we obtain the stated result by induction from $P_{e} (\cC_0) \leq 1-\delta$ (which suffices because $x/(1-x)$ is increasing).
\end{proof}

\begin{thm} \label{thm:bec}
For $p\in (0,1)$, consider an erasure channel with capacity $C=1-p$.
For any $s\in \mathbb{N}$, there is a $t_0 \in \mathbb{N}$ such that, for all $t\geq t_0$, one can choose $m'=(st)^2$, $R = C-2/\sqrt{2\pi m'}$, and $r =\lfloor m'/2+ \sqrt{m'}\,\Phi^{-1} (R) /2 \rfloor$ to define the code sequence $\cC_k = \RM(r,m'+k)$.
For $k=2t$, $m=m'+k$, the code $\cC_k=\RM(r,m)$ satisfies $R(\cC_k)\geq C - 1/s - 2/(st)$ and
\[ P_e (\cC_{k}) \leq \left( \frac{2}{3} \right)^{(2\sqrt{m}/s)/\sqrt{1+2/(s^2t)}} \sqrt{2\pi m}. \]
Thus, we achieve a bit-erasure probability decaying exponentially in $\sqrt{m}$ for a code whose rate is within roughly $1/s$ of channel capacity.
\end{thm}

\begin{proof}
For this setup, we need $R = C-2/\sqrt{2\pi m'} \in (0,1)$.
This holds because $C\in (0,1)$ implies $R< 1$ while \[t\geq t_0 \coloneqq \lceil 4/(sC\sqrt{2\pi}) \rceil\] gives $2/\sqrt{2\pi m'} \leq C/2$ and the lower bound $R\geq C/2 > 0$.

Next, we note that the EXIT area theorem for codes on the BEC implies that $(1-p) P_e (\cC_0) \leq R(\cC_0)$ because $(1-p) P_e (\cC_0)$ is the smallest possible area under an EXIT function passing through $(p,P_e (\cC_0))$ and the area under the EXIT function must equal the rate~\cite{Kudekar-stoc16,Ashikhmin-it04}.
Thus, we can lower bound $\delta$ by noting that
\begin{align*}
\delta
&= 1-P_e (\cC_0) \\
&\overset{(a)}{\geq} \frac{1-p-R(\cC_0)}{1-p} \\
&\overset{(b)}{\geq} \frac{R+2/\sqrt{2\pi m'} - R(\cC_0)}{1-p} \\
&\overset{(c)}{\geq} \frac{1}{\sqrt{2\pi m'}},
\end{align*}
where $(a)$ is due to the EXIT area bound $P_e (\cC_0) \leq R(\cC_0)/(1-p)$, $(b)$ holds because $1-p=C=R+2/\sqrt{2\pi m'}$ by assumption, and $(c)$ follows from $R(\cC_0) \leq R + 1/\sqrt{2\pi m'}$ via~\eqref{lem:rate_vs_floor} and neglecting the division by $1-p$.
Then, Lemma~\ref{lem:nested_sym_err_bound} shows that 
\[ P_e (\cC_{k}) \leq \left( \frac{2}{3} \right)^{2t} \frac{1-\delta}{\delta} \leq \left( \frac{2}{3} \right)^{2t} \frac{1}{\delta}. \]
Since $m=m'+2t=m'(1+2/(s^2t))$, we have $m\geq m'$ and
\[ t=\frac{1}{s}\sqrt{m'}=\frac{1}{s}\sqrt{m}/\sqrt{1+2/(s^2 t)}.\]
Thus, the stated bound follows.

For the lower bound on $R(\cC_k)$,   we can write
\begin{align*}
C - R(\cC_k)
&= R  + \frac{2}{\sqrt{2\pi m'}} - R(\cC_k) \\
& \leq R + \frac{2}{\sqrt{2\pi m'}} - R(\cC_0) +  \frac{k}{2\sqrt{m'}} \\
&\leq \frac{5}{\sqrt{2\pi m'}} + \frac{1}{s} \\
&\leq \frac{2}{st} + \frac{1}{s},
\end{align*}
where $R(\cC_k)\geq R(\cC_0) - k/(2\sqrt{m'})$ (by Lemma~\ref{lem:rate}) implies the first inequality, the second inequality follows from~\eqref{lem:rate_vs_floor}, and $5/\sqrt{2\pi}\leq 2$ gives the last inequality.
\end{proof}

\begin{rem}
Thus, we see how the simple nesting property identified in~\cite[Lemma~9]{Reeves-it23} can be paired with Lemma~\ref{lem:f_bool_restrict_var} to achieve the same exponential order as~\cite{Abbe-focs23,Abbe-arxiv23b} without requiring floral constructions based on sunflowers and camellias.
We will see in Section~\ref{sec:bms} that the same idea can be generalized to the BMS channels.
\end{rem}

Without using other code properties, however, the bit-erasure probability shown above decays too slowly to show that the block error vanishes.
Still, with high probability, it does localize the correct codeword within a relatively small Hamming ball.
For any transitive binary code $\cC$ on a BMS channel, consider the case where each bit is estimated separately by a (possibly extrinsic) bit-MAP decoder with bit-error probability $Q$.
The hard decisions are then collected into a candidate vector for possible further processing.
The following lemma shows that the candidate vector will be $\sqrt{Q\vphantom{)}}$-close to the codeword with probability at least $1-\sqrt{Q\vphantom{)}}$.
\begin{lem} \label{lem:list_ball}
Let the random variable $\Delta \in [0,1]$ denote the relative distance from the candidate vector to the transmitted codeword.
Then, we have
\[ \Pr \left(\Delta \geq \sqrt{Q} \, \right) \leq \sqrt{Q}.
\]
\end{lem}
\begin{proof}
By symmetry, the error probability of each bit equals $Q$. 
Since $\Delta$ is a non-negative random variable, Markov's inequality implies that
\begin{align*}
\Pr \left(\Delta \geq \sqrt{Q} \, \right)
\leq \frac{\ex{\Delta}}{\sqrt{Q\vphantom{)}}}
\leq \frac{Q}{\sqrt{Q\vphantom{)}}},
\end{align*}
where $\ex{\Delta} = Q$ by linearity of expectation.
\end{proof}

\begin{rem}
For the erasure case studied in this section, one can achieve $Q = \frac{1}{2} P_e (\cC_{k})$ by returning a uniform random bit when the decoder reports a bit erasure.
\end{rem}

\section{Two-Look Analysis for BMS Channels}
\label{sec:bms}

The recursive two-look approach can be applied to BMS channels by focusing on the conditional expectation decoding function~\eqref{eq:cond_exp_dec_fun}.
In this case, one gets a recursive bound on the extrinsic minimum mean-squared error (MMSE).

\subsection{Setup}

As noted in Definition~\ref{def:bms}, the output of a BMS channel can be written as $Y = X Z$ where $X\in \cX = \{\pm 1\}$ is the channel input and $Z \in \cY = \mathbb{R}$ is a multiplicative noise term independent of $X$. For multiple channel uses, this extends naturally to $Y_i = X_i Z_i$ (denoted by $\bY = \bX \odot \bZ$) where $\bX \in \cX^N$ is the input sequence, $\bZ \in \cY^N$ is an i.i.d.\ sequence of multiplicative noise, and $\bY$ is the output sequence.

To transmit a codeword over a BMS channel, each binary codeword $\bu \in \cC$ is mapped to a channel input sequence $\bx \in \{\pm 1\}^N$ via the binary phase-shift keying (BPSK) mapping defined by $x_i = (-1)^{u_i}$.
The resulting set of BPSK-modulated codeword sequences is denoted by  $\cC_x$.
In this work, we always assume that the automorphism group $\cG$ of the code is transitive and this property is automatically inherited by the distribution of $\bX$ because $P_{\bX} (\bx) \propto \ind_{\cC_x}(\bx) = \ind_{\cC_x}(\pi \bx)$ for all $\pi \in \cG$.

For BMS channels, the key measure of performance will be the extrinsic minimum mean-squared error (MMSE) of $X_0$ given $Y_{\sim 0}$ denoted by $\mmse(X_0 | Y_{\sim 0})$.
Since $X_0 \in \{\pm 1\}$, this satisfies
\begin{align}
\tca \mmse(X_0 | Y_{\sim 0})
\oca = \ex{(X_0 - \ex{X_0|Y_{\sim 0}})^2} \nonumber \\
&\tc{\qquad} = \ex{X_0^2 - 2 X_0 \ex{X_0|Y_{\sim 0}} + \ex{X_0|Y_{\sim 0}}^2 } \nonumber \\
&\tc{\qquad} = 1 - \ex{\ex{X_0|Y_{\sim 0}}^2}. \label{eq:ext_mmse}
\end{align}

Now, we will follow roughly the same path as the BEC analysis in Section~\ref{sec:bec} for BMS channels.
Let $\cC$ be a transitive binary linear code of length-$N$ where bit-0 is distinguished without loss of generality.
Let the $N-1$ non-zero bit indices be partitioned into 4 disjoint sets $A,B,C,D \subseteq [N] \setminus \{0\}$ and define the code projections $\cC' = \cC|_{\{0\}\cup A\cup B}$ and $\cC'' = \cC|_{\{0\} \cup A \cup C}$.
Let $\bm{X} = (X_0,\ldots,X_{N-1}) \in \cX^N$ be a uniform random codeword from $\cC$ and let $\bm{Y} = (Y_0,\ldots,Y_{N-1}) \in \cY^N$ be an observation of $\bm{X}$ through the BMS channel $W$.
As an example of~\eqref{eq:cond_exp_dec_fun}, we define
\begin{equation}
g(y_{\sim 0}) \coloneqq \ex{X_0|Y_{\sim 0}=y_{\sim 0}} \label{eq:cond_mean_g}
\end{equation}
to be the extrinsic conditional mean estimator for $\cC$.
As a function of $\cC$, the extrinsic MMSE of $X_0$ from $Y_{\sim 0}$ is defined to be
\begin{align*}
M (\cC) \coloneqq
& \, \mmse(X_0 | Y_{\sim 0}) 
=  \; 1 - \ex{g(Y_{\sim 0})^2},
\end{align*}
where the second expression follows from~\eqref{eq:ext_mmse} and~\eqref{eq:cond_mean_g}.

Let $\bm{Z} = (Z_0,\ldots,Z_{N-1}) \in \cY^N$ be the i.i.d.\ multiplicative noise sequence. When transmitting a uniform random codeword from a linear code over a BMS channel, the codeword independence property says that any performance metric $\ex{\smash{\phi(\bX \odot \hat{\bX})}}$, where $\hat{\bX}$ is drawn from the conditional distribution of $\bX$ given $\bY$, is independent of the true codeword $\bX$~\cite[p.~190]{RU-2008}.
The same property also implies that
\begin{equation} \label{eq:nishi}
g(x_{\sim 0} \odot z_{\sim 0}) = x_0 g(z_{\sim 0}),
\end{equation}
for all $\bx \in \cC_x$ and $\bz \in \cY^N$~\cite[Lemma~24]{Reeves-it23}.
The key idea is that, for any $\bx' \in \cC_x$, we find that $\bx' \odot \bY = \bx' \odot (\bX \odot \bZ) = (\bx' \odot \bX) \odot \bZ$ has the same distribution as $\bY$ because $\bx' \odot \bX \smash{\overset{d}{=}} \bX$.
The following lemma uses~\eqref{eq:nishi} to characterize the performance in a manner that is independent of the transmitted codeword.
\begin{lem} \label{lem:mmse_cw_ind}
For this setup, we have
$\mmse(X_0 | Y_{\sim 0}) = 1 - \ex{ g(Z_{\sim 0})^2}$
and $\ex{ g(Z_{\sim 0})^2} = \ex{g(Z_{\sim 0})}$.
\end{lem}

 \begin{proof}
Using~\eqref{eq:ext_mmse} and~\eqref{eq:nishi}, we find that
\begin{align*}
\mmse(X_0 | Y_{\sim 0}) &= 1 - \ex{g(Y_{\sim 0})^2} \\ &= 1 - \ex{(X_0 g(Z_{\sim 0}))^2} \\ &= 1 - \ex{ g(Z_{\sim 0})^2}
\end{align*}
and
\begin{align*}
\mmse(X_0 | Y_{\sim 0}) &= \ex{(X_0 - \ex{X_0|Y_{\sim 0}})^2} \\
&= \ex{(X_0 - X_0 g(Z_{\sim 0}))^2} \\
&= 1-2 \ex{g(Z_{\sim 0})} + \ex{g(Z_{\sim 0})^2}.
\end{align*}
Subtracting these two expressions gives the second result.
\end{proof}
 Abusing notation, we define the decoding function $g: \cY^{|A|} \times \cY^{|B|} \times \cY^{|C|} \times \cY^{|D|} \to [-1,1] $ by
\begin{align*}
g \tca (y_A,y_B,y_C,y_D) \tcb
& = \ex{X_0|Y_A = y_A,Y_B = y_B, Y_C = y_C, Y_D = y_D}.
\end{align*}
By Lemma~\ref{lem:mmse_cw_ind}, it follows that
\[ M(\cC) = 1-\ex{g(Z_{A},Z_B,Z_C,Z_D)}. \]
Likewise, let $f\colon \cY^{|A|} \times \cY^{|B|} \to [-1,1]$ be a real function of $n=|A|+|B|$ variables defined by
\begin{align*}
f(y_A,y_B) &\coloneqq \ex{X_0|Y_A=y_A,Y_B=y_B} \\
&\;= g(y_A,y_B,\bm{0},\bm{0}),
\end{align*}
where the second expression holds because a BMS channel output of 0 corresponds to an erasure.
Again, Lemma~\ref{lem:mmse_cw_ind} implies
\[ M (\cC') = 1 - \ex{f(Z_{A},Z_B)}. \]

\subsection{Recursion}

Just like the BEC case, we let $G'$ be the automorphism group of $\cC'$ and $G_0' \subset G'$ be the subgroup that stabilizes bit 0 (i.e., $G_0' = \{\sigma \in G'\,|\, \sigma(0)=0\}$).
The symmetry group of the decoding function $f$ (e.g., see Definition~\ref{def:symf}) is inherited from the code $\cC'$ and each $\sigma \in G_0'$ defines a $\pi \in \mathrm{Sym}{(f)}$ .
One annoying issue is that the code $\cC'$ has length $N'=|A|+|B|+1$ and the decoding function $f$ has $n=N'-1 = |A|+|B|$ variables because the extrinsic decoder does not have access to $Y_0$.
Thus, mapping the code automorphism $\sigma \in G'$ to a functional symmetry $\pi \in \mathrm{Sym}(f)$ requires the awkward definition $\pi(i) = \sigma(i+1)-1$ for $i\in [n]$.

Let $F$ denote the subgroup of $\mathrm{Sym}(f)$ generated by $G_0'$.
In the sequel, $\Pi$ denotes a uniform random element of a subgroup $H \subseteq F$ that we are free to choose and, unless stated otherwise, $H=F$.
This freedom is justified because Lemma~\ref{lem:f_bool_restrict_var} holds for an arbitrary probability distribution $q$ on $\mathrm{Sym}(f)$ and the results below depend on the distribution of $\Pi$ only through the coefficient $\rho = \max_{S\subseteq [n]: S\neq \emptyset} \Pr(\Pi(S) \subseteq A )$.
For the product $\RM$ codes of Section~\ref{sec:product_rm}, choosing a proper subgroup $H$ makes $\rho$ easy to evaluate.

\begin{lem} \label{lem:code_mmse_rho}
Let $\rho = \max_{S\subseteq [n]: S\neq \emptyset} \Pr(\Pi(S) \subseteq A )$.
If the two code projections are equal (i.e., $\cC'=\cC''$), then we have the bound
\begin{align*}
\frac{M(\cC)}{1-M(\cC)} &\leq \left(\frac{1+\rho}{2}\right) \frac{M(\cC')}{1-M(\cC')}.
\end{align*}
\end{lem}

 \begin{proof}
As noted in Remark~\ref{rem:anova}, Lemma~\ref{lem:f_bool_restrict_var} actually holds for all real functions of i.i.d.\ real random variables.
But, the proofs given in Section~\ref{sec:aobf} only hold for real functions of i.i.d.\ boolean random variables.
For the general case, one can rely on more general decomposition results.

To establish the result, we consider the estimator defined by
\[ \tilde{f} (y_A,y_B,y_C) = \alpha \big( f(y_A,y_B) + f(y_A,y_C) \big) \]
where $\alpha$ is a real number that will be specified later on. 
It follows that
\begin{align*}
M(\cC)
&= \mmse(X_0|Y_{\sim 0}) \\
&\leq \mmse(X_0|Y_A,Y_B,Y_C) \\
&\leq \ex{\big(X_0-\tilde{f}(Y_A,Y_B,Y_C)\big)^2}.
\end{align*}
Denoting the last term by $\tilde{M}$ and expanding gives
\begin{align*}
\tilde{M} 
&= \ex{\big(X_0-\alpha f(Y_A,Y_B)-\alpha f(Y_A,Y_C)\big)^2} \\
&= \ex{\big(X_0-\alpha X_0 f(Z_A,Z_B)-\alpha X_0 f(Z_A,Z_C)\big)^2} \\
&=  \ex{X_0^2 \big(1- \alpha f(Z_A,Z_B)-  \alpha f(Z_A,Z_C)\big)^2} \\
&=  \mathbb{E}\Big[ 1+\alpha^2 f(Z_A,Z_B)^2+ \alpha^2 f(Z_A,Z_C)^2 \tcb \tca \tc{\quad\quad} -2 \alpha  f(Z_A,Z_B)  -2 \alpha  f(Z_A,Z_C) \tcb \tca \tc{\quad\quad} +2 \alpha^2  f(Z_A,Z_B) f(Z_A,Z_C)  \Big] \\
&= 1 + 2\alpha( \alpha -2) (1-M(\cC')) \tcb \tca \tc{\quad\quad} + 2\alpha^2 \ex{ f(Z_A,Z_B) f(Z_A,Z_C) },
\end{align*}
where the last step follows from $f(Z_A,Z_C)\overset{d}{=} f(Z_A,Z_B)$ and $\ex{f(Z_A,Z_B)} = \ex{f(Z_A,Z_B)^2} = 1-M(\cC')$.
Next, we invoke the inequality
\begin{align*}
&\ex{ f(Z_A,Z_B) f(Z_A,Z_C) } \\
& \qquad \leq \rho \ex{ f(Z_A,Z_B)^2} + (1-\rho) \ex{f(Z_A,Z_B)}^2\\
& \qquad = \rho (1-M(\cC')) + (1-\rho) (1-M(\cC'))^2,
\end{align*}
where the inequality follows from Lemma~\ref{lem:f_bool_corr} after combining $\rho = \max_{S\subseteq [n]: S\neq \emptyset} \Pr(\Pi(S) \subseteq A )$ with  Lemma~\ref{lem:f_bool_restrict_var} (with $q$ uniform over the chosen subgroup $H \subseteq F$) to see that $\var{\ex{f(Z_A,Z_B)|Z_A}} \leq \rho \var{f(Z_A,Z_B)}$.
Thus, for all $\alpha \geq 0$, we obtain 
\begin{align*}
\tilde{M} 
& \le  1 + 2\alpha( \alpha -2) (1-M(\cC')) \tcb \tca \tc{\quad} + 2\alpha^2 \big[ \rho  (1-M(\cC')) + (1- \rho)  (1-M(\cC'))^2 \big]  \\
& \tc{\!\!} = 1 -  2 (1 - M(\cC')) \left[  2 \alpha   - \alpha^2 ( 2 - (1-\rho) M(\cC'))\right] \!.   
\end{align*}
Evaluating this at $\alpha = 1 /( 2 - (1- \rho) M(\cC'))$ gives the minimum value and shows that
\[ M(\cC) \leq \frac{(1+\rho) M(\cC')}{ 2 - (1-\rho) M(\cC')}. \]
The stated result is given by applying $\phi(x)=x/(1-x)$ to both sides of this expression and one obtains a valid inequality because $\phi$ is increasing on $[0,1)$.
\end{proof}
 
The following lemma shows that one can start with $M (\cC_0) = 1-\delta$ for any $\delta > 0$ and still achieve the best exponential rate of decay (i.e., $(-\ln M(\cC_k))/k \approx \ln (2/(1+\rho))$) allowed by the correlation inequality when $k$ is large enough.
It can be applied to any sequence of transitive codes satisfying the strong nesting property with a correlation bound of $\rho$.

\begin{lem} \label{lem:nested_sym_mmse_bound}
Consider a sequence of transitive codes $\cC_k$ on a fixed BMS channel $W$ with capacity $C$.
Assume, for each $k\geq 0$, that there are subsets $A,B,C,D$ satisfying Lemma~\ref{lem:code_mmse_rho} with $\rho = \rho_k$ and that $\rho_k \leq \rho_0 < 1$ for all $k\geq 0$.
Specifically, we require that $\cC = \cC_{k+1}$ supports $\cC' = \cC_{k}$ on the indices $\{0\}\cup A \cup B$ and $\cC''=\cC_{k}$ on the indices $\{0\}\cup A \cup C$.
If $R(\cC_0) < C$, then $M(\cC_0)\leq 1-\delta$ for $\delta = C-R(\cC_0) > 0$ and
\[ 2 P_b (\cC_k) \leq M (\cC_k) \leq \left(\frac{1+\rho_0}{2}\right)^k \left(\frac{1-\delta}{\delta} \right). \]
\end{lem}

\begin{proof}
First, based on the EXIT area theorem (Theorem~\ref{thm:exit_area}), we observe that Theorem~\ref{thm:exit_mmse_H} implies
\[ M(\cC_0) = \mmse(X_0|Y_{\sim 0}) \leq 1-(C-R(\cC_0)) \leq 1-\delta. \]
For this setup, $M (\cC_k) \leq 1$ and Lemma~\ref{lem:code_mmse_rho} implies
\[ M (\cC_{k+1}) \leq \frac{M(\cC_{k+1})}{1-M(\cC_{k+1})} \leq \left(\frac{1+\rho_0}{2}\right) \frac{M(\cC_{k})}{1-M(\cC_{k})}. \]
Thus, the upper bound on $M(\cC_k)$ is given by induction and the bound on bit-error probability follows from Theorem~\ref{thm:exit_mmse_H}.
\end{proof}

\begin{thm} \label{thm:bms}
Consider a fixed BMS channel $W$ with capacity $C\in(0,1)$.
For any fixed $s\in \mathbb{N}$, there is a $t_0\in \mathbb{N}$ such that, for all $t\geq t_0$, one can choose $m'=(st)^2$, $R = C-2/\sqrt{2\pi m'}$, and $r =\lfloor m'/2+ \sqrt{m'}\,\Phi^{-1} (R) /2 \rfloor$ to define the code sequence $\cC_k = \RM(r,m'+k)$.
Choosing $k=2t$ and $m=m'+k$, the code $\cC_k = \RM(r,m)$ satisfies $R(\cC_k)\geq C - 1/s - 2/(st)$ and
\[ 2 P_b (\cC_{k}) \leq M (\cC_{k}) \leq \left( \frac{3}{4} \right)^{(2\sqrt{m}/s)/\sqrt{1+2/(s^2t)}} \sqrt{2\pi m}. \]
Thus, we achieve an MMSE (and bit-error probability) decaying exponentially in $\sqrt{m}$ for a code whose rate is roughly within $1/s$ of channel capacity.
\end{thm}

 \begin{proof}
For this setup, we need $R = C-2/\sqrt{2\pi m'} \in (0,1)$.
This holds because $C\in (0,1)$ implies $R< 1$ while \[t\geq t_0 \coloneqq \lceil 4/(sC\sqrt{2\pi}) \rceil\] gives $2/\sqrt{2\pi m'} \leq C/2$ and the lower bound $R\geq C/2 > 0$.

Next, based on the EXIT area theorem (Theorem~\ref{thm:exit_area}), Theorem~\ref{thm:exit_mmse_H} shows that
\[ M(\cC_0) = \mmse(X_0|Y_{\sim 0}) \leq 1-(C-R(\cC_0)), \]
where $C$ is the channel capacity of the BMS channel.
Since $r \leq m'/2+ \sqrt{m'}\,\Phi^{-1} (R) /2$, the bound in~\eqref{lem:rate_vs_floor} implies  $R(\cC_0) \leq R + 1/\sqrt{2\pi m'}$ and we find that
\begin{align*}
\delta
&= C-R(\cC_0) \\
&\geq C - R - \frac{1}{\sqrt{2\pi m'}} \\
& = \frac{1}{\sqrt{2\pi m'}}.
\end{align*}
Then, the error bound
\[ 2 P_b (\cC_{k}) \leq M (\cC_{k}) \leq \left( \frac{3}{4} \right)^{2\sqrt{m'}/s} \sqrt{2\pi m'} \]
follows directly from Lemma~\ref{lem:nested_sym_mmse_bound}.
Since $m=m'+2t=m'(1+2/(s^2t))$, we have $m\geq m'$ and 
\[ t=\frac{1}{s}\sqrt{m'}=\frac{1}{s}\sqrt{m}/\sqrt{1+2/(s^2 t)}.\]
Thus, the stated bound follows.
Finally, for the lower bound on $R(\cC_k)$, the calculation is identical to that of Theorem~\ref{thm:bec}.
\end{proof}
 
\section{Analysis of Boolean Functions}
\label{sec:aobf}

In this section, we review the Fourier analysis of boolean functions because it plays a crucial role in the proofs of correlation inequalities given in Lemmas~\ref{lem:f_bool_restrict_var} and~\ref{lem:gl2_boolean_bound}.

\subsection{Biased Fourier Decomposition}
\label{sec:fourier}

Fix $p\in(0,1)$ and let $L^{2}(\{0,1\}^n,\mathbb{R})$ denote the Hilbert space of functions mapping $\{0,1\}^n$ to the real numbers equipped with the inner product
\[
\left\langle f,g\right\rangle _{\mu}\triangleq\sum_{\bx\in\left\{ 0,1\right\} ^{n}}\mu(\bx)f(\bx)g(\bx),
\]
where $\mu\colon\left\{ 0,1\right\} ^{n}\to\mathbb{R}_{\geq0}$ is the non-negative weight function defined by $|\bx| = \sum_{i=0}^{n-1} |x_i|$ and
\[ \mu(\bx) = p^{|\bx|} (1-p)^{n-|\bx|}. \]
In this space, the Fourier transform of $f$ is a set of real numbers $\hat{f} (S)$ indexed by $S\subseteq [n]$ and defined by
\[ \hat{f}(S) = \left\langle f, u_S \right\rangle_{\mu}, \]
where $\{u_S (\bx)\}_{S\subseteq [n]}$ is an orthonormal basis for $L^{2}(\{0,1\}^n,\mathbb{R})$ satisfying
\begin{enumerate}
\item $u_S (\bx)$ only depends on variables whose indices are contained in $S$ (e.g., $u_{\emptyset}(\bx) = 1$),
\item $u_{S \cup T} (\bx) = u_S (\bx) u_T (\bx)$ if $S\cap T = \emptyset$,
\end{enumerate}
A brief introduction to this subject, along with the definition of the orthonormal basis, is included in Appendix~\ref{app:bfa}.
For a more detailed introduction, see~\cite{ODonnell-2014}.

For the purpose of Fourier analysis, the restriction $f_A$, which is defined in~\eqref{eq:fA}, is treated as function on the original space that is independent of all variables whose indices are not contained in $A$.
Let $\bX,\bX'$ be independent random vectors distributed according to $\mu$.
Then, the Fourier transform of $f_A$  equals
\begin{align}
\tc{\phantom{a}} \tca \tc{\!\!\!} \widehat{f}_A (S)
= \left\langle f_A, u_S \right\rangle_{\mu} \tc{\nonumber} \tcb 
&= \ex{f_A (\bX) u_S (\bX)} \nonumber \\
&= \ex{f_A (\bX) u_{S\cap A} (\bX) u_{S\setminus A}(\bX) } \nonumber \\
&= \ex{f_A (\bX) u_{S \cap A} (\bX) } \ex{u_{S \setminus A}(\bX)} \nonumber \\
&= \ex{\ex{ f (\bX') \,|\, X_A ' \!=\! X_A}  u_{S \cap A} (\bX) } \ex{u_{S \setminus A}(\bX)} \nonumber \\
&= \ex{ f (\bX') u_{S \cap A} (\bX') } \ex{u_{S \setminus A}(\bX)} \nonumber \\
&= \widehat{f}(S\cap A) \ex{ u_{S\setminus A} (\bX)} \nonumber \\
&= \begin{cases} \widehat{f}(S) & \text{if } S\subseteq A \\ 0 & \text{otherwise}, \end{cases} \label{eq:fAhat_S}
\end{align}
where the third step follows from $u_S (\bx) = u_{S\cap A} (\bX) u_{S\setminus A}(\bX)$, the fourth step holds because $f_A (\bx)$ only depends on $x_A$, the sixth step holds because $u_{S\cap A}(\bx)$ only depends on $x_A$, and the last step holds because $\ex{ u_{S\setminus A} (\bX)}=0$ unless  $S \setminus A = \emptyset$ (or equivalently $S\subseteq A$).

From Definition~\ref{def:symf}, the symmetry group of $f$ is denoted by $\mathrm{Sym}(f)$ and contains the permutations of the input variables that preserve the function.
This symmetry group plays an important role in our analysis and one key property is that, for all $\pi \in \mathrm{Sym}(f)$, we have
\begin{align}
\hat{f} (\pi(S))
&= \left\langle f, u_{\pi(S)} \right\rangle_{\mu} \nonumber \\
&= \ex{f (\bX) u_{\pi(S)} (\bX)} \nonumber \\
&= \ex{f (\pi \bX) u_{\pi(S)} (\pi \bX)} \nonumber \\
&= \ex{f (\bX) u_S (\bX)} \nonumber \\
&= \hat{f} (S), \label{eq:fhatpi_eq_fhat}
\end{align}
where $f(\pi\bx)=f(\bx)$ by invariance of $f$ under $\pi$ and $u_{\pi(S)}(\pi\bx)=u_S(\bx)$ by definition.

\subsection{Symmetry Bounds on Fourier Weight}

In this section, we apply Fourier analysis to boolean functions to prove correlation inequalities that leverage the symmetry group of the function.
To establish vanishing bit-error probability, only the simplest form in Lemma~\ref{lem:f_bool_restrict_var} is required.
For faster decay rates, we split the Fourier coefficients into two groups: those indexed by sets of size $\lesssim \log N$ and those indexed by sets of size $\gtrsim \log N$.
The first group is bounded using hypercontractivity via the biased ``level-$k$'' inequality (e.g., Lemma~\ref{lem:biased_level_k}) and the second group is bounded by applying the same correlation inequality with a larger symmetry group (e.g., Lemma~\ref{lem:f_bool_restrict_var} and Lemma~\ref{lem:gl_sym}).

The following lemma uses the symmetry group of $f$ to obtain a strong bound on the variance of a restriction in terms of the variance of the original function.
This lemma can be seen as a slightly more general statement of the symmetry bound introduced in~\cite[p.~186]{Abbe-focs23}.
In particular, it applies to symmetry groups other than the general linear (GL) group. 
This idea was first applied (with great effect) to decoding in~\cite{Abbe-focs23} though similar techniques have been applied before to boolean functions~\cite{Bourgain-gafa97}. 

In this section, we will use the notation $\cA = 2^A$ to denote the power set of $A$.
We also note that the variable $k$ is reused here to denote the Fourier level and this does not represent the index of the code sequence.
On a positive note, this choice does match the terminology of the well-known ``level-$k$'' inequality. The following lemma is proven for an arbitrary probability distribution $q$ on the symmetry group $\mathrm{Sym}(f)$.
Later, we will typically choose $q$ to be uniform over a subgroup of $\mathrm{Sym}(f)$ that is inherited from the automorphism group of a code.

\begin{lem} \label{lem:f_bool_restrict_var}
Let $f\colon \{0,1\}^n \to \mathbb{R}$ be a real function of boolean variables and $q: \mathrm{Sym}(f) \to \mathbb{R}$ be a probability distribution over $\mathrm{Sym}(f)$. Choose $\Pi$ to be random element of the symmetry group $\mathrm{Sym}(f)$ drawn according to $q$.
Then, for any $k\in [n+1]$ and subset $A \subseteq [n]$, we have
\begin{equation} \label{eq:wk_sym_bound}
\sum_{S \subseteq [n]: |S| = k} \widehat{f_A} (S)^2 = \sum_{S \subseteq [n]: |S|=k} \hat{f}(S)^2  \Pr (\Pi(S) \in \cA)
\end{equation}
and
\begin{equation}  \label{eq:var_sym_bound}
\var {f_A} = \sum_{S \subseteq [n]: S \neq \emptyset} \hat{f}(S)^2  \Pr (\Pi(S) \in \cA),
\end{equation}
where $\cA=2^A$ and $\Pr(\Pi(S)\in\cA)$ is the probability, under $\Pi\sim q$, that the permuted set is contained in $A$.
\end{lem}

 \begin{proof}
Using the results from Section~\ref{sec:fourier}, the Fourier representation of $f$ shows that
\begin{align*}
    \sum_{S \subseteq A : |S|=k} \widehat{f_A} (S)^2
    &\overset{(a)}{=} \sum_{S \in \cA : |S|=k} \hat{f}(S)^2 \\
    &= \sum_{S \subseteq [n] : |S|=k} \bm{1}_{\cA}(S) \hat{f}(S)^2 \\
    &\overset{(b)}{=} \sum_{S \subseteq [n] : |S|=k} \bm{1}_{\cA}(\pi(S)) \hat{f}(\pi(S))^2 \\
    &\overset{(c)}{=} \sum_{S \subseteq [n] : |S|=k} \bm{1}_{\cA}(\pi(S)) \hat{f}(S)^2,
\end{align*}
where $(a)$ follows from~\eqref{eq:fAhat_S}, $(b)$ holds because the permutation $\pi$ only reorders terms in the sum and $(c)$ is given by~\eqref{eq:fhatpi_eq_fhat}. Thus, for any probability distribution $q$ on the symmetry group $\mathrm{Sym}(f)$, we find that
\begin{align*}
        \tc{\phantom{AA}} \tca \tc{\!\!\!\!\!\!\!\!\!\!} \sum_{S \subseteq A : |S|=k} \widehat{f_A} (S)^2 \tcb
        &= \sum_{\pi \in \mathrm{Sym}(f)} q(\pi) \sum_{S \subseteq [n] : |S|=k} \bm{1}_{\cA}(\pi(S)) \hat{f}(S)^2 \\
        &= \sum_{S \subseteq [n] : |S|=k } \hat{f}(S)^2 \sum_{\pi \in \mathrm{Sym}(f)} q(\pi) \bm{1}_{\cA}(\pi(S))  \\
        &= \sum_{S \subseteq [n] : |S|=k} \hat{f}(S)^2 \Pr (\Pi(S) \in \cA),
\end{align*}
where $\Pi$ is a random element of the symmetry group $\mathrm{Sym}(f)$ drawn from the probability distribution $q$.
Summing this bound over $k\in [n+1] \setminus \{0\}$ gives the stated variance bound.
\end{proof}
 
If one can bound $\Pr (\Pi(S) \in \cA) \leq \rho$ uniformly over all non-empty $S \subseteq [n]$, then this naturally gives the bound $\var{f_A} \leq \rho \var{f}$.
The following corollary presents two particular cases of Lemma~\ref{lem:f_bool_restrict_var}.
\begin{cor} \label{cor:symmetry_bounds}
If $G$ is a transitive subgroup of $\mathrm{Sym}(f)$ and $q$ is uniform over $G$, then
\begin{equation} \label{eq:trans_rho}
\Pr (\Pi(S) \in \cA) \leq \frac{|A|}{n}.
\end{equation}
If $G$ is a subgroup of $\mathrm{Sym}(f)$ and $\cO$ is the set of orbits of $G$, then choosing $q$ to be uniform over $G$ gives
\begin{equation} \label{eq:orbit_rho}
\Pr (\Pi(S) \in \cA) \leq \max_{O \in \cO} \frac{|A \cap O|}{|O|}.
\end{equation}
\end{cor}

 \begin{proof}
Although the first condition is a special case of the second, we treat them separately because the first proof builds intuition for the second.

If $G$ is transitive and we choose $q$ to be uniform over $G$, then $\Pi$ maps any fixed element $i\in [n]$ to a uniform random element.
This is true because, for each $i\in [n]$, the group $G$ can be partitioned into disjoint subsets identified by $\pi(i)$ (i.e., the location where $\pi \in G$ maps $i$).
These subsets must have the same size because they are cosets of the stabilizer subgroup $\{\pi \in G \, | \, \pi(i) = i \}$ of the $i$th element.
Thus, for any nonempty $S\subseteq[n]$ and $i\in S$, we find that
\[ \Pr (\Pi(S) \in \cA) \leq \Pr (\Pi (i) \in A) = \frac{|A|}{n}. \]
This establishes~\eqref{eq:trans_rho}.

More generally, if $\cO$ is the set of orbits of $G$, then applying the above argument to each gives
\[ \Pr (\Pi(S) \in \cA) \leq \min_{O \in \cO : S\cap O \neq \emptyset} \frac{|A \cap O|}{|O|}. \]
This is true because, for any orbit $O \in \cO$ and any $i \in O$, the group $G$ can be partitioned into (equal size) cosets $H_j = \{\pi \in G \, | \, \pi(i) = j \}$ of the stabilizer subgroup $H_i$.
But, now the upper bound on the probability depends on $S \subseteq [n]$.
To get a uniform bound, we observe that every non-empty set must intersect some orbit.

Thus, we get~\eqref{eq:orbit_rho} by maximizing over all orbits. 
We note that this type of decomposition is related to the approach taken in~\cite{Kumar-itw16}.
\end{proof}
 
For any optimal hard-output (e.g., $\{0,1\}$) extrinsic decoding function, applying an element of the code's symmetry group clearly produces another optimal decoding function which may differ only in how ties are broken.
But, is there a single optimal decoding function that is invariant under the symmetry group?
The answer is yes because the set of all possible channel output vectors also decomposes into disjoint orbits under the symmetry group.
Thus, within each orbit, ties can be broken consistently to produce a symmetric optimal decoding function.
In particular, returning 0 for every tie produces an optimal decoding function invariant under the symmetry group.

The following corollary derives the general linear group symmetry bound introduced in~\cite[p.~186]{Abbe-focs23} from Lemma~\ref{lem:f_bool_restrict_var}.
In our notation, the setup corresponds to $n=2^m - 1$ where each index $i\in [n]$ is associated with the length-$m$ binary expansion \[\bv=(v_0,\ldots,v_{m-1})=\theta_m (i+1)\in \mathbb{F}_2^m.\]
Later, we use this to analyze the restriction of $f\colon \{0,1\}^n \to \mathbb{R}$ to the first $2^{m-\ell}-1$ variables with indices in the set $A = \{0,1,\ldots,2^{m-\ell}-2\}$ for some $\ell \in [m]$.
This matches the coding applications we target, where the lengths of the small codes $\cC',\cC''$ are $2^\ell$ times smaller than the big code $\cC$.

For $S \subseteq [n]$, let $\theta_m (S+1)$ be the image of $S$ (with 1 added to each entry) under the mapping $\theta_m$.
Following~\cite{Abbe-focs23}, we use the shorthand $\dim(S)=\dim (\theta_m (S+1))$ to denote the dimension of the $\mathbb{F}_2^m$-subspace spanned by elements of $\theta_m (S+1)$.
Let $\cG$ be the group of coordinate permutations $\pi$ where $\pi(i)=\theta_m^{-1}(A\, \theta_m (i+1))-1$ is induced by some invertible linear transformation $A\colon \mathbb{F}_2^m \to \mathbb{F}_2^m$.
Then, we say that $f$ has $\mathrm{GL}(m,2)$ symmetry and $\cG\subseteq \mathrm{Sym}(f)$

\begin{cor}[{\cite{Abbe-focs23}}] \label{lem:gl_sym}
Let $A = \{0,1,\ldots,2^{m-\ell}-2\}$ for some $\ell \in [m]$ with $\cA=2^A$ and let $\Pi$ be a uniform random permutation drawn from $\cG$.
For any $S \subseteq [n]$, we have
\begin{equation} \label{eq:gl_sym_var_bound}
\Pr (\Pi(S) \in \cA) \leq 2^{-\ell \dim (S)}.
\end{equation}
\end{cor}

 \begin{proof}
First, we note that $\theta_m (\cA+1)$ is the punctured subspace of $\mathbb{F}_2^m$ where $0$ is missing and the last $\ell$ coordinates of all vectors are fixed to 0.
Using this identification, a uniform random element from $\cG \subseteq \mathrm{Sym}(f)$ maps the smallest subspace containing $S$ to a uniform random subspace of the same dimension and thus $\Pr (\Pi(S) \in \cA)$ equals the number of $\dim(S)$-dimensional subspaces of $\mathbb{F}_2^{m-\ell}$ divided by the number $\dim(S)$-dimensional subspaces of $\mathbb{F}_2^{m}$.
Since one can count these subspaces with Gaussian binomial coefficients, we find that
\begin{align*}
\tc{\phantom{a}} \tca \tc{\!\!} \Pr ( \Pi(S) \in \cA)
\oca = \qbinom{m-\ell}{\dim(S)}_2 \bigg/ \qbinom{m}{\dim(S)}_2 \\
&= \frac{(2^{m-\ell} \!-\! 1)(2^{m-\ell-1}\!-\!1)\cdots(2^{m-\ell-\dim(S)+1}\!-\!1)}{(2^{m} - 1)(2^{m-1}-1)\cdots(2^{m-\dim(S)+1}-1)} \\
&\leq 2^{-\ell \dim(S)}. \tag*{\qedhere}
\end{align*}
\end{proof}
 
\begin{rem} \label{rem:anova}
While the results and proofs above are presented for real functions of boolean variables, they actually hold much more generally.
For functions on finite sets, Fourier methods extend naturally~\cite{ODonnell-2014,Abbe-focs23}.
For the functions of arbitrary i.i.d.\ random variables, the same results hold via the orthogonal ANOVA decomposition~\cite{Anantharam-arxiv25}.
\end{rem}

\subsection{Hypercontractivity and the Biased \texorpdfstring{Level-$k$}{Level-k} Inequality}

The level-$k$ inequality for boolean functions allows one to bound the total mass of Fourier coefficients of low weight in terms of the expected value~\cite[p.~264]{ODonnell-2014}.
In particular, if $\bX \in \{0,1\}^n$ is uniformly distributed and $\alpha = \ex{f(\bX)}$, then it implies that
\[ \sum_{S \subseteq [n]: |S|\leq \frac{1}{8}\ln (1/\alpha)} \hat{f}(S)^2 \leq \alpha^{3/2}, \]
where $\hat{f}(S)$ is the Fourier transform with respect to the uniform distribution.

For highly symmetric functions, we can use this idea to improve the bounds in the previous section.
But, first we need to generalize the result to biased i.i.d.\ Bernoulli distributions. In Appendix~\ref{app:thm1025}, we use results from~\cite[Sec.~10.3]{ODonnell-2014} to obtain a straightforward generalization of the small-set expansion theorem to biased distributions. From this, we can derive the following biased level-$k$ inequality.
\begin{lem} [Biased level-$k$ inequality] \label{lem:biased_level_k}
Let $\bX \in \{0,1\}^n$ have an i.i.d.\ distribution with $\Pr(X_0=1)=p\in(0,1)$ and let $\lambda = \min\{p,1-p\}$.
If $f\colon\{0,1\}^n\to\{0,1\}$ is a function with $\alpha = \ex{f(\bX)}\in (0,1)$, then
\begin{equation} \label{eq:biased_level_k}
\sum_{S \subseteq [n]: |S| \leq (1-\epsilon) \ln \alpha / \ln \lambda} \hat{f}(S)^2 \leq \alpha^{2\epsilon /(1+\lambda) }.
\end{equation}
In particular, since $\lambda\leq 1/2$, we can choose $\epsilon=7/8$ to see that
\[ \sum_{S \subseteq [n]: |S| \leq \frac{1}{8} \ln \alpha /\ln \lambda} \hat{f}(S)^2 \leq \alpha^{7/6}.  \]
\end{lem}

 \begin{proof}
For any $\rho \in (0,1]$ and $k\geq 0$, we have
\begin{align*}
\sum_{S \subseteq [n]: |S|\leq k} \hat{f}(S)^2
&\leq \rho^{-k}
\sum_{S \subseteq [n]} \hat{f}(S)^2 \rho^{|S|} \\
&\overset{(a)}{\leq} \left(\frac{1}{q-1} \lambda^{1-2/q}\right)^{-k} \alpha^{2-2/q} \\
&\overset{(b)}{=} \left( \lambda^{2-2\lambda/(1+\lambda)}\right)^{-k} \alpha^{2-2\lambda/(1+\lambda)},
\end{align*}
where $(a)$ is shown in Appendix~\ref{app:thm1025} and $(b)$ follows from choosing $q =  1+1/\lambda$.
Next, we choose $k= (1-\epsilon)\ln \alpha / \ln \lambda $ and observe that
\begin{align*}
\left( \lambda^{2-2\lambda/(1+\lambda)}\right)^{-k} &=
\left( \lambda^{2-2\lambda/(1+\lambda)}\right)^{-(1-\epsilon)\ln \alpha / \ln\lambda} \\
&= \alpha^{-(1-\epsilon) (2-2\lambda/(1+\lambda))} \\
&= 
 \alpha^{-2(1-\epsilon)/(1+\lambda)}.
\end{align*}
Substituting this into the previous equation and simplifying gives~\eqref{eq:biased_level_k}.
\end{proof}
 
For highly symmetric functions, the biased level-$k$ inequality can be used to improve the restriction variance bound when the expectation is small.
For example, if $n=2^m - 1$ and the function has $\mathrm{GL}(m,2)$  symmetry (like decoding functions for $\RM$ codes), then the following lemma gives an improved result.

\begin{lem} \label{lem:gl2_boolean_bound}
Let $f\colon \{0,1\}^n \to \{0,1\}$ with $n=2^m - 1$ be a boolean function with $\mathrm{GL}(m,2)$ symmetry and let $A = \{0,1,\ldots,2^{m-\ell}-2\}$ for some $\ell \in [m]$.
Then, we have
\[ \sum_{S \subseteq [n]} \widehat{f_A} (S)^2 \leq \alpha^2 + 2^{-\ell} \alpha^{7/6} + \alpha (c(\lambda) \ln (1/\alpha))^{-\ell}, \]
where $\bX\in\{0,1\}^n$ is i.i.d.\ with $\Pr(X_0=1)=p\in(0,1)$, $\alpha = \ex{f(\bX)}$, $\lambda = \min\{p,1-p\}$,  and $c(\lambda) = 1/(8\ln(1/\lambda))$.
\end{lem}

 \begin{proof}
We set up the calculation by combining the Fourier restriction bound~\eqref{eq:var_sym_bound} from Lemma~\ref{lem:f_bool_restrict_var} with the $\mathrm{GL}(m,2)$ symmetry bound~\eqref{eq:gl_sym_var_bound} from Lemma~\ref{lem:gl_sym}.
This gives
\begin{align*}
\sum_{S \subseteq [n]} \tca \widehat{f_A} (S)^2 \oca = \hat{f}(\emptyset)^2 + \sum_{S \subseteq [n]:S\neq \emptyset} \hat{f}(S)^2  \Pr (\Pi(S) \in \cA) \\
&\leq \hat{f}(\emptyset)^2 + \sum_{S \subseteq [n]:S \neq \emptyset} \hat{f}(S)^2  2^{-\ell \dim(S)} \\
&= \hat{f}(\emptyset)^2 + \sum_{k=1}^{\lfloor k_0 \rfloor} \sum_{S \subseteq [n]: |S|=k} \hat{f}(S)^2  2^{-\ell \dim(S)} \tcb \tca \tc{\qquad} + \sum_{k=\lfloor k_0 \rfloor+1}^n \sum_{S \subseteq [n]: |S|=k} \hat{f}(S)^2  2^{-\ell \dim(S)} \\
&\stackrel{(a)}{\leq} \alpha^2 + 2^{-\ell}\sum_{k=1}^{\lfloor k_0 \rfloor} \sum_{S \subseteq [n]: |S|=k} \hat{f}(S)^2 \tcb \tca \tc{\qquad} + \sum_{k=\lfloor k_0 \rfloor+1}^n \sum_{S \subseteq [n]: |S|=k} \hat{f}(S)^2  2^{-\ell \log_2 k} \\
&\stackrel{(b)}{\leq} \alpha^2 +  2^{-\ell} \alpha^{7/6} \tcb \tca \tc{\qquad} + 2^{-\ell \log_2 k_0} \sum_{k=\lfloor k_0 \rfloor+1}^n \sum_{S \subseteq [n]: |S|=k} \hat{f}(S)^2 \\
&\stackrel{(c)}{\leq} \alpha^2 + 2^{-\ell} \alpha^{7/6} + (c \ln (1/\alpha))^{-\ell} \alpha \\
&= \alpha ( \alpha + 2^{-\ell} \alpha^{1/6} + (c \ln (1/\alpha))^{-\ell}),
\end{align*}
where $(a)$ holds because $\dim(S)\geq \log_2 |S|$ (i.e., $|S| \leq 2^{\dim(S)}$), $(b)$ follows from Lemma~\ref{lem:biased_level_k} with $k_0 = c \ln(1/\alpha)$, and $(c)$ holds because the sum is less than $\ex{f(\bX)^2}=\ex{f(\bX)}$.
\end{proof}

\begin{rem}
This bound is considerably weakened by the inequality $|S| \leq 2^{\dim(S)}$.
If one could find another way to bound the terms where $\dim(S)\ll|S|$, then the remaining terms might be bounded by $O(\alpha^{1+\delta})$ for some $\delta>0$ and the overall result would be much stronger.
\end{rem}

\begin{lem} \label{lem:rm_level_k}
For $\cC= \RM(r,m+\ell)$ with $\cC'=\cC''= \RM(r,m)$, we find that
\begin{align*} P_e (\cC) &\leq P_e (\cC')^2 + 2^{-\ell}  P_e (\cC')^{7/6} \tcb \tca \tc{\qquad} +  (-c \ln P_e (\cC'))^{-\ell} P_e (\cC') .
\end{align*}
Moreover, for $\ell=1$, we have
\begin{equation} \label{eq:rm_level_k}
P_e (\cC) \leq P_e (\cC') \frac{2}{c \ln ( 1/P_e (\cC'))}.
\end{equation}
\end{lem}

 \begin{proof}
Our calculation starts by using the monotonicity of $g$ to upper bound the error probability of $g(Z_A,Z_B,Z_C,Z_D)$ for $\cC$ and then applies Lemma~\ref{lem:gl2_boolean_bound} to the Fourier expansion of $f_A$ (for $\cC'$) to get
\begin{align*}
P_e (\cC)
&= \ex{g(Z_A,Z_B,Z_C,Z_D)} \\
&\leq \ex{f(Z_A,Z_B) f(Z_A,Z_C)} \\
&= \ex{\ex{f(Z_A,Z_B)|Z_A}^2} \\
&= \sum_{S \subseteq [n]} \widehat{f_A} (S)^2 \\
&\leq P_e (\cC')^2 + 2^{-\ell}  P_e (\cC')^{7/6} \tcb \tca \tc{\qquad} + (-c \ln P_e (\cC'))^{-\ell} P_e (\cC') \\
&\leq P_e (\cC') \Big(P_e (\cC') + 2^{-\ell}  P_e (\cC')^{1/6} \tcb \tca \tc{\qquad} + (-c \ln P_e (\cC'))^{-\ell} \Big) .
\end{align*}
For $\ell=1$, we can apply Lemma~\ref{lem:alpha_rel_bound} to obtain~\eqref{eq:rm_level_k}.
\end{proof}

\begin{lem} \label{lem:alpha_rel_bound}
For $\alpha,p\in(0,1)$, let $\lambda=\min\{p,1-p\}$ and $c=1/(8\ln(1/\lambda))$. Then, we have
\[ \alpha + \frac{1}{2} \alpha^{1/6} + \frac{1}{c \ln (1/\alpha)} \leq \frac{2}{c \ln (1/\alpha)}. \]
\end{lem}

 \begin{proof}
For $\delta \in (0,1]$, let $\psi: [0,1] \to \mathbb{R}$ be defined by
\[ \psi(\alpha) = \frac{\alpha^{\delta}}{\frac{1}{c\ln(1/\alpha)}}. \]
By taking derivatives, one can verify that $\ln \psi(\alpha)$ is strictly concave on $(0,1)$ and that its unique maximum equals $\psi(e^{-1/\delta}) = c e^{-1} / \delta$.
While $c(p) = 1/(8\ln(1/\min\{p,1-p\}))$ depends implicitly on $p$, we note that $c(p) \leq c(1/2) = 1/(8\ln 2)$.  
The stated result follows from upper bounding each term to get
\begin{equation*}
\alpha + \frac{1}{2} \alpha^{1/6} \leq \bigg(\underbrace{\frac{e^{-1}}{8\ln 2} + \frac{3 e^{-1}}{8\ln 2}}_{\leq 1}\bigg) \frac{1}{c \ln (1/\alpha)}. \qedhere
\end{equation*}
\end{proof}

 \subsection{Block Error Probability for RM Codes on the BEC}

The two-look recursion used by Theorem~\ref{thm:bec} achieves $\ln 1/P_e (\RM(r_m,m))= \Omega(\sqrt{m})$ which is too slow to apply the list-decoding approach in~\cite{Abbe-focs23} for block error probability.
We note that the block error probability question for $\RM$ codes on the BEC was settled in~\cite{Kudekar-it17} and simplified in~\cite{Mondelli-izsc16}.
Now, we apply an improved argument for the BEC to recover that result as a precursor to a more complicated argument for the BSC in Section~\ref{sec:bsc}.
In particular, we use the ``level-$k$'' inequality from the theory of boolean functions to improve this decay rate to $\ln 1/P_e (\RM(r_m,m))= \Omega(\sqrt{m} \ln m)$.
Then, Lemma~\ref{lem:list_ball} uses Markov's inequality to show that the actual fraction of errors is not too much larger than its average with relatively high probability.
Given that the fraction of errors is not too big, the conditional probability of block error can be bounded using the weight enumerator and it is much smaller than the Markov failure probability.

\begin{thm}
\label{thm:fast_rm_err_bound_bec}
For $p\in (0,1)$, consider a BEC with capacity $C=1-p$.
For any $s\in \mathbb{N}$, there is a $t_0 \in \mathbb{N}$ such that, for all $t\geq t_0$, one can choose $m'=(st)^2$, $R = C-2/\sqrt{2\pi m'}$, and $r =\lfloor m'/2+ \sqrt{m'}\,\Phi^{-1} (R) /2 \rfloor$ to define the code sequence $\cC_k = \RM(r,m'+k)$.
Choosing $k=2t$ and $m=m'+k$, the code $\cC_k = \RM(r,m)$ satisfies $R(\cC_k)\geq C - 1/s - 2/(st)$ and
\[ P_e (\cC_{k})
\leq \exp\left(-\frac{1}{3s} \sqrt{m}\ln (m)/\sqrt{1+2/(s^2 t)} \right). \]
In addition, the block-error probability satisfies
\[ P_B (\cC_{k}) \leq \sqrt{P_e (\cC_{k})} + O\left(2^{-2^{m/3}}\right) \]
and we achieve a block-error probability decaying exponentially in $\sqrt{m} \ln m$ for a code whose rate is roughly within $1/s$ of channel capacity.
\end{thm}
 
 \begin{proof}
For this setup, we need $R = C-2/\sqrt{2\pi m'} \in (0,1)$.
This holds because $C\in (0,1)$ implies $R< 1$ while \[t\geq t' \coloneqq \max\{\lceil 4/(sC\sqrt{2\pi}) \rceil,2\}\] gives $2/\sqrt{2\pi m'} \leq C/2$ and the lower bound $R\geq C/2>0$.

To analyze the error rate, we follow the same approach as before but the $k$ steps of recursive bounding are split evenly between the bounds from Lemma~\ref{lem:nested_sym_err_bound} and Lemma~\ref{lem:rm_level_k}.
This is identical to the proof of Theorem~\ref{thm:bec} except that we apply Lemma~\ref{lem:nested_sym_err_bound} for $t$ steps rather than $2t$ steps to get
\[ P_e (\cC_{t}) \leq \left( \frac{2}{3} \right)^{t} \sqrt{2\pi s^2 t^2} . \]

Next, we apply the bound from Lemma~\ref{lem:rm_level_k} for $t$ more steps.
Since $-\ln(2/3) \geq 0.4 \geq 1/3$, there is a $t'' \in \mathbb{N}$ such that
\[ -\ln P_e (\cC_{t}) \geq \frac{t}{3} \]
for all $t\geq t''$.
Thus, for $t\geq t''$, Lemma~\ref{lem:rm_level_k} implies that
\begin{align*}
P_e (\cC_{t+j}) &\leq P_e (\cC_{t}) \left(\frac{2}{c \ln ( 1/P_e (\cC_t))}\right)^j \\
&\leq P_e (\cC_{t}) \left(\frac{6}{c t } \right)^{j}.
\end{align*}
These bounds apply recursively and remain valid at every stage because, as we will see next, the factors on the right are at most one for $t \geq t_0$.
If $t \geq s^2 (6/c)^3$, then $6/(ct) \leq 1/(st)^{2/3} = (1/m')^{1/3}$.
Thus, for $t \geq t_0 = \max\{s^2(6/c)^3,t',t''\}$,  we can choose $j=t$ to see that
\begin{align*}
P_e (\cC_{k})
&\leq \exp\left(-\frac{1}{3} t + t \ln\frac{6}{ct} \right) \\
&\leq \exp\left(-\frac{1}{3} t - \frac{1}{3} t \ln m' \right) \\
&\leq \exp\left(-\frac{1}{3s} \sqrt{m'} \ln(e m') \right).
\end{align*}
Since $m=m'+2t=m'(1+2/(s^2t))$, we have $m'=m/(1+2/(s^2t))$.
Moreover, with $t\geq 2$, we find $em'=em/(1+2/(s^2t))\geq em/2\geq m$.

The resulting decay rate is fast enough for the list-decoding argument of Section~\ref{sec:list_dec}.
For $\epsilon=\frac{1}{2} P_e (\cC_{k})$, we first apply Lemma~\ref{lem:list_ball} to generate a candidate vector whose relative distance $\Delta$ from the transmitted codeword satisfies $\Delta \leq \sqrt{\epsilon}$ with probability at least $1- \sqrt{\epsilon}$.
Ranking the codewords within distance $h = \sqrt{\epsilon} \, 2^m$ of this candidate by their likelihoods, Lemma~\ref{lem:list_success} bounds the failure probability of the resulting two-stage decoder by $\sqrt{\epsilon} + T_{\cC_k,\beta}(2h)$, where $\beta = p$ is the Bhattacharyya parameter of the BEC($p$).
Since $\Phi^{-1}(R)$ is bounded and $m = m'+2t$ with $t=\sqrt{m'}/s$, we see that $t$ grows without bound as $m'$ increases and that the code sequence satisfies $|m-2r| = O(\sqrt m)$.
Thus, we can restrict our attention to large enough $t$ so that $\sqrt{1+2/(s^2 t)} \leq 4/3$.
By the bit-error probability bound in the theorem statement, we have $P_e (\cC_k) \leq \exp(-\sqrt{m} \ln m / (4s))$.
Now, we can choose $h \leq 2^{m-\nu \sqrt m \log_2 m - 1}$ for $\nu = 1/(8 s)$ and apply Lemma~\ref{lem:as_weight_enum} with $\tilde m = m$ and $c = \nu/2$ because $h \leq 2^{m-\nu \sqrt m \log_2 m-1} \leq 2^{m - c\sqrt m \log_2 m -2}$ for sufficiently large $m$.
This gives $T_{\cC_k,\beta}(h) = O(2^{-2^{m/3}})$.
Since block-MAP decoding performs at least as well as this two-stage decoder, the stated block-error bound follows.
Finally, the lower bound on $R(\cC_k)$ is identical to the one in Theorem~\ref{thm:bec}.
\end{proof}

\begin{rem}
For $\RM$ codes on the BEC, a vanishing block-error probability was established first in~\cite{Kudekar-it17} based on a sophisticated analysis of boolean functions with large symmetry groups~\cite{Bourgain-gafa97}.
A simpler technique using weight enumerators was introduced in~\cite{Kudekar-isit16}.
Now, there is an even simpler argument based on support weights~\cite[Remark~2]{Pfister-isit25} that gives a drop-in replacement for the list decoding argument referenced above.
\end{rem}

\section{Block Recovery for RM Codes on the BSC}
\label{sec:bsc}

\subsection{Three-Look Bit-MAP Setup on the BSC}

For the BSC, one natural extension is to analyze the extrinsic bit-MAP decoding function $f(y_A,y_B)$ for bit-0 from a special subset of the received bits.
With only two looks, however, there is no simple way to resolve ties between the two decisions.
So instead, we consider a majority vote decoder from three looks based on overlapping sets.

Consider a transitive code $\cC$ of length-$N$ with a special distinguished bit.
Since $\cC$ is transitive, we assume without loss of generality that the distinguished bit has index 0.
Let the $N-1$ non-zero bit indices be partitioned into 5 disjoint sets $A,B,C,D,E \subseteq [N] \setminus \{0\}$ and define the code projections $\cC' = \cC|_{\{0\}\cup A\cup B}$, $\cC'' = \cC|_{\{0\} \cup A \cup C}$, and $\cC''' = \cC|_{\{0\} \cup A \cup D}$.
Let $\bm{X} = (X_0,\ldots,X_{N-1}) \in \cX^N$ be a uniform random codeword from $\cC$ and let $\bm{Y} = (Y_0,\ldots,Y_{N-1}) \in \cY^N$ be an observation of $\bm{X}$ through a BSC with error probability $p$.
The extrinsic bit-error probability for code $\cC$ from BSC observations is denoted by $P_b (\cC) $ and equals the probability that bit-0 of $\cC$ is decoded incorrectly from $Y_{\sim 0}$.

Let $g(y_{A},y_B,y_C,y_D,y_E)$ be an optimal extrinsic bit-MAP decoding function for the code $\cC$ on the BSC that respects the code symmetry and only depends on the error pattern.
By linearity and BSC symmetry, assume without loss of generality that the all-zero codeword is transmitted.
As noted above, an optimal extrinsic bit-MAP rule can be made invariant under the stabilizer of bit 0 by breaking ties consistently on each orbit (for example, always deciding 0 on a tie).
Equivalently, a syndrome decoder may choose coset leaders consistently on syndrome orbits.
The resulting error indicator depends only on the error pattern and inherits the required code symmetry.

With this form of tie breaking, whether or not bit-0 is decoded in error depends only on the error pattern.
Let $\bm{Z} = (Z_0,\ldots,Z_{N-1}) \in \{0,1\}^N$ be the error vector for the received sequence $\bm{Y}$.
For a realization $\bm{z}$ of $\bm{Z}$, it follows that $g(z_{A},z_B,z_C,z_D,z_E)$ is the indicator function that bit-0 of $\cC$ is decoded incorrectly by the extrinsic bit-MAP decoder from $Y_{\sim 0}$.
It follows that
\[ P_b (\cC) = \ex{g(Z_{A},Z_B,Z_C,Z_D,Z_E)}. \]

Likewise, for a realization $\bm{z}$ of $\bm{Z}$, let $f\colon \{0,1\}^n \to \{0,1\}$ be the extrinsic bit-MAP decoding function for $\cC'$ with $n=|A|+|B|$.
For a realization $\bm{z}$ of $\bm{Z}$, it follows that $f(z_{A},z_B)$ is the indicator function that bit-0 of $\cC'$ is decoded incorrectly by the extrinsic bit-MAP decoder from $Y_{A},Y_B$.
Thus,
\[ P_b (\cC') = \ex{f(Z_{A},Z_B)}. \]
For the majority vote decoder, the error indicator function satisfies
\begin{align}
\tca g_{\text{Maj}} (z_A,z_B,z_C,z_D) \tc{\nonumber} \tcb
&\;=  \text{Majority}\big(f(z_A,z_B),f(z_A,z_C),f(z_A,z_D)\big) \nonumber \\
&\;= f(z_A,z_B) f(z_A,z_C) + f(z_A,z_B) f(z_A,z_D) \tc{\nonumber} \tcb \tca \tc{\quad} - 2 f(z_A,z_B) f(z_A,z_C) f(z_A,z_D) \tc{\nonumber} \tcb \tca \tc{\quad} + f(z_A,z_C) f(z_A,z_D)  \nonumber \\
& \leq f(z_A,z_B) f(z_A,z_C) + f(z_A,z_B) f(z_A,z_D) \tc{\nonumber} \tcb \tca \tc{\quad}  + f(z_A,z_C) f(z_A,z_D), \label{eq:majority_union}
\end{align}
where the function $\text{Majority} \colon \{0,1\}^3 \to \{0,1\}$ outputs the majority vote of its three inputs.

Let $G$ be the automorphism group of $\cC'$ and $G_0 \subset G$ be the subgroup that stabilizes bit 0 (i.e., $G_0 = \{\sigma \in G\,|\, \sigma(0)=0\}$).
Since the decoding function $f$ naturally inherits all symmetries of the code, each $\sigma \in G_0$ defines a $\pi \in \mathrm{Sym}{(f)}$.
In particular, for any $\sigma \in G_0$, we can define $\pi(i) = \sigma(i+1)-1$ and observe $\pi \in \mathrm{Sym}(f)$.

Let $G_0 '$ denote the subgroup of $\mathrm{Sym}(f)$ generated by $G_0$.
In the sequel, $\Pi$ denotes a uniform random element of a subgroup $H \subseteq G_0'$ that we are free to choose and, unless stated otherwise, $H = G_0'$.
This freedom is justified because Lemma~\ref{lem:f_bool_restrict_var} holds for an arbitrary probability distribution $q$ on $\mathrm{Sym}(f)$ and the results below depend on the distribution of $\Pi$ only through the coefficient $\rho = \max_{S\subseteq [n]: S\neq \emptyset} \Pr(\Pi(S) \in \cA )$.
For the $\RM$ product codes discussed in Section~\ref{sec:product_rm}, choosing a convenient subgroup $H$ makes $\rho$ easy to evaluate.

\begin{lem} \label{lem:bsc_code_bound_rho}
Let $\rho = \max_{S\subseteq [n]: S\neq \emptyset} \Pr(\Pi(S) \in \cA )$.
If the three code projections are equal (i.e., $\cC'=\cC''=\cC'''$), then we find that
\[ P_{b} (\cC) \leq 3(1-\rho) P_b (\cC')^2 + 3\rho P_b (\cC'). \]
\end{lem}

 \begin{proof}
First, combining $\rho = \max_{S\subseteq [n]: S\neq \emptyset} \Pr(\Pi(S) \in \cA )$ with  Lemma~\ref{lem:f_bool_restrict_var} (where $q$ is uniform over the chosen subgroup $H \subseteq G_0'$) shows \[\var{\ex{f(Z_A,Z_B)|Z_A}} \leq \rho \var{f(Z_A,Z_B)}.\]
Second, using the above setup, we have
\begin{align*}
P_b \tca (\cC)
\oca = \ex{g(Z_A,Z_B,Z_C,Z_D,Z_E)} \\
&\leq \ex{g_{\text{Maj}} (Z_A,Z_B,Z_C,Z_D)} \\
&\leq \ex{f(Z_A,Z_B) f(Z_A,Z_C)} \tcb \tca \tc{\qquad} + \ex{f(Z_A,Z_B) f(Z_A,Z_D)} \tcb \tca \tc{\qquad} + \ex{f(Z_A,Z_C) f(Z_A,Z_D) } \\
&\leq 3\rho \underbrace{\ex{f(Z_A,Z_B)^2}}_{\ex{f(Z_A,Z_B)}} + 3(1-\rho) \ex{f(Z_A,Z_B)}^2  \\
&= 3\rho P_b (\cC') + 3(1-\rho) P_b (\cC')^2,
\end{align*}
where the first inequality holds by the data-processing inequality, the second is given by~\eqref{eq:majority_union}, and the third follows from Lemma~\ref{lem:f_bool_corr}.
\end{proof}
 
\begin{rem}
Three looks yield a significantly stronger bound in the small-error regime, but the resulting recursion is contractive only below a critical error threshold. Indeed, when the error rate is close to 1/2, the factor of 3 prevents contraction even as $\rho \to 0$, whereas for $\rho =1/4$ contraction occurs once the extrinsic error probability is below 1/9. Thus, one must first drive the error probability into this regime using a two-look argument or a slower-decaying approach such as~\cite{Reeves-it23,Reeves-isit23}. The three-look recursion can then be used to obtain the stronger decay. Alternatively, Lemma~\ref{lem:code_mmse_rho} provides an improved result using only two looks.
\end{rem}

\subsection{Fast Decay for Reed--Muller Codes on the BSC}

Now, we use the ``level-$k$'' inequality to improve the decay rate of our bound on the bit-error probability of RM codes on the BSC.

\begin{lem} \label{lem:rm_level_k_bsc}
For $\cC= \RM(r,m+2\ell)$ with $\cC'=\cC''=\cC'''= \RM(r,m)$, we find that
\begin{align*}
P_b (\cC) &\leq 3 P_b (\cC') \Big(P_b (\cC') + 2^{-\ell}  P_b (\cC')^{1/6} \tcb \tca \tc{\qquad\qquad\qquad} +  (-c \ln P_b (\cC'))^{-\ell} \Big).
\end{align*}
Moreover, for $\ell=2$, we have
\begin{equation} \label{eq:rm_level_k_bsc}
P_b (\cC) \leq P_b (\cC') \left(\frac{2}{c \ln ( 1/P_b (\cC'))} \right)^2 .
\end{equation}
\end{lem}

 \begin{proof}
From Section~\ref{sec:multiple_looks}, we know that $\RM(r,m+2\ell)$ allows 3 $\RM(r,m)$ looks with fractional overlap $\rho=2^{-\ell}$.
To bound the expectation of $g_{\text{Maj}} (z_A,z_B,z_C,z_D)$ for $\cC$, we use~\eqref{eq:majority_union} to bound by the sum of three pairwise products and then apply Lemma~\ref{lem:gl2_boolean_bound} to $f_A$ for $\cC'$.
This gives
\begin{align*}
P_b (\cC)
&= \ex{g (Z_A,Z_B,Z_C,Z_D,Z_E)} \\
&\leq \ex{g_{\text{Maj}} (Z_A,Z_B,Z_C,Z_D)} \\
&\leq 3 \ex{f(Z_A,Z_B) f(Z_A,Z_C)} \\
&= 3 \, \ex{\ex{f(Z_A,Z_B)|Z_A}^2} \\
&= 3 \sum_{S \subseteq [n]} \widehat{f_A} (S)^2 \\
&\leq 3 \Big(P_b (\cC')^2 + 2^{-\ell}  P_b (\cC')^{7/6} \tcb \tca \tc{\qquad\qquad} +  (-c \ln P_b (\cC'))^{-\ell} P_b (\cC') \Big) \\
&\leq 3 P_b (\cC') \gamma,
\end{align*}
where $\gamma = P_b (\cC') + 2^{-\ell}  P_b (\cC')^{1/6} +  (-c \ln P_b (\cC'))^{-\ell}$.
For $\ell\!=\!2$, we can use the same approach as Lemma~\ref{lem:alpha_rel_bound}.
In this case, we want to upper bound $\psi(\alpha)^2$ instead of $\psi(\alpha)$.
Although $c$ depends on $p$, this provides a uniform bound for $p\in (0,1)$ of
\[ \psi(\alpha)^2 = \frac{\alpha^{2\delta}}{\left(\frac{1}{c \ln (1/\alpha)}\right)^2} \leq \left(\frac{c e^{-1}}{\delta}\right)^2 \leq \left(\frac{e^{-1}}{8 \delta \ln 2}\right)^2. \]
For $\ell=2$, this implies that
\begin{align*}
\gamma &= \alpha + \frac{1}{4} \alpha^{1/6} + \left(\frac{1}{c \ln (1/\alpha)}\right)^2 \\
&\leq \Bigg( \bigg(\underbrace{\frac{e^{-1}}{4\ln 2}\bigg)^2 \!\!+\! \frac{1}{4}\bigg(\frac{12 e^{-1}}{8\ln 2}\bigg)^2 \!\!+\!1}_{\leq 1.2} \Bigg) \left(\frac{1}{c \ln (1/\alpha)} \right)^2 \!\!.
\end{align*}
Plugging this back into the bound gives~\eqref{eq:rm_level_k_bsc}.
\end{proof}

\begin{thm}\label{thm:fast_rm_err_bound_bsc}
Reed--Muller codes achieve capacity on the BSC.
Specifically, for a BSC with capacity $C\in(0,1)$ and maximum gap to capacity $\epsilon >0$, choose $s\in\mathbb{N}$ such that $s\geq 2/\epsilon$.
Then, there exists $t_0\in\mathbb{N}$ such that the following holds for every $t\geq t_0$ divisible by $4$. Set $m'=(st)^2$, $R = C-2/\sqrt{2\pi m'}$,  $r =\lfloor m'/2+ \sqrt{m'}\,\Phi^{-1} (R) /2 \rfloor$, and define $\cC_k=\RM(r,m'+k)$. For $k=2t$ and $m=m'+k$, the rate satisfies
\[
R(\cC_k)\geq C-\frac{1}{s}-\frac{2}{st},
\]
the bit-error probability satisfies
\[
P_b(\cC_k)
\leq
\exp\left(
-\frac{1}{8s}
\frac{\sqrt{m}\ln m}{\sqrt{1+2/(s^2t)}}
\right).
\]
and the block-error probability satisfies
\[
P_B(\cC_k)
\leq
\sqrt{P_b(\cC_k)}
+O\left(2^{-2^{m/3}}\right).
\]
Thus,  one achieves a block-error probability decaying exponentially in $\sqrt{m} \ln m$ for a code whose rate is roughly within $1/s$ of channel capacity.
\end{thm}

\begin{proof}
Let $\delta_t = C-R(\cC_0)$ and, for the chosen parameters, observe that $\delta_t \geq 1/\sqrt{2\pi m'} = 1/(st\sqrt{2\pi})$.
For the construction to work, we need $R = C-2/\sqrt{2\pi m'} \in (0,1)$.
This holds because $C\in (0,1)$ implies $R<1$ while
\[t\geq t' \coloneqq \max\{\lceil 4/(sC\sqrt{2\pi}) \rceil,2\}\] gives $2/\sqrt{2\pi m'} \leq C/2$ and hence $R\geq C/2 >0$.

To analyze the error rate, we revisit the BEC argument in Theorem~\ref{thm:fast_rm_err_bound_bec} with appropriate modifications.
The main difference is Lemma~\ref{lem:rm_level_k_bsc} bounds $P_b (\cC_{j+4})$ in terms of $P_b (\cC_{j})$.
Thus, we get 4 times fewer steps for the same $k$.
This is because each step in the recursion requires three $\RM(r,m')$ looks with a fractional overlap of $\rho=1/4$.
As outlined in Section~\ref{sec:multiple_looks}, these looks are provided by $\RM(r,m'+4)$.

The recursive bounds and the improvement enabled by the  ``level-$k$'' inequality are essentially the same as in the BEC case.
The main difference is the change in the number of steps mentioned above and that the ``level-$k$'' bound inherits the factor of 3 increase from Lemma~\ref{lem:bsc_code_bound_rho} in order to bound the error probability of the majority vote.
As before, we split the $2t$ steps in the code sequence evenly between the two stages.

We first apply the two-look recursion for $t$ stages.
Since each step bounds $P_b (\cC_{j+1})$ in terms of $P_b (\cC_{j})$, we get a bound on $P_b (\cC_{t})$.
Then, for the uniform RM overlap bound $\rho_j\leq1/2$, we can apply Lemma~\ref{lem:nested_sym_mmse_bound} with $\delta_t$ to see that
\begin{align*}
P_b(\cC_t)
&\leq \left(\frac{3}{4}\right)^t
\frac{1-\delta_t}{\delta_t} \leq \left(\frac{3}{4}\right)^t st\sqrt{2\pi}.
\end{align*}
Since $\ln(4/3)>1/4$, there is a $t'' \in \mathbb{N}$ such that
\begin{equation} \label{eq:bsc_cor_initial_decay}
\ln\frac{1}{P_b(\cC_t)}\geq\frac{t}{4}
\end{equation}
for all $t\geq t''$.

Let $p=h^{-1}(1-C) \in (0,1/2)$ be the crossover probability of the fixed BSC and set $c=c(p)>0$.
Starting from $\cC_t$, apply Lemma~\ref{lem:rm_level_k_bsc} a total of $t/4$ times.
For sufficiently large $t$, Equation~\eqref{eq:bsc_cor_initial_decay} gives
\[ \frac{2}{c\ln(1/P_b(\cC_t))}\leq\frac{8}{ct}\leq1. \]
Consequently, the error probability cannot increase during these applications, so the same upper bound on the recursive factor remains valid at every stage.
It follows that
\begin{equation} \label{eq:bsc_cor_level_recursion}
P_b(\cC_{2t})
\leq P_b(\cC_t)\left(\frac{8}{ct}\right)^{t/2}.
\end{equation}
Choosing $t_0 = \max\{t',t'',\lceil 64s/c^2 \rceil \}$ ensures that the preceding inequalities hold and additionally that $t\geq64s/c^2$ for all $t\geq t_0$.
Multiplying the latter inequality by $c^2 t / 64$ and taking the square root shows this is equivalent to
\[ \frac{8}{ct}\leq (m')^{-1/4}. \]
Combining this inequality with~\eqref{eq:bsc_cor_initial_decay} and~\eqref{eq:bsc_cor_level_recursion} yields
\begin{align*}
P_b(\cC_k)
&\leq e^{-t/4}(m')^{-t/8} \\
&=\exp\left(-\frac{t}{8}\ln(e^2m')\right) \\
&\leq\exp\left(-\frac{1}{8s}\sqrt{m'}\ln(e m')\right),
\end{align*}
since $t=\sqrt{m'}/s$.
The stated bit-error bound follows because $m=m'(1+2/(s^2t))$, so $m'=m/(1+2/(s^2t))$.
Moreover, for $t\geq2$, $em'=em/(1+2/(s^2t))\geq em/2\geq m$.

This decay rate is fast enough to apply the list-decoding argument of Section~\ref{sec:list_dec}.
To do so, we let $\epsilon=P_b (\cC_{k})$ and apply Lemma~\ref{lem:list_ball} to generate a candidate vector whose relative distance $\Delta$ from the transmitted codeword satisfies $\Delta \leq \sqrt{\epsilon}$ with probability at least $1- \sqrt{\epsilon}$.
Ranking the codewords within distance $h = \sqrt{\epsilon}\, 2^m$ of this candidate by their likelihoods, Lemma~\ref{lem:list_success} bounds the failure probability of the resulting two-stage decoder by $\sqrt{\epsilon} + T_{\cC_k,\beta}(2h)$, where $\beta = 2\sqrt{p(1-p)}$ is the Bhattacharyya parameter of the BSC($p$).

Since $\Phi^{-1}(R)$ is bounded and $m = m'+2t$ with $t=\sqrt{m'}/s$, we see that $t$ grows without bound as $m'$ increases and that the code sequence satisfies $|m-2r| = O(\sqrt m)$.
Thus, we can restrict our attention to large enough $t$ so that $\sqrt{1+2/(s^2 t)} \leq 9/8$.
By the bit-error probability bound in the theorem statement, we have $P_b (\cC_k) \leq \exp(-\sqrt{m} \ln m / (9s))$.
Now, we can choose $h \leq 2^{m -\nu \sqrt m \log_2 \! m - 1}$ for $\nu = 1/(18 s)$ and apply Lemma~\ref{lem:as_weight_enum} with $\tilde m = m$ and $c = \nu/2$ because $h \leq 2^{m-\nu \sqrt m \log_2 m - 1} \leq 2^{m - c\sqrt m \log_2 m -2}$ for sufficiently large $m$.
This gives $T_{\cC_k,\beta}(2h) = O(2^{-2^{m/3}})$.
Since block-MAP decoding performs at least as well as this two-stage decoder, we conclude that $P_B(\cC_k)\leq\sqrt{P_b(\cC_k)}+O(2^{-2^{m/3}})$, as claimed.
Finally, the lower bound on $R(\cC_k)$ is identical to that of Theorem~\ref{thm:bec}.
\end{proof}
 
\begin{rem}
This achieves the same qualitative error bound as~\cite{Abbe-focs23} using a recursive argument with only three looks per stage.
But, it is still unclear exactly how the two methods are related at a deeper level.
While we simply use hypercontractivity, they combine sunflower boosting and a bespoke analysis to achieve a similar recursive bound.
\end{rem}

\section{RM Product Codes Achieve Capacity}
\label{sec:product_rm}

This section extends the recursive multiple-look analysis to product codes with binary $\RM$ component codes.
These codes are also known as tensor products of $\RM$ codes.
Product codes~\cite{Elias-ire54} are among the oldest constructions in coding theory and $\RM$-product codes form a natural test case for the methods of this paper: they are transitive but \emph{not} doubly transitive in general by Lemma~\ref{lem:product_not_dbl}.

For RM product codes, this lack of doubly-transitive symmetry causes no difficulty because Lemma~\ref{lem:f_bool_restrict_var} retains, for each Fourier index set, the exact probability that a random symmetry maps it into the common observation set, and the recursion lemmas (e.g., Lemmas~\ref{lem:code_mmse_rho}, \ref{lem:nested_sym_mmse_bound}, and~\ref{lem:bsc_code_bound_rho}) depend on the code only through the coefficient $\rho = \max_{S \subseteq [n]: S \neq \emptyset} \Pr(\Pi(S)\in\cA)$.
For $\RM$-product codes, this probability is evaluated exactly by applying independent general-linear transformations to the two coordinate factors in Corollary~\ref{cor:product_gl_sym}.

The section is organized as follows.
We first record the standard structure of product codes and show that rectangular projections and bilateral nesting behave exactly as in the single-$\RM$ case.
Then, we evaluate the Fourier membership probability of Lemma~\ref{lem:f_bool_restrict_var} for the product symmetry group to get the two new symmetry parameters needed: the coefficient $\rho$ from Corollary~\ref{cor:product_gl_sym} and the restriction bound required for the level-$k$ argument in Corollary~\ref{cor:product_boolean_bound}.
Next, we verify that the two-look recursion applies with $\rho\leq1/2$ so that Theorem~\ref{thm:product_bms} yields capacity-approaching bit-MAP performance on every BMS channel with bit-error probability decaying exponentially in $\sqrt{\ln N}$.
On the BSC, Corollary~\ref{cor:product_boolean_bound} shows the three-look level-$k$ recursion applies with the product restriction bound and provides a decay rate that is exponential in $\sqrt{\ln N} \ln\ln N$ as reported in Theorem~\ref{thm:product_capacity}.
Finally, Theorem~\ref{thm:product_block} shows that the fast bit-error decay allows a two-stage row--column list-redecoding procedure to recover the complete transmitted array with probability tending to one.

\subsection{Product Codes and Rectangular Nesting}
\label{sec:pc_rect}

\begin{defn} \label{defn:product_code}
For binary linear codes $\cC_1\subseteq\mathbb{F}_2^{N_1}$ and $\cC_2\subseteq\mathbb{F}_2^{N_2}$, the \emph{tensor product code} is
\[ \cP=\cC_1\otimes\cC_2 \triangleq \Span\big\{\ba \bb^{\mathsf T} \,\big|\, \ba\in\cC_1,\ \bb\in\cC_2\big\} \subseteq\mathbb{F}_2^{N_1\times N_2}, \]
where codewords are $N_1 \times N_2$ binary matrices and $\ba\bb^{\mathsf T}$ denotes the outer product.
\end{defn}

\begin{lem} \label{lem:product_char}
Let $\cP=\cC_1\otimes\cC_2$, where $\cC_1,\cC_2$ are nonzero with dimensions $k_1,k_2$ and minimum distances $d_1,d_2$.  Then:
\begin{enumerate}
\item $\cP$ consists of all matrices $\bM\in\mathbb{F}_2^{N_1\times N_2}$ whose columns all lie in $\cC_1$ and whose rows all lie in $\cC_2$;
\item $\dim(\cP)=k_1k_2$, so that $R(\cP)=R(\cC_1)R(\cC_2)$;
\item $d(\cP)=d_1d_2$.
\end{enumerate}
\end{lem}
These well-known product-code identities are established, for example, in~\cite{Elias-ire54}.  To denote RM-product codes, we write
\[ \PRM(r_1,m_1,r_2,m_2) \triangleq \RM(r_1,m_1)\otimes\RM(r_2,m_2) \]
and index the bits directly by pairs $(u,v)\in\mathbb{F}_2^{m_1}\times\mathbb{F}_2^{m_2}$, with the columns (fixed $v$) lying in $\RM(r_1,m_1)$ and the rows (fixed $u$) lying in $\RM(r_2,m_2)$.
The bijection $\theta_{m_1+m_2}$ from Definition~\ref{def:binary_rm} identifies this index set with $[2^{m_1+m_2}]$, where the distinguished bit 0 corresponds to $(0,0)$, so all earlier conventions apply.

\begin{rem} \label{rem:rm_subcode}
Expanding both factors in the monomial basis shows that $\PRM(r_1,m_1,r_2,m_2)$ is the evaluation code, over $(\bx,\by)\in\mathbb{F}_2^{m_1}\times\mathbb{F}_2^{m_2}$, of the multilinear polynomials $h(\bx,\by)$ whose degree in the $\bx$-variables is at most $r_1$ and whose degree in the $\by$-variables is at most $r_2$.
Indeed, the span of the products $f(\bx)g(\by)$ with $f\in\cF(r_1,m_1)$ and $g\in\cF(r_2,m_2)$ is exactly the span of the monomials $\bx^{\bi}\by^{\bj}$ with $|\bi|\leq r_1$ and $|\bj|\leq r_2$.
In particular, $\PRM(r_1,m_1,r_2,m_2)$ is a \emph{subcode} of $\RM(r_1+r_2,\,m_1+m_2)$.
A key point of this section is that the recursive method applies with only small changes to a code family whose symmetry group not doubly transitive. \end{rem}

\begin{lem} \label{lem:rect_projection}
For a product code $\cP=\cC_1\otimes\cC_2$ and coordinate subsets $I\subseteq[N_1]$ and $J\subseteq[N_2]$, the \emph{rectangular projection} is
\[ \cP\big|_{I\times J} = (\cC_1|_I)\otimes(\cC_2|_J). \]
\end{lem}

 \begin{proof}
The projection $\bM\mapsto \bM_{I\times J}$ is linear and maps the generator $\ba\bb^{\mathsf T}$ to $a_I^{\vphantom{T}} b_J^{\mathsf T}$.
Since the image of a span is the span of the images, the projection of $\cP$ is the span of $\{a_I^{\vphantom{T}} b_J^{\mathsf T} \,|\, \ba\in\cC_1, \bb\in\cC_2\}$, which is exactly $(\cC_1|_I)\otimes(\cC_2|_J)$ because every generator of the latter has the form $a_I^{\vphantom{T}} b_J^{\mathsf T}$.
\end{proof}
 
\begin{cor} \label{cor:bilateral_nesting}
For $i\in\{1,2\}$, choose $m_i\in\mathbb N_0$ and $\ell_i\in\mathbb N_0$, and let $U_i\subseteq\mathbb F_2^{m_i+\ell_i}$ be an affine subspace of dimension $m_i$ with affine embedding $\varphi_i\colon\mathbb F_2^{m_i}\to U_i$.
Then, the projection of $\PRM(r_1,m_1+\ell_1,r_2,m_2+\ell_2)$ onto $U_1\times U_2$, with coordinates relabeled through $(\varphi_1,\varphi_2)$, identifies $\PRM(r_1,m_1,r_2,m_2)$ via rectangular nesting.
\end{cor}

 \begin{proof}
By Remark~\ref{rem:rm_subcode}, a codeword of the longer code is the evaluation of a polynomial $h(\bx,\by)$ of bidegree at most $(r_1,r_2)$.
Substituting the affine maps $\bx=\varphi_1(\bx')$ and $\by=\varphi_2(\by')$ sends each monomial $\bx^{\bi}\by^{\bj}$ to a product of $|\bi|$ affine forms in $\bx'$ and $|\bj|$ affine forms in $\by'$, so $h(\varphi_1(\bx'),\varphi_2(\by'))$ again has bidegree at most $(r_1,r_2)$.
Hence the relabeled projection is contained in $\PRM(r_1,m_1,r_2,m_2)$.
Conversely, since $\varphi_i$ is an injective affine map, there is an affine map $\psi_i\colon\mathbb{F}_2^{m_i+\ell_i}\to\mathbb{F}_2^{m_i}$ with $\psi_i\circ\varphi_i=\mathrm{id}$.
Thus, given any $h'$ of bidegree at most $(r_1,r_2)$, the polynomial $h(\bx,\by)=h'(\psi_1(\bx),\psi_2(\by))$ has bidegree at most $(r_1,r_2)$ and restricts to $h'$.
This proves the reverse inclusion.
\end{proof}
 
\subsection{Product Code Symmetry and Transitivity}
\label{sec:pc_sym}

\begin{lem} \label{lem:product_aut}
Let $\cP=\cC_1\otimes\cC_2$.  If $\pi_1$ is a permutation automorphism of $\cC_1$ and $\pi_2$ is a permutation automorphism of $\cC_2$, then the permutation $(\pi_1\times\pi_2)(u,v)=(\pi_1(u),\pi_2(v))$ of the index set is a permutation automorphism of $\cP$.
\end{lem}

 \begin{proof}
Let $\bM' = (\pi_1\times\pi_2)\bM$, i.e., $M'_{u,v}=M_{\pi_1^{-1}(u),\pi_2^{-1}(v)}$.
Column $v$ of $\bM'$ is the permutation $\pi_1$ applied to column $\pi_2^{-1}(v)$ of $\bM$ and hence lies in $\cC_1$; similarly every row of $\bM'$ lies in $\cC_2$.
By Lemma~\ref{lem:product_char}, $\bM'\in\cP$.
\end{proof}
 
Since the affine group $\mathrm{AGL}(m_i,2)$ is contained in the automorphism group of $\RM(r_i,m_i)$, Lemma~\ref{lem:product_aut} shows that the automorphism group of $\PRM(r_1,m_1,r_2,m_2)$ contains $\mathrm{AGL}(m_1,2)\times\mathrm{AGL}(m_2,2)$, which acts transitively on $\mathbb{F}_2^{m_1}\times\mathbb{F}_2^{m_2}$.
Hence, $\RM$-product codes are transitive and all coordinates have the same marginal bit-MAP performance and the EXIT-area initialization in Theorem~\ref{thm:exit_mmse_H} applies.
Moreover, the stabilizer of the distinguished coordinate $(0,0)$ contains the subgroup induced by $\mathrm{GL}(m_1,2)\times\mathrm{GL}(m_2,2)$, acting on $\Omega=(\mathbb{F}_2^{m_1}\times\mathbb{F}_2^{m_2})\setminus\{(0,0)\}$ by $(u,v)\mapsto(G_1u,G_2v)$.
This action is analyzed in Section~\ref{sec:pc_bool}.

\begin{rem} \label{rem:not_2trans}
For $m_1,m_2\geq1$, the subgroup $\mathrm{GL}(m_1,2)\times\mathrm{GL}(m_2,2)$ has exactly three orbits on $\Omega$: the row axis $\{(u,0):u\neq0\}$, the column axis $\{(0,v):v\neq0\}$, and the off-axis pairs $\{(u,v):u\neq0,v\neq0\}$.
We do not consider the full automorphism group of the product code because our analysis does not require it.
\end{rem}

\begin{lem} \label{lem:product_not_dbl}
Let $\cC=\cC_1\otimes \cC_2$ be a binary product code where $\cC_i$ has minimum distance $d_i \geq 2$ and $\dim(\cC)>1$.
Then, $\cC$ is not doubly transitive.
\end{lem}

\begin{proof}
It is well-known that every minimum-weight codeword of $\cC$ has support $S_1\times S_2$, where $S_1$ and $S_2$ are supports of minimum-weight codewords of $\cC_1$ and $\cC_2$, respectively. 

For a coordinate $i$ of $\cC_1$, let $N_1 (i)$ be the number of minimum weight codewords in $\cC_1$ with a 1 in position $i$.
For distinct coordinates $i,i'$, let $N_1 (i,i')$ be the number of minimum weight codewords in $\cC_1$ with 1's in positions $i$ and $i'$.
Define $N_2(j,j')$ similarly for $\cC_2$.

Since $\dim(\cC)>1$, at least one component, say $\cC_1$, is not a
repetition code. Hence, $d_1$ is less than the length of $\cC_1$.
Choose a minimum-weight support $S_1$ of $\cC_1$ and coordinates
$i\in S_1$ and $i'\notin S_1$.
Then, $N_1(i)>N_1(i,i')$ because every minimum-weight support counted by $N_1(i,i')$ is also counted by $N_1(i)$, while the chosen support $S_1$ is counted only by $N_1(i)$.

Since $d_2\geq 2$, choose distinct coordinates $j,j'$ contained
in some minimum-weight support $S_2$ of $\cC_2$.
Thus, $N_2(j,j')>0$.

Now count minimum-weight codewords of $\cC$ containing two specified
coordinates.
A minimum-weight support $S_1'\times S_2'$ contains both
$(i,j)$ and $(i,j')$ if and only if $i\in S_1'$ and $j,j'\in S_2'$.
Therefore, the number of minimum-weight codewords containing
${(i,j),(i,j')}$ is $N_1(i)N_2(j,j')$.
Similarly, $S_1'\times S_2'$ contains both $(i,j)$ and $(i',j')$ if
and only if $i,i'\in S_1'$ and $j,j'\in S_2'$, so the number containing ${(i,j),(i',j')}$ is $N_1(i,i')N_2(j,j')$.
Since $N_1(i)>N_1(i,i')$ and $N_2(j,j')>0$,
these two numbers are different.

Any automorphism of $\cC$ permutes the minimum-weight codewords.
Hence, if $\cC$ were doubly transitive, every pair of distinct
coordinates would be contained in the same number of minimum-weight
codewords. The two pairs above violate this condition, so $\cC$ is
not doubly transitive.
\end{proof}

\subsection{Symmetry Bounds on Fourier Weight for Product Codes}
\label{sec:pc_bool}

The analysis in the remainder of this section requires only one new symmetry input: an evaluation of the membership probability $\Pr(\Pi(S)\in\cA)$ appearing in Lemma~\ref{lem:f_bool_restrict_var} for the group $\mathrm{GL}(m_1,2)\times\mathrm{GL}(m_2,2)$, which transforms the two coordinate factors independently.
The following corollary provides this and can be seen as the analogue of Lemma~\ref{lem:gl_sym} for the product action.

\begin{cor}[Product GL symmetry] \label{cor:product_gl_sym}
For $m_1,m_2\in\mathbb{N}$, let
\[ \Omega = \big(\mathbb{F}_2^{m_1}\times\mathbb{F}_2^{m_2}\big) \setminus\{(0,0)\}, \qquad n = |\Omega| = 2^{m_1+m_2}-1, \]
and suppose that $\mathrm{Sym}(f)$ for $f\colon \{0,1\}^{\Omega}\to\mathbb{R}$ contains the permutations of $\Omega$ induced by $\mathrm{GL}(m_1,2)\times\mathrm{GL}(m_2,2)$ acting as $(u,v)\mapsto(G_1u,G_2v)$.
Let $W_i\subseteq\mathbb{F}_2^{m_i}$ be a subspace of codimension $\ell_i\in[m_i+1]$ for $i \in \{1,2\}$ and let
\[ A=(W_1\times W_2)\setminus\{(0,0)\}. \]
For $S\subseteq\Omega$, define
\begin{align*}
d_1(S) &= \dim\Span\{u \in \mathbb{F}_2^{m_1} :(u,v)\in S\},\\
d_2(S) &= \dim\Span\{v \in \mathbb{F}_2^{m_2} :(u,v)\in S\}.
\end{align*}
If $\Pi$ is induced by independent uniform random elements $G_1 \in \mathrm{GL}(m_1,2)$ and $G_2 \in \mathrm{GL}(m_2,2)$, then
\begin{align}
\Pr\big(\Pi(S)\in\cA\big)
&= \frac{\qbinom{m_1-\ell_1}{d_1(S)}_2}{\qbinom{m_1}{d_1(S)}_2} \frac{\qbinom{m_2-\ell_2}{d_2(S)}_2}{\qbinom{m_2}{d_2(S)}_2} \nonumber\\
&\leq 2^{-\ell_1d_1(S)-\ell_2d_2(S)}, \label{eq:product_containment}
\end{align}
where a Gaussian binomial coefficient is zero if its lower index exceeds its upper index.
In particular, if $\ell_1=\ell_2=\ell$, then, for every nonempty $S\subseteq\Omega$,
\[ \Pr\big(\Pi(S)\in\cA\big) \leq \min\big\{2^{-\ell},\,|S|^{-\ell}\big\}. \]
\end{cor}

 \begin{proof}
Let $L_i(S)\subseteq\mathbb{F}_2^{m_i}$ be the span of the projection of $S$ onto its $i$-th coordinate, so that $\dim L_i(S)=d_i(S)$.
Since $W_1\times W_2$ is closed under the coordinate projections and $(0,0)\notin\Pi(S)$, the event $\Pi(S)\subseteq A$ is equivalent to the pair of events
\[ G_1L_1(S)\subseteq W_1, \qquad G_2L_2(S)\subseteq W_2, \]
because $G_i$ is linear and $W_i$ is a subspace.
These events are independent because $G_1$ and $G_2$ are.
Moreover, since $\mathrm{GL}(m_i,2)$ acts transitively on the $d$-dimensional subspaces of $\mathbb{F}_2^{m_i}$ and the sets $\{G : G L_i(S) = V\}$ are cosets of the setwise stabilizer of $L_i(S)$, the subspace $G_iL_i(S)$ is uniformly distributed over the $d_i(S)$-dimensional subspaces of $\mathbb{F}_2^{m_i}$.
The fraction of such subspaces contained in $W_i$ is $\qbinom{m_i-\ell_i}{d_i(S)}_2 \big/ \qbinom{m_i}{d_i(S)}_2$ and multiplying the two fractions proves the equality in~\eqref{eq:product_containment}.
For the inequality, termwise comparison in
\[ \frac{\qbinom{m_i-\ell_i}{d}_2}{\qbinom{m_i}{d}_2} = \prod_{j=0}^{d-1} \frac{2^{m_i-\ell_i-j}-1}{2^{m_i-j}-1} \leq 2^{-\ell_i d} \]
suffices, since $(2^{a-\ell}-1)/(2^{a}-1)\leq 2^{-\ell}$ for all $a \geq \ell \geq 0$ (and the left side is zero when $d > m_i-\ell_i$).

Now suppose $\ell_1=\ell_2=\ell$.
A nonempty $S$ contains a pair $(u,v)\neq(0,0)$, so $d_1(S)+d_2(S)\geq1$ and~\eqref{eq:product_containment} gives the bound $2^{-\ell}$.
Also, $S\subseteq (L_1(S)\times L_2(S))\setminus\{(0,0)\}$ implies $|S|\leq2^{d_1(S)+d_2(S)}-1$ and hence
\[ 2^{-\ell(d_1(S)+d_2(S))} \leq \big(|S|+1\big)^{-\ell} \leq |S|^{-\ell}, \]
which completes the proof.
\end{proof}
 
\begin{rem}
By Remark~\ref{rem:not_2trans}, the group $\mathrm{GL}(m_1,2)\times\mathrm{GL}(m_2,2)$ has only three orbits on $\Omega$.
Hence, the uniform bound $\rho\leq2^{-\ell}$ (with near-equality attained by a singleton $S$ on an axis) also follows from the orbit bound~\eqref{eq:orbit_rho} and the two-look product recursion in Section~\ref{sec:pc_bms} requires no new machinery.
The dimension-dependent decay in~\eqref{eq:product_containment} is needed only for the level-$k$ argument in Corollary~\ref{cor:product_boolean_bound}.
\end{rem}

For the level-$k$ analysis on the BSC in Section~\ref{sec:pc_bsc}, we also need the analogue of Lemma~\ref{lem:gl2_boolean_bound}.
The next lemma isolates exactly what that proof requires from the symmetry group: an upper bound on the membership probability $\Pr(\Pi(S) \in \cA)$ that decays with the size of $S$.
It applies to any distribution $q$ on $\mathrm{Sym}(f)$ and, in particular, to symmetry groups that are not doubly transitive.

\begin{lem} \label{lem:containment_profile}
Let $f\colon\{0,1\}^n\to\{0,1\}$ be a boolean function, let $\bX\in\{0,1\}^n$ be i.i.d.\ with $\Pr(X_0=1)=p\in(0,1)$, define $\lambda=\min\{p,1-p\}$, and write
\[ \alpha=\ex{f(\bX)} \in (0,1), \qquad c(\lambda)=\frac{1}{8\ln(1/\lambda)}. \]
Let $A\subseteq[n]$ and let $\Pi$ be a random element of $\mathrm{Sym}(f)$ drawn according to some distribution $q$.
If there are $\rho\in[0,1]$ and $\ell\geq1$ such that
\[ \Pr\big(\Pi(S)\in\cA\big) \leq \min\big\{\rho,\,|S|^{-\ell}\big\} \quad \text{for all nonempty } S\subseteq[n], \]
then we have
\[ \sum_{S\subseteq [n]}\widehat{f_A}(S)^2 \leq \alpha^2 +\rho\,\alpha^{7/6} +\alpha\big(c(\lambda)\ln(1/\alpha)\big)^{-\ell}. \]
\end{lem}

 \begin{proof}
Let $k_0=c(\lambda)\ln(1/\alpha)>0$.
Summing~\eqref{eq:wk_sym_bound} over all levels gives
\[ \sum_{S\subseteq[n]}\widehat{f_A}(S)^2 = \sum_{S\subseteq[n]} \hat{f}(S)^2\,\Pr\big(\Pi(S)\in\cA\big), \]
and we bound the right-hand side by splitting the sum into three parts.
The term indexed by the empty set equals $\hat{f}(\emptyset)^2=\alpha^2$.
For the nonempty sets with $1\leq |S|\leq k_0$, the membership probability is at most $\rho$, so Lemma~\ref{lem:biased_level_k} (with $\epsilon = 7/8$, i.e., $|S| \leq \frac18 \ln\alpha/\ln\lambda = k_0$) gives
\[ \sum_{S: 1\leq |S|\leq k_0} \hat{f}(S)^2\,\Pr\big(\Pi(S)\in\cA\big) \leq \rho\,\alpha^{7/6}. \]
For $|S|>k_0$, the membership probability is at most $|S|^{-\ell}\leq k_0^{-\ell}$.
Since $f$ is boolean, Parseval's identity gives $\sum_{S}\hat{f}(S)^2=\ex{f(\bX)^2}=\alpha$, so these terms contribute at most $\alpha\, k_0^{-\ell}$.
Adding the three bounds proves the claim.
\end{proof}
 
Combining Lemma~\ref{lem:containment_profile} with Corollary~\ref{cor:product_gl_sym} shows that the bound in Lemma~\ref{lem:gl2_boolean_bound} holds verbatim for functions whose symmetry group contains only the product of two general linear groups.

\begin{cor} \label{cor:product_boolean_bound}
In the setting of Corollary~\ref{cor:product_gl_sym} with $\ell_1=\ell_2=\ell\geq1$, let $f\colon \{0,1\}^{\Omega}\to\{0,1\}$ be invariant under the permutations induced by $\mathrm{GL}(m_1,2)\times\mathrm{GL}(m_2,2)$ on $\Omega$ and let $A=(W_1\times W_2)\setminus\{(0,0)\}$.
If $\bX$ is i.i.d.\ with $\Pr(X_0 = 1) = p \in (0,1)$ and $\alpha=\ex{f(\bX)} \in (0,1)$, then we have
\[ \sum_{S\subseteq \Omega}\widehat{f_A}(S)^2 \leq \alpha^2 +2^{-\ell}\alpha^{7/6} +\alpha\big(c(\lambda)\ln(1/\alpha)\big)^{-\ell}, \]
where $\lambda = \min\{p,1-p\}$ and $c(\lambda) = 1/(8\ln(1/\lambda))$.
\end{cor}

 \begin{proof}
Draw $\Pi$ uniformly from the subgroup of $\mathrm{Sym}(f)$ induced by $\mathrm{GL}(m_1,2)\times\mathrm{GL}(m_2,2)$.
Corollary~\ref{cor:product_gl_sym} gives $\Pr(\Pi(S)\in\cA)\leq\min\{2^{-\ell},|S|^{-\ell}\}$ for all nonempty $S$, so the result follows from Lemma~\ref{lem:containment_profile} with $\rho=2^{-\ell}$.
\end{proof}
 
\subsection{Rectangular Two- and Three-Look Constructions}
\label{sec:pc_looks}

Rectangular looks are obtained by applying the multiple-look construction of Definition~\ref{def:mult_look} independently in the two coordinate factors and pairing the resulting subspaces.

\emph{Two looks.}
For $\ell \in \mathbb{N}$ satisfying $\ell\leq\min\{m_1,m_2\}$ and $i\in\{1,2\}$, apply Definition~\ref{def:mult_look} with $(s,t)=(2,\ell)$ in $\mathbb{F}_2^{m_i+\ell}$ to obtain the $m_i$-dimensional subspaces
\begin{align*}
U_{i,0} &= \mathbb{F}_2^{m_i-\ell}\times\mathbb{F}_2^{\ell} \times\{0\}^{\ell}, \\
U_{i,1} &= \mathbb{F}_2^{m_i-\ell}\times\{0\}^{\ell} \times\mathbb{F}_2^{\ell},
\end{align*}
with $W_i \triangleq U_{i,0}\cap U_{i,1} =\mathbb{F}_2^{m_i-\ell}\times\{0\}^{2\ell}$ of codimension $\ell$ in each.
The rectangles
\[ L_j = U_{1,j}\times U_{2,j}, \qquad j\in\{0,1\}, \]
have intersection $L_0\cap L_1 = W_1\times W_2$, which contains $(0,0)$.
Each $U_{i,j}$ carries the natural linear parameterization by $\mathbb{F}_2^{m_i}$ (identity on the $\mathbb{F}_2^{m_i-\ell}$ block and placing the last $\ell$ coordinates in the $j$-th slot) and these parameterizations agree on $W_i$.
By Corollary~\ref{cor:bilateral_nesting}, under these identifications the projections of $\PRM(r_1,m_1+\ell,r_2,m_2+\ell)$ onto $L_0$ and $L_1$ are \emph{equal} to $\PRM(r_1,m_1,r_2,m_2)$, with a common labeling of the shared coordinates $W_1\times W_2$.

\emph{Three looks.}
Similarly, applying Definition~\ref{def:mult_look} with $(s,t)=(3,\ell)$ in $\mathbb{F}_2^{m_i+2\ell}$ gives subspaces $U_{i,0},U_{i,1},U_{i,2}$ of dimension $m_i$ whose pairwise intersections all equal a common subspace $W_i$ of codimension $\ell$.
Pairing indices gives three rectangles $L_j=U_{1,j}\times U_{2,j}$ for $j\in\{0,1,2\}$.  Every pair of rectangles has intersection $W_1\times W_2$ and their projections are all equal to $\PRM(r_1,m_1,r_2,m_2)$ under the natural identifications.
This is exactly the common-set structure required by Lemma~\ref{lem:bsc_code_bound_rho}.

\emph{The coefficient $\rho$.}
Relative to any one look, the shared observation set is $A = (W_1\times W_2)\setminus\{(0,0)\}$ with each $W_i$ of codimension $\ell$.
Drawing $\Pi$ uniformly from the subgroup induced by $\mathrm{GL}(m_1,2)\times\mathrm{GL}(m_2,2)$ (as permitted by the setup), Corollary~\ref{cor:product_gl_sym} gives
\begin{equation} \label{eq:product_rho}
\rho = \max_{S \subseteq \Omega: S\neq \emptyset} \Pr\big(\Pi(S)\in\cA\big) \leq 2^{-\ell},
\end{equation}
together with the refined profile $\Pr(\Pi(S)\in\cA)\leq\min\{2^{-\ell},|S|^{-\ell}\}$.
Thus, bilateral codimension-one restrictions give $\rho\leq1/2$ and bilateral codimension-two restrictions give $\rho\leq1/4$, exactly as in the single-$\RM$ recursions.

\begin{rem}[Overlap versus $\rho$] \label{rem:overlap_vs_rho}
The normalized overlap of the rectangles is $|L_0\cap L_1|/|L_0| = 2^{-2\ell}$, but this is \emph{not} the correct correlation coefficient.
Indeed, a singleton $S=\{(u,0)\}$ on the row axis has
\[ \Pr\big(\Pi(S)\in\cA\big) = \frac{2^{m_1-\ell}-1}{2^{m_1}-1} \approx 2^{-\ell} \gg 2^{-2\ell}. \]
For doubly-transitive codes, the normalized overlap bounds $\rho$ via Corollary~\ref{cor:symmetry_bounds}.
But, for product codes, the axis orbits are the bottleneck and~\eqref{eq:product_rho} is tight up to the factor $(1-2^{-(m_1-\ell)})/(1-2^{-m_1})$.
This is precisely where the Fourier-membership formulation of Lemma~\ref{lem:f_bool_restrict_var} pays off because it applies without modification to stabilizer actions that are not transitive.
\end{rem}

\subsection{Bit-Error Decay for $\PRM$ Codes on BMS Channels}
\label{sec:pc_bms}

Throughout this subsection, fix a BMS channel $W$ with capacity $C\in(0,1)$ and let $M(\cC)$ denote the extrinsic MMSE of bit 0 for the transitive code $\cC$, as in Section~\ref{sec:bms}.
For fixed $r_1,r_2,m_1,m_2$, define the nested sequence
\begin{equation} \label{eq:product_nested_sequence}
\cP_k \triangleq \PRM(r_1,m_1+k,r_2,m_2+k),\qquad k\geq0.
\end{equation}

\begin{lem} \label{lem:product_two_look}
For every $k\geq0$, on any BMS channel, we have the product two-look recursion
\[ \frac{M(\cP_{k+1})}{1-M(\cP_{k+1})} \leq \frac34\, \frac{M(\cP_k)}{1-M(\cP_k)}. \]
\end{lem}

 \begin{proof}
Write $\cC=\cP_{k+1}$ and use the two-look rectangular construction with $\ell=1$, so that the rectangles $L_0,L_1$ have intersection $W_1\times W_2$, which contains $(0,0)$.
Take $A=(W_1\times W_2)\setminus\{(0,0)\}$, $B=L_0\setminus(W_1\times W_2)$, $C=L_1\setminus(W_1\times W_2)$, and let $D$ be the remaining coordinates.
The code $\cC$ is transitive (Lemma~\ref{lem:product_aut}) and the projections of $\cC$ onto $\{0\}\cup A\cup B = L_0$ and $\{0\}\cup A\cup C = L_1$ are equal to $\cC'=\cC''=\cP_k$ under the identifications of Section~\ref{sec:pc_looks}.
The extrinsic conditional-mean decoding function $f$ of $\cC'$ inherits the symmetries of $\cC'$, so $\mathrm{Sym}(f)$ contains the permutations of $\Omega_k=(\mathbb F_2^{m_1+k}\times\mathbb F_2^{m_2+k})\setminus\{(0,0)\}$ induced by $\mathrm{GL}(m_1+k,2)\times\mathrm{GL}(m_2+k,2)$.
Drawing $\Pi$ uniformly from this subgroup, Corollary~\ref{cor:product_gl_sym} with $\ell_1=\ell_2=1$ gives $\rho\leq1/2$ in~\eqref{eq:product_rho}.
Lemma~\ref{lem:code_mmse_rho} (whose proof applies to real-valued functions of the channel outputs by Remark~\ref{rem:anova}) then gives the stated bound because $(1+\rho)/2\leq3/4$.
\end{proof}
 
We now fix the code family.
Given $C\in(0,1)$ and $\eta\in(0,C)$, choose $Q_1,Q_2\in(0,1)$ with
\begin{equation} \label{eq:Q_choice}
Q_1Q_2 = C-\frac{\eta}{4} \qquad\Big(\text{e.g., } Q_1=Q_2=\sqrt{C-\eta/4}\,\Big),
\end{equation}
and, for $i\in\{1,2\}$, define
\begin{equation} \label{eq:family_choice}
r_i(m) = \Big\lfloor \frac m2 + \frac{\sqrt m}{2}\, \Phi^{-1}(Q_i) \Big\rfloor , \qquad k_m = 8\Big\lfloor \frac{\eta\sqrt m}{16} \Big\rfloor .
\end{equation}
The codes of interest are the nested family
\begin{equation} \label{eq:family_def}
\cP_j \triangleq \PRM\big(r_1(m),m+j,r_2(m),m+j\big), \quad j\geq0,
\end{equation}
and the code $\cP_{k_m}$ has blocklength $N_m = 2^{2(m+k_m)}$.

\begin{lem}\label{lem:product_rate}
Let $A_j = R(\RM(r_1(m),m+j))$ and $B_j = R(\RM(r_2(m),m+j))$, so that $R(\cP_j)=A_jB_j$ by Lemma~\ref{lem:product_char}.
If $\sqrt m \geq \max\{32/\eta,\;6/Q_1,\;6/Q_2,\; |\Phi^{-1}(Q_1)|,\;|\Phi^{-1}(Q_2)|\}$, then
\[ R\big(\cP_{k_m}\big) \geq C-\eta, \quad 0 < R(\cP_0) \leq C - \frac{\eta}{8}, \quad k_m \geq \frac{\eta\sqrt m}{4}. \]
\end{lem}

 \begin{proof}
The condition $\sqrt m\geq|\Phi^{-1}(Q_i)|$ ensures $0\leq r_i(m)\leq m$.
By~\eqref{lem:rate_vs_floor}, we have $Q_i - 3/\sqrt{2\pi m} \leq A_0,B_0 \leq Q_i + 1/\sqrt{2\pi m}$ (with $Q_1$ for $A_0$ and $Q_2$ for $B_0$) and, by the last claim of Lemma~\ref{lem:rate}, $A_0 - A_j \leq j/(2\sqrt m)$ and $B_0 - B_j \leq j/(2\sqrt m)$.
Since all rates lie in $[0,1]$, this implies
\[ A_0B_0-A_jB_j = (A_0-A_j)B_0 + A_j(B_0-B_j) \leq \frac{j}{\sqrt m}. \]
Using $6/\sqrt{2\pi}\leq3$ and $k_m\leq\eta\sqrt m/2$, we find that
\begin{align*}
R\big(\cP_{k_m}\big)
&\geq A_0B_0 - \frac{k_m}{\sqrt m} \\
&\geq Q_1Q_2 - \frac{3}{\sqrt m} - \frac{\eta}{2} \\
&\geq C - \frac{\eta}{4} - \frac{3\eta}{32} - \frac{\eta}{2} \; \geq \; C-\eta,
\end{align*}
where $3/\sqrt m\leq3\eta/32$ follows from $\sqrt m\geq32/\eta$.
For the second claim, with $a = 1/\sqrt{2\pi m}\leq1$, we have
\begin{align*}
R(\cP_0)
&\leq (Q_1+a)(Q_2+a) \; \leq \; Q_1Q_2 + 3a \\
&\leq C - \frac{\eta}{4} + \frac{2}{\sqrt m} \; \leq \; C - \frac{\eta}{8},
\end{align*}
since $3/\sqrt{2\pi}\leq2$ and $2/\sqrt m\leq\eta/16$.
Positivity of $R(\cP_0)$ follows from $A_0 \geq Q_1 - 3/\sqrt{2\pi m} \geq Q_1 - Q_1/3 > 0$ (using $\sqrt m\geq6/Q_1$ and $3/\sqrt{2\pi}\leq2$, so that $3/\sqrt{2\pi m} \leq 2/\sqrt{m} \le Q_1/3$) and similarly for $B_0$.
Finally, $k_m \geq \eta\sqrt m/2 - 8 \geq \eta\sqrt m/4$ because $8\leq\eta\sqrt m/4$.
\end{proof}
 
The following theorem shows that $\RM$-product codes achieve vanishing bit-error probability for all rates less than channel capacity.

\begin{thm}\label{thm:product_bms}
Fix a BMS channel with capacity $C\in(0,1)$ and $\eta\in(0,C)$, and define $\cP_{k_m}$ by~\eqref{eq:Q_choice}--\eqref{eq:family_def}.
Then there is an $m_\eta<\infty$, depending only on $\eta$ and $Q_1,Q_2$, such that, for all $m\geq m_\eta$, we have $R(\cP_{k_m}) \geq C-\eta$ and
\[ 2P_b\big(\cP_{k_m}\big) \leq M\big(\cP_{k_m}\big) \leq \exp\Big(-\frac{\eta}{16}\sqrt m\Big). \]
Since $\ln N_m \leq 3 m\ln 2$, the bit-error probability also decays exponentially in $\sqrt{\ln N_m}$.
\end{thm}

 \begin{proof}
Let $m_\eta$ be large enough that the hypothesis of Lemma~\ref{lem:product_rate} holds and, in addition, $\sqrt m\geq 128\ln(8/\eta)/\eta$.
The rate claim is Lemma~\ref{lem:product_rate}.
The sequence $\{\cP_j\}_{j\geq0}$ is a strongly nested sequence of transitive codes because, for each $j$, the proof of Lemma~\ref{lem:product_two_look} exhibits subsets $A,B,C,D$ satisfying the hypotheses of Lemma~\ref{lem:code_mmse_rho} with $\rho_j\leq1/2$.
By Lemma~\ref{lem:product_rate}, $\delta \triangleq C-R(\cP_0) \geq \eta/8 > 0$, so Lemma~\ref{lem:nested_sym_mmse_bound} gives
\[ 2P_b\big(\cP_{k_m}\big) \leq M\big(\cP_{k_m}\big) \leq \Big(\frac34\Big)^{k_m}\,\frac{1-\delta}{\delta} \leq \Big(\frac34\Big)^{k_m}\,\frac{8}{\eta}. \]
Taking logarithms and using $k_m\geq\eta\sqrt m/4$ and $\ln(4/3)\geq 0.287$, we find that
\begin{align*}
\ln\frac{1}{M(\cP_{k_m})}
&\geq k_m\ln\frac43 - \ln\frac8\eta \\
&\geq 0.0717\,\eta\sqrt m - \ln\frac8\eta \; \geq \; \frac{\eta}{16}\sqrt m ,
\end{align*}
where the last step uses $(0.0717-0.0625)\,\eta\sqrt m \geq 0.0092\,\eta\sqrt m \geq \ln(8/\eta)$, which holds because $\sqrt m\geq128\ln(8/\eta)/\eta$.
Finally, $\ln N_m = 2(m+k_m)\ln2 \leq 3m\ln2$ because $k_m\leq\eta\sqrt m/2\leq m/2$.
\end{proof}
 
\subsection{Fast Bit-Error Decay for $\PRM$ Codes on the BSC}
\label{sec:pc_bsc}

We now specialize to a BSC with capacity $C\in(0,1)$, crossover probability $p = h^{-1}(1-C)\in(0,1/2)$, and constant $c = c(p) = 1/(8\ln(1/p))$.
Let $\bZ$ denote the i.i.d.\ Bernoulli($p$) error pattern.
As in Section~\ref{sec:bsc}, we fix, for each look code, an optimal extrinsic bit-MAP decoding function for bit $(0,0)$ whose ties are broken consistently on the orbits of the stabilizer subgroup induced by the product of row and column stabilizer subgroups.
By linearity and channel symmetry, the resulting error indicator $f\colon\{0,1\}^{\Omega}\to\{0,1\}$ depends only on the error pattern and is invariant under the induced permutations.
For the recursion from $\cP_k$ to $\cP_{k+2\ell}$, the index set is $\Omega_k=(\mathbb F_2^{m_1+k}\times\mathbb F_2^{m_2+k})\setminus\{(0,0)\}$ and Corollary~\ref{cor:product_boolean_bound} is applied with ambient dimensions $m_1+k,m_2+k$. 

\begin{lem} \label{lem:product_level_k_bsc}
Fix $\ell\geq1$ and let $\alpha = P_b(\cP_k)$.
If $\alpha\in(0,1)$, then we have the product level-$k$ recursion
\begin{align}
P_b(\cP_{k+2\ell}) \leq 3\alpha\left( \alpha+2^{-\ell}\alpha^{1/6} +\big(c\ln(1/\alpha)\big)^{-\ell} \right). \label{eq:product_level_recursion}
\end{align}
Moreover, for $\ell=2$, we have
\begin{equation} \label{eq:product_level_recursion_l2}
P_b(\cP_{k+4}) \leq \alpha \left(\frac{2}{c\ln(1/\alpha)}\right)^{2}.
\end{equation}
\end{lem}

 \begin{proof}
Write $\cC=\cP_{k+2\ell}$ and use the three-look rectangular construction of Section~\ref{sec:pc_looks}: rectangles $L_0,L_1,L_2$ whose pairwise intersections equal $W_1\times W_2$, which contains $(0,0)$.
Take $A=(W_1\times W_2)\setminus\{(0,0)\}$, $B=L_0\setminus(W_1\times W_2)$, $C=L_1\setminus(W_1\times W_2)$, and $D=L_2\setminus(W_1\times W_2)$, with $E$ the remaining coordinates.
Let $f$ be the invariant extrinsic bit-MAP error indicator of the common projection $\cC'=\cP_k$, so that $\alpha = \ex{f(Z_A,Z_B)}$.
By the data-processing inequality and the majority-vote bound~\eqref{eq:majority_union} (exactly as in the proof of Lemma~\ref{lem:bsc_code_bound_rho}), we have
\begin{align*}
P_b(\cC)
&\leq \ex{g_{\text{Maj}}(Z_A,Z_B,Z_C,Z_D)} \\
&\leq 3\,\ex{f(Z_A,Z_B)f(Z_A,Z_C)},
\end{align*}
where the three pairwise expectations coincide because the three looks are copies of the same code observed through identically distributed noise that is conditionally independent given $Z_A$.
Since $Z_B$ and $Z_C$ are independent given $Z_A$, we have
\begin{align*}
\ex{f(Z_A,Z_B)f(Z_A,Z_C)}
&= \ex{\ex{f(Z_A,Z_B)\,|\,Z_A}^2} \\
&= \sum_{S\subseteq \Omega}\widehat{f_A}(S)^2,
\end{align*}
by Parseval's identity applied to the restriction $f_A$ (whose Fourier support lies in the subsets of $A$).
Corollary~\ref{cor:product_boolean_bound} bounds the right side by $\alpha^2 + 2^{-\ell}\alpha^{7/6} +\alpha(c\ln(1/\alpha))^{-\ell}$ and multiplying by 3 proves~\eqref{eq:product_level_recursion}.
For $\ell=2$, the scalar estimate in the proof of Lemma~\ref{lem:rm_level_k_bsc} applies verbatim to the right-hand side of~\eqref{eq:product_level_recursion} and gives~\eqref{eq:product_level_recursion_l2}.
\end{proof}
 
The following theorem shows $\RM$-product codes achieve the fast bit-error probability decay, required for vanishing block-error probability, at all rates less than channel capacity.

\begin{thm} \label{thm:product_capacity}
Fix a BSC with capacity $C\in(0,1)$ and $\eta\in(0,C)$, and define $\cP_{k_m}$ by~\eqref{eq:Q_choice}--\eqref{eq:family_def}.
Then there is an $m_\eta<\infty$, depending only on $\eta$, $Q_1,Q_2$, and $c=c(p)$, such that, for all $m\geq m_\eta$, we have $R(\cP_{k_m})\geq C-\eta$ and
\[ P_b\big(\cP_{k_m}\big) \leq \exp\Big(-\frac{\eta}{128}\,\sqrt m\,\ln m\Big). \]
\end{thm}

 \begin{proof}
Let $m_\eta$ be large enough that, for all $m\geq m_\eta$:
\begin{enumerate}
\item[(i)] the hypothesis of Lemma~\ref{lem:product_rate} holds (so $k_m\geq\eta\sqrt m/4$, $R(\cP_{k_m})\geq C-\eta$, and $\delta = C-R(\cP_0)\geq\eta/8$);
\item[(ii)] $k_m \geq 64\ln(4/\eta)$;
\item[(iii)] $k_m \geq (16/c)^2$;
\item[(iv)] $m \geq (4/\eta)^4$ (so that $\ln k_m \geq \tfrac12\ln m - \ln(4/\eta) \geq \tfrac14\ln m$).
\end{enumerate}
The rate claim is contained in (i).
Recall that $k_m$ is a multiple of 8 and write $k=k_m$.

\emph{Coarse phase.}
As in the proof of Theorem~\ref{thm:product_bms}, Lemma~\ref{lem:nested_sym_mmse_bound} applied to the first $k/2$ steps of the nested sequence~\eqref{eq:family_def} gives
\[ P_b(\cP_{k/2}) \leq \frac12\Big(\frac34\Big)^{k/2}\frac{8}{\eta} = \Big(\frac34\Big)^{k/2}\frac{4}{\eta}, \]
and hence, using $\ln(4/3)\geq0.287$ and $k\big(\tfrac12\ln\tfrac43 - \tfrac18\big) \geq k/64 \geq \ln(4/\eta)$ by (ii),
\begin{equation} \label{eq:coarse_phase_out}
\ln\frac{1}{P_b(\cP_{k/2})} \geq \frac k2\ln\frac43 - \ln\frac4\eta \geq \frac k8 .
\end{equation}

\emph{Fast phase.}
Starting from $\cP_{k/2}$, apply Lemma~\ref{lem:product_level_k_bsc} with $\ell=2$ exactly $k/8$ times.
If $P_b(\cP_j)=0$ at any stage, the theorem is immediate.
Otherwise, the required condition $\alpha\in(0,1)$ holds at every application.
Each application advances both constituent dimensions by 4, so the final index is $k/2 + 4\cdot(k/8) = k$.
By~\eqref{eq:coarse_phase_out} and (iii), at the first application the recursive factor satisfies
\[ \left(\frac{2}{c\ln(1/P_b(\cP_{k/2}))}\right)^{2} \leq \Big(\frac{16}{ck}\Big)^{2} \leq 1 . \]
Since the factor is at most one, the bit-error probability does not increase along the recursion, so $\ln(1/P_b(\cP_j)) \geq k/8$ at every stage $j\geq k/2$ and the same bound on the factor remains valid throughout.
Consequently, we obtain
\begin{equation} \label{eq:fast_phase_out}
P_b\big(\cP_{k}\big) \leq P_b(\cP_{k/2})\,\Big(\frac{16}{ck}\Big)^{k/4} \leq e^{-k/8}\Big(\frac{16}{ck}\Big)^{k/4}.
\end{equation}
By (iii), $16/(ck)\leq k^{-1/2}$, so $\big(16/(ck)\big)^{k/4}\leq \exp(-\tfrac k8\ln k)$ and
\begin{align*}
P_b\big(\cP_{k_m}\big)
&\leq \exp\Big(-\frac k8 - \frac k8\ln k\Big) \; \leq \; \exp\Big(-\frac k8\ln k\Big) \\
&\leq \exp\Big(-\frac{\eta}{128}\sqrt m\,\ln m\Big),
\end{align*}
where the last step uses $k\geq\eta\sqrt m/4$ from (i) and $\ln k\geq\tfrac14\ln m$ from (iv).
\end{proof}
 
\begin{cor} \label{cor:product_blocklength}
In the setting of Theorem~\ref{thm:product_capacity}, with $N_m = 2^{2(m+k_m)}$ the product blocklength, there is an $m_\eta'\geq m_\eta$ such that, for all $m \geq m_\eta'$,
\[ P_b\big(\cP_{k_m}\big) \leq \exp\Big( -\frac{\eta}{500}\,\sqrt{\ln N_m}\,\ln\ln N_m \Big). \]
\end{cor}

 \begin{proof}
Since $k_m\leq m/2$, we have $2m\ln2 \leq \ln N_m \leq 3m\ln2$, so $m \geq \ln N_m/(3\ln 2) \geq \tfrac13\ln N_m$ and hence $\sqrt m\geq\tfrac{1}{\sqrt3}\sqrt{\ln N_m}$.
Moreover, $\ln m \geq \ln\ln N_m - \ln(3\ln2) \geq \tfrac12\ln\ln N_m$ once $\ln\ln N_m \geq 2\ln(3\ln2)$.
Substituting into Theorem~\ref{thm:product_capacity} gives the claim because $\tfrac{1}{128}\cdot\tfrac{1}{\sqrt3}\cdot\tfrac12 \geq \tfrac{1}{500}$.
\end{proof}
 
\begin{rem}
The constants in Theorem~\ref{thm:product_capacity} were not optimized.
The structure of the argument shows the general principle: bilateral nesting changes only constant factors because every recursive increment is paid in \emph{both} constituents while the product-rate loss is bounded by the \emph{sum} of the two constituent rate losses (Lemma~\ref{lem:product_rate}).
The same argument should apply almost without change to $s$-fold tensor powers with $s\geq 3$,
\[\RM(r_1,m_1)\otimes\cdots\otimes\RM(r_s,m_s).\]
The membership probability factors over the $s$ coordinate groups, the bound $\rho\leq2^{-\ell}$ and the profile $\min\{2^{-\ell},|S|^{-\ell}\}$ are unchanged, and only the constants in the rate bookkeeping depend on $s$.
\end{rem}

\subsection{Two-Pass Block Recovery}
\label{sec:pc_block}

We now convert the fast bit-error estimate into a vanishing block-error probability.
The decoder uses the hard bit-MAP decisions only to define short lists and it ranks the codewords in those lists using the original BSC observations.
This distinction is essential because the candidate array is a data-dependent function of the entire received word and its errors may be strongly correlated.

We state the argument for a rectangular product.  Let
\[ \cC_1\subseteq\{0,1\}^{N_1},\qquad \cC_2\subseteq\{0,1\}^{N_2},\qquad \cP=\cC_1\otimes\cC_2, \]
where columns belong to $\cC_1$ and rows belong to $\cC_2$.
Let $\bX\in\cP$ be transmitted over a BSC($p$) with $p\in(0,1/2)$ and write
\[ \bY=\bX\oplus \bZ, \qquad Z_{u,v}\overset{\mathrm{i.i.d.}}{\sim}\text{Bernoulli}(p). \]
As in Section~\ref{sec:list_dec}, we write $W^{N} (\by \,|\, \bc)$ for the probability of receiving $\by$ when $\bc$ is transmitted over $N$ uses of the BSC($p$) and we let
\[ \beta \triangleq 2\sqrt{p(1-p)} \in (0,1) \]
denote the Bhattacharyya parameter of the BSC($p$), so that $T_{\cC,\beta}(h)$ in~\eqref{eq:competition_function} is the truncated union bound of $\cC$ through Hamming weight $h$ for this channel.
Suppose that an arbitrary first-stage decoder produces a candidate array $\widehat \bX=\widehat \bX(\bY)$ and define its average bit-error probability by
\begin{equation} \label{eq:product_candidate_quality}
\varepsilon_{\rm cand} \triangleq \frac{1}{N_1N_2}\ex{\dH(\widehat \bX,\bX)}.
\end{equation}
No code constraint or independence assumption is imposed on $\widehat \bX$.

Fix relative radii $\tau_1,\tau_2\in(0,1/2)$ and put $t_1=\tau_1N_2$ and $t_2=\tau_2N_1$.  For every row $u$, form the list
\[ \cL_u =\{\bb\in\cC_2:\dH(\bb,\widehat X_{u,*})\leq t_1\} \]
and choose
\begin{equation} \label{eq:row_redecoder}
\widetilde X_{u,*} \in\arg\max_{\bb\in\cL_u}W^{N_2}(Y_{u,*}\mid \bb).
\end{equation}
An empty list or an unresolved tie is declared a failure.
Here, the word ``list'' denotes the intersection of the code with a Hamming ball and neither an upper bound on its cardinality nor an efficient enumeration procedure is needed.

\begin{lem} \label{lem:row_refinement}
For the expected fraction of row errors
\[ r_{\rm row} =\frac1{N_1}\sum_{u=0}^{N_1-1} \pr{\widetilde X_{u,*}\neq X_{u,*}}, \]
we have
\begin{equation} \label{eq:row_refinement_bound}
r_{\rm row} \leq \frac{\varepsilon_{\rm cand}}{\tau_1}+T_{\cC_2,\beta}(2t_1).
\end{equation}
\end{lem}

 \begin{proof}
Set $D_u=\dH(\widehat X_{u,*},X_{u,*})$.  From~\eqref{eq:product_candidate_quality} and Markov's inequality,
\begin{align}
\frac1{N_1}\sum_{u=0}^{N_1-1}\pr{D_u>t_1}
&\leq\frac{1}{N_1t_1}\sum_{u=0}^{N_1-1}\ex{D_u} \leq\frac{\varepsilon_{\rm cand}}{\tau_1}. \label{eq:true_row_missing}
\end{align}
Thus, the average probability that the transmitted row is absent from its list is at most $\varepsilon_{\rm cand}/\tau_1$.

For each row, the decoder in~\eqref{eq:row_redecoder} is exactly the list decoder of Lemma~\ref{lem:list_success} applied to $\cC_2$ with the estimate $\widehat X_{u,*}$ and radius $t_1$, so
\[ \pr{\widetilde X_{u,*}\neq X_{u,*}} \leq \pr{D_u > t_1} + T_{\cC_2,\beta}(2 t_1). \]
Averaging over $u$ and applying~\eqref{eq:true_row_missing} proves~\eqref{eq:row_refinement_bound}.

We emphasize again that the rows of $\widehat \bX$ may be arbitrarily correlated and that Lemma~\ref{lem:list_success} never conditions on the random list $\cL_u$, which may depend on the same received row used to rank it.
\end{proof}
 
Let the number of incorrectly refined rows be
\[ B_{\rm row}=\big|\{u:\widetilde X_{u,*}\neq X_{u,*}\}\big| \]
and define the event $\cE=\{B_{\rm row}\leq t_2\}$.
Since $\ex{B_{\rm row}}=N_1r_{\rm row}$, Markov's inequality and Lemma~\ref{lem:row_refinement} give
\begin{equation} \label{eq:bad_row_count}
\pr{\cE^c} \leq\frac{r_{\rm row}}{\tau_2} \leq \frac{\varepsilon_{\rm cand}}{\tau_1\tau_2} +\frac{T_{\cC_2,\beta}(2 t_1)}{\tau_2}.
\end{equation}

For each column $v$, form the second-pass list
\[ \cK_v =\{\ba\in\cC_1:\dH(\ba,\widetilde X_{*,v})\leq t_2\} \]
and choose
\begin{equation} \label{eq:column_redecoder}
\overline X_{*,v} \in\arg\max_{\ba\in\cK_v}W^{N_1}(Y_{*,v}\mid \ba).
\end{equation}
Again, an empty list or an unresolved tie is counted as a failure.

\begin{thm} \label{thm:two_pass_bound}
The probability that the two-pass list decoder defined by~\eqref{eq:row_redecoder}--\eqref{eq:column_redecoder} fails to recover the entire product codeword satisfies
\begin{equation} \label{eq:master_block_bound}
P_{L}(\cP) \leq \frac{\varepsilon_{\rm cand}}{\tau_1\tau_2} +\frac{T_{\cC_2,\beta}(2\tau_1N_2)}{\tau_2} +N_2\,T_{\cC_1,\beta}(2\tau_2N_1).
\end{equation}
Thus, the block-error probability satisfies $P_B (\cP) \leq P_L (\cP)$.
\end{thm}

 \begin{proof}
On $\cE$, the refined array $\widetilde \bX$ differs from $\bX$ in at most $t_2$ rows.
Therefore, simultaneously for every column $v$, we have
\[ \dH(\widetilde X_{*,v},X_{*,v})\leq B_{\rm row}\leq t_2, \]
so the true column belongs to $\cK_v$.
If the column decoder nevertheless fails, then there exists $\ba\in\cC_1\setminus\{X_{*,v}\}$ such that
\[ \dH(\ba,X_{*,v})\leq2t_2, \quad W^{N_1}(Y_{*,v}\mid \ba) \geq W^{N_1}(Y_{*,v}\mid X_{*,v}). \]
Lemma~\ref{lem:low_weight_competition} bounds this unconditional event by $T_{\cC_1,\beta}(2 t_2)$ and a union bound over the $N_2$ columns yields
\[ P_{L} (\cP)\leq\pr{\cE^c}+N_2\,T_{\cC_1,\beta}(2 t_2). \]
Substitution of~\eqref{eq:bad_row_count} proves~\eqref{eq:master_block_bound}.
The block-MAP decoder minimizes block-error probability, so it performs at least as well as this two-pass decoder.

We note that, as in the row argument, no conditional channel law is used.
In particular, we do not condition on $\cE$, the set of bad rows, or the column lists.
We use $\cE$ only for the deterministic implication that the true columns lie in their lists and then enlarge every column failure to an unconditional competition event governed by the original i.i.d.\ BSC noise.
\end{proof}
 
We finally specialize the master bound to the $\RM$-product sequence of Theorem~\ref{thm:product_capacity}.
The two competition terms are controlled by the weight-enumerator estimate in Lemma~\ref{lem:as_weight_enum}.

\begin{thm} \label{thm:product_block}
Consider the setting of Theorem~\ref{thm:product_capacity} and write $\bar m = m+k_m$ for the dimension of each constituent code, so that $\cP_{k_m} = \PRM(r_1(m),\bar m,r_2(m),\bar m)$.
Set $\nu = \eta/128$ so that $\varepsilon_m \triangleq P_b(\cP_{k_m}) \leq 2^{-\nu\sqrt m \log_2 \! m}$ and choose the radii
\[ \nu_1=\nu_2=\frac{\nu}{3}, \quad \tau_1=2^{-\nu_1\sqrt m \log_2 \! m}, \quad \tau_2=2^{-\nu_2\sqrt m \log_2 \! m}. \]
Then, the two-pass decoder of Theorem~\ref{thm:two_pass_bound} satisfies
\begin{align}
P_B(\cP_{k_m}) &\leq 2^{-(\nu\sqrt m \log_2 \! m)/3} +O\big(2^{(\nu \sqrt{m} \log_2 \! m)/3} 2^{-2^{\bar m/3}}\big) \tc{\nonumber} \tcb \tca \tc{\qquad} +O\big(2^{\bar m}2^{-2^{\bar m/3}}\big) \; \longrightarrow \; 0. \label{eq:product_block_decay}
\end{align}
In particular, the $\RM$-product code sequence achieves the capacity of the BSC under block-MAP decoding.
\end{thm}

 \begin{proof}
Use the hard bit-MAP decisions as the candidate $\widehat \bX$.
Transitivity gives~\eqref{eq:product_candidate_quality} with $\varepsilon_{\rm cand}=\varepsilon_m$ and we apply Theorem~\ref{thm:two_pass_bound} with the displayed radii.
Since $\nu_1+\nu_2=2\nu/3<\nu$, the first term is
\[ \frac{\varepsilon_m}{\tau_1\tau_2} \leq 2^{-(\nu-2\nu/3)\sqrt m \log_2 \! m} = 2^{-(\nu\sqrt m \log_2 \! m)/3}=o(1). \]
Both constituents are $\RM$ codes with parameter $\bar m = m+O(\sqrt m)$ and length $N=2^{\bar m}$ and, by~\eqref{eq:family_choice}, they satisfy $|\bar m - 2 r_i(m)| = O(\sqrt{\bar m})$.

Next, we bound the two competition terms using Lemma~\ref{lem:as_weight_enum} with the Bhattacharyya parameter $\beta = 2\sqrt{p(1-p)}\in (0,1)$ of the BSC($p$).
The required cutoffs are $2\tau_1 N = 2^{\bar m - \nu_1 \sqrt m \log_2 m + 1}$ and $2\tau_2 N = 2^{\bar m - \nu_2 \sqrt m \log_2 m + 1}$.  Since $\bar m /m \to 1$, we have $\nu_i \sqrt m \log_2 m -1 \geq \frac{\nu_i}{2} \sqrt{\bar m} \log_2 \bar m$ for sufficiently large $m$.
Hence, $2\tau_i N \leq 2^{\bar m-(\nu_i/2)\sqrt{\bar m}\log_2 \bar m}$ and the lemma applies to each term with $\tilde m=\bar m$ and $c=\nu_i/2$.

Thus, the remaining two terms in~\eqref{eq:master_block_bound} are those displayed in~\eqref{eq:product_block_decay} and they vanish doubly exponentially in $\bar m$ up to subexponential and exponential prefactors.
This proves that the explicit two-pass decoder has a vanishing block-error probability and the same is therefore true for block-MAP decoding.
\end{proof}
 
\begin{rem}
The decoder is information-theoretic in the sense that it need not enumerate either component list efficiently.
The result is correspondingly a block-MAP capacity statement and not a polynomial-time decoding theorem.
The argument extends without change to rectangular products, with separate radii and separate truncated union bounds for the two constituents.
\end{rem}

\begin{cor} \label{cor:product_block_bms}
$\RM$ product codes achieve capacity under block-MAP decoding on every BMS channel.
\end{cor}

 \begin{proof}
Theorem~\ref{thm:product_block} implies a vanishing block-error probability on the BSC.
By Lemma~\ref{lem:product_char}, the minimum distance of the balanced product family is
\[ d_m =2^{\bar m-r_1(m)}2^{\bar m-r_2(m)} =2^{\bar m+O(\sqrt{\bar m})}, \]
whereas its blocklength is $N_m=2^{2\bar m}$.
Hence $d_m/\ln N_m\to\infty$ and Theorem~\ref{thm:bsc_to_bms} applies.
Given a BMS channel of capacity $C$ and a target rate below $C$, choose an intermediate BSC capacity strictly between the target rate and $C$, construct the product family for that BSC, and apply the transfer theorem.
This gives a vanishing block-error probability on the original BMS channel.
\end{proof}
 
\subsection{Discussion}
\label{sec:pc_outlook}

The results of this section illustrate the division of labor in the recursive framework.
Everything that depends only on transitivity and the coefficient $\rho$ (i.e., the EXIT initialization, the two-look MMSE recursion, and the three-look majority argument) applies to product codes without change.
The only genuinely new inputs are the membership-probability bound of Corollary~\ref{cor:product_gl_sym} and the observation in Remark~\ref{rem:overlap_vs_rho} that, for nontransitive stabilizer actions, the correct coefficient is the membership probability and not the normalized overlap.
Three further points deserve mention.
First, Theorem~\ref{thm:product_bms} holds on every BMS channel, while the faster decay of Theorem~\ref{thm:product_capacity} is proved only for the BSC, thus mirroring the situation for $\RM$ codes in this paper.
Second, for the bit-error probability bounds, the key is that the product structure is compatible with the previous argument rather than providing any real gain.
Third, Theorem~\ref{thm:two_pass_bound} is a genuinely product-specific bit-to-block principle.
Its proof uses the row constraints to reduce the residual uncertainty to a small set of row locations and then uses the column constraints to remove that uncertainty.
The unconditional competition-event argument may be useful for other array-code constructions whose component codes admit sufficiently strong low-weight enumerator bounds.

\section{From BSC Block Error to BMS Block Error} \label{sec:tzs}

This section shows that, for a code sequence whose minimum distance grows sufficiently fast, achieving a non-trivial block-error probability on the BSC with capacity $C$ is sufficient to achieve vanishing block-error probability on any BMS channel with capacity at least $C$.
To do this, we combine two results.
First, we use a well-known result from~\cite{Tillich-cpc00} to amplify a weak bound on the block-error probability for the BSC($p$) channel into a strong bound for the BSC($p'$) with $p'$ slightly smaller than $p$.
Then, we apply a little-known result from~\cite[p.~88]{Sasoglu-phd11} that extends a block-error probability bound on the BSC($p$) to a similar bound on any BMS channel with the same or higher capacity.
This gives a sort of folk theorem that is known to experts but not written anywhere explicitly.

Specifically, consider a sequence $\cC_m$ of codes where the $m$th code has length $N_m$, rate $R(\cC_m) $, and minimum distance $d_m$.
We assume that $N_m$ is increasing and that $R(\cC_m) \to R \in (0,1)$.
Let $B_m (p)$ denote the block-error probability of block-MAP decoding for $\cC_m$ on the BSC($p$).
For this setup, we have the following theorem.
\begin{thm} \label{thm:bsc_to_bms} Consider a sequence of codes $\cC_m$ with length $N_m$, minimum distance $d_m$, and block-error probability $B_m (p)$ on the BSC($p$) with $p\in(0,1/2)$.
For fixed $\kappa\geq 8\sqrt{\ln 2}/p$, assume communication over any BMS channel with capacity at least $1-h(p-8\sqrt{\ln 2}/\kappa)$.
If $N_m \geq 8$, $d_m \geq \kappa^2 \ln N_m$, and $B_m (p) \leq 1-1/N_m^2$, then
$B_m (p-8\sqrt{\ln 2}/\kappa) \leq 1/N_m^2$ and block-MAP decoding of $\cC_m$ has a block-error probability at most $2/N_m$.
\end{thm}

 \begin{proof}
For the code $\cC_m$, assume the all-zero codeword is sent over a BSC($p$) and received as $\bY$.
In this case, syndrome decoding can implement block-MAP decoding and we choose the syndrome decoder where each syndrome is assigned the minimal weight error pattern in the associated coset that appears first in the lexicographical ordering on $\{0,1\}^{N_m}$.
Let $f_m\colon\{0,1\}^{N_m}\to\{0,1\}$ be the boolean function for which $f_m(\by)=0$ if the syndrome decoder returns the all-zero codeword and $f_m(\by)=1$ otherwise.

For $p\leq 1/2$, we have $B_m (p) = \ex{f_m (\bY)}$ and we define $\theta_m = B_{m}^{-1} (1/2)$.  
The results of~\cite{Tillich-cpc00} show that $B_m (p)$ satisfies
\begin{align*}
B_m (p) &\leq 1 - \Phi\big(\alpha_m (p)\big) \quad \text{for } 0<p<\theta_m \\
B_m (p) &\geq 1 - \Phi\big(\alpha_m (p)\big) \quad \text{for } \theta_m<p \leq 1/2,
\end{align*}
where $\alpha_m (p) \coloneqq \sqrt{d_m}\big( \sqrt{-\ln(1-\theta_m)}-\sqrt{-\ln (1-p)} \big)$ and $\Phi\colon \mathbb{R} \to [0,1]$ is the c.d.f.\ of a standard Gaussian. 
For non-trivial codes, we have $\theta_m \leq 1/2$ because, for $p=1/2$, $\bY$ is uniform and $B_m (1/2) =1- 2^{-N_m R(\cC_m)}$.

Since $\sqrt{x}-\sqrt{x'} = (x-x') / (\sqrt{x}+\sqrt{x'})$, it follows that
\begin{align} \label{eq:alpha_m_p}
\alpha_m (p)
&= \sqrt{d_m}\frac{-\ln(1-\theta_m)+\ln (1-p)}{ \sqrt{-\ln(1-\theta_m)}+\sqrt{-\ln (1-p)}}.
\end{align}
For $0\leq a \leq b \leq 1/2$, we note that
\[ \ln \frac{1-a}{1-b} = \ln(1-a) - \ln(1-b) = \int_{a}^b \frac{1}{1-z} dz \geq b-a \] because $\frac{1}{1-z}\geq 1$ for $z\in [0,1/2]$.  
If we test the $m$th code on a BSC with error probability $p_m$ and use $\big| \ln\frac{1-p_m}{1-\theta_m} \big| \geq |p_m-\theta_m| $ for $p_m,\theta_m \in [0,1/2]$, then we can solve~\eqref{eq:alpha_m_p} for $\ln\frac{1-p}{1-\theta_m}$ to see that
\begin{align}
|p_m-\theta_m|
&\leq \left| \ln \frac{1-p_m}{1-\theta_m} \right| \nonumber \\
&= \frac{|\alpha_m (p_m)|}{\sqrt{d_m}} \tc{\nonumber} \tcb \tca \tc{\qquad \cdot}  \left(  \sqrt{-\ln(1-\theta_m)}+\sqrt{-\ln (1-p_m)} \right) \nonumber\\
& \leq \frac{2\sqrt{\ln 2}|\alpha_m (p_m)|}{\sqrt{d_m}}, \label{eq:p_theta_alpha_bound}
\end{align}
where the second inequality follows from $\sqrt{-\ln (1-x)} \leq \sqrt{\ln 2}$ for $x\leq 1/2$.

Since $B_m (p)$ is strictly increasing for $p\in[0,1/2]$, we can bound the transition width $B_{m}^{-1} \left(1-\delta_m \right) - B_{m}^{-1} \left(\delta_m \right)$.
For $z \geq 0$, we use the standard bound
\[  1-\Phi(z) \leq \frac{1}{\sqrt{2 \pi z^2}}   e^{-z^2 / 2} \]
to see that $\alpha_m (p_m) = \sqrt{2\ln (1/\delta_m)}$ implies
\begin{align*}
B_m (p_m) &= 1-\Phi\big(\alpha_m (p_m) \big) \\
&\leq \sqrt{\frac{1}{2 \pi  \ln(1/\delta_m)}} e^{- \ln (1/\delta_m)} \\
&\leq \delta_m.
\end{align*}
and $B_{m}^{-1} \left(\delta_m \right) \geq p_m$.
For the other side, we can choose $\alpha_m (p_m ') = -\sqrt{2 \ln (1/\delta_m)}$ to see that
\begin{align*}
B_m (p_m ') &= 1-\Phi\big(\alpha_m (p_m ') \big) \\
&\geq 1-\sqrt{\frac{1}{2 \pi  \ln (1/\delta_m)}} e^{-\ln (1/\delta_m)} \\
&\geq 1-\delta_m
\end{align*}
and $B_{m}^{-1} \left(1-\delta_m \right) \leq p_m '$ because, for $z\geq 0$,
\[ 1-\Phi(-z) = \Phi(z) \geq 1- \frac{1}{\sqrt{2 \pi z^2}}   e^{-z^2 / 2}.  \]
Thus, it follows from~\eqref{eq:p_theta_alpha_bound} that
\begin{align*}
B_{m}^{-1} \left(1-\delta_m \right) - B_{m}^{-1} \left(\delta_m \right) &\leq p_m' - \theta_m + \theta_m - p_m \\
&\leq 4\sqrt{2\ln 2} \sqrt{ \frac{\ln(1/\delta_m)}{d_m}}.
\end{align*}

If we choose $\delta_m = 1/N_m^2$, then $B_m (p_m) \leq 1/N_m^2$ and we can apply~\cite[Proposition~7.1]{Sasoglu-phd11} to see that this code sequence has a block-error probability of at most
\[ \frac{1}{N_m} + h\left(\frac{1}{N_m^2}\right)  \underset{\text{if }N_m \geq 8}{\leq} \frac{2}{N_m}\]
on any BMS channel whose capacity is at least $1-h(p_m)$.
Finally, if $d_m \geq \kappa^2 \ln N_m$ and $B_{m} (p) \leq 1-1/N_m^2$, then it follows that
\[ p - p_m \leq p_m ' - p_m \leq 4\sqrt{2\ln 2} \sqrt{ \frac{2\ln N_m}{d_m}} \leq \frac{8\sqrt{\ln 2}}{\kappa}. \qedhere \]
\end{proof}

\begin{cor}
If $d_m = \omega(\ln N_m)$ and there is an $m_0 < \infty$ such that $B_m (p) \leq 1-1/N_m^2$ for all $m\geq m_0$, then the code sequence achieves a vanishing block-error probability under block-MAP decoding on any BMS channel whose capacity is greater than $1-h(p)$.
\end{cor}
\begin{proof}
    Fix a BMS channel of capacity $C>1-h(p)$. By continuity, choose a fixed $\kappa\geq8\sqrt{\ln2}/p$ large enough that $C\geq1-h(p-8\sqrt{\ln2}/\kappa)$. Since $d_m/\ln N_m\to\infty$, the conditions $N_m\geq8$ and $d_m\geq\kappa^2\ln N_m$ hold for all sufficiently large $m$. Theorem~\ref{thm:bsc_to_bms} then gives block-error probability at most $2/N_m\to0$.
\end{proof}

\section{Conclusions and Outlook}

In this paper, we revisit and extend several techniques used to analyze the performance of structured codes on BMS channels.
The strategy outlined in this work unifies a number of previous results, simplifies key arguments, and requires weaker assumptions.
By leveraging code symmetry and techniques such as two-look or three-look recursive nesting schemes, we show how performance bounds for short codes can be systematically transferred to longer codes.
An important element in these results is the connection between soft-decision decoding functions for BMS channels and powerful techniques for the analysis of boolean functions.

For the BSC, hypercontractivity and ``level-\(k\)'' inequalities are used to improve the decay rate of the bit-error probability bounds.
By combining the fast bit-error decay with list-decoding arguments and known bounds on the weight enumerator of RM codes, the same techniques guarantee vanishing block-error probability at rates arbitrarily close to capacity.
Finally, a result of Tillich and Z\'{e}mor is also combined with a theorem of \c{S}a\c{s}o\u{g}lu to show that nontrivial bounds on block-error probability for the BSC can be transferred to arbitrary BMS channels, provided the minimum distance grows moderately fast with block length.
Essentially the same program is applied to the product of $\RM$ codes in Section~\ref{sec:product_rm} with a slight twist to prove vanishing block-error probability.

In summary, connections between code symmetry, nesting, and functional analysis continue to provide new insight about structured codes and their ability to approach capacity on a range of channels.
In terms of outlook, we now note some natural extensions and open questions:
\begin{itemize}
\item The recursive two-look approach has been extended to show that RM codes on binary-input symmetric classical-quantum channels can reliably recover some bits at any rate below the Holevo capacity~\cite{Mandal-isit25}.
However, to achieve the faster decay rate and extension to block-error probability, some new ideas are required due to the measurement restrictions imposed by quantum mechanics.
\item For generalized RM (GRM) codes on classical non-binary channels, we believe the techniques in this paper can be combined with the results of~\cite{Reeves-isit23} to establish the ``faster decay rate'' for the symbol-error probability.
However, as far as we know, the extension to block-error probability does not immediately follow because current bounds on the weight enumerators of GRM codes are not strong enough.
\item As mentioned in Remark~\ref{rem:floral_block}, if the correlation bound can be strengthened for multiple looks, then it may be possible to achieve a super-exponential decay rate in $k$ which would establish the vanishing block-error probability result without the list-decoding argument.
\item While the two-look bound in this work does not explicitly require a code sequence based on RM codes, all known cases where it applies are built from RM codes.
For example, Section~\ref{sec:product_rm} shows that the tensor product of two $\RM$ codes, which is transitive but not doubly transitive in general, also falls within the reach of the method.
Extending the analysis to code families not built from RM codes would be quite interesting.
\end{itemize}

\section*{Acknowledgements}

While all the arguments for this paper were developed in 2024 without the use of AI, the authors would like to acknowledge the use of ChatGPT as an editorial assistant to improve the clarity and identify small errors in the previous version of this manuscript.
All text was independently checked and validated by the authors, who take full responsibility for the content of the manuscript.

\appendices

\makeatletter
\renewcommand{\thesection}{\Alph{section}}
\renewcommand{\thesubsection}{\thesection.\arabic{subsection}}
\renewcommand\thesubsectiondis{\thesection.\arabic{subsection}}
\renewcommand{\thesubsubsection}{\thesection.\arabic{subsection}.\arabic{subsubsection}}
\renewcommand{\thesubsubsectiondis}{\thesection.\arabic{subsection}.\arabic{subsubsection}}
\makeatother

\section{Deferred Proofs}
\label{app:deferred}

\subsection{Proof of Lemma~\ref{lem:rate}}
\label{proof:lem:rate}

\begin{proof}
For $\cC = \RM(r,m)$, it is well-known that $R(\cC) = \pr{\mathrm{Bin}(m,1/2) \leq r} $ with
\[ \pr{\mathrm{Bin}(m,1/2) \leq r} \coloneqq \frac{1}{2^{m}} \sum_{i=0}^r \binom{m}{i} \]
and an absolute bound on $R(\cC_k)$ follows directly from the Berry-Esseen central limit theorem.
Using an improved form of this bound for the symmetric binomial distribution~\cite[Corollary~1.2]{Hipp-siamtpa08}, one gets
\[ \left| \pr{\mathrm{Bin}(m,1/2) \leq r} - \Phi\left( \frac{2r-m}{\sqrt{m}} \right) \right| \leq \frac{1}{\sqrt{2\pi m}} \]
for all $r,m\in \mathbb{N}_0$ satisfying $0 \leq r \leq m$.
If $r =\lfloor m/2+ \sqrt{m}\,\Phi^{-1} (R) /2 \rfloor$, then the monotonicity of $\Phi$ implies
\[ \Phi\left( \frac{2r-m}{\sqrt{m}} \right) \leq R\]
and one gets the upper bound $R(\cC) \leq R + 1/\sqrt{2\pi m}$ in~\eqref{lem:rate_vs_floor}.
Likewise, the lower bound in~\eqref{lem:rate_vs_floor} follows from
\begin{align*}
| R(\cC) & -  R | 
\leq \Big| \pr{\mathrm{Bin}(m,1/2) \leq r} - \Phi\Big( \frac{2r-m}{\sqrt{m}} \Big) \tcb \tca \tc{\qquad \qquad \qquad \qquad \qquad \;} + \Phi\Big( \frac{2r-m}{\sqrt{m}} \Big) - R \Big| \\
&\leq \left| \pr{\mathrm{Bin}(m,1/2) \leq r} - \Phi\left( \frac{2r-m}{\sqrt{m}}  \right) \right| \tcb \tca \tc{\qquad \qquad \qquad \quad \;\;\,} + \left| \Phi\left( \frac{2r-m}{\sqrt{m}} \right) - R \right| \\
&\leq \frac{1}{\sqrt{2\pi m}} + \frac{2}{\sqrt{2\pi m}},
\end{align*}
where the second term follows from $(2r-m)/\sqrt{m} \geq \Phi^{-1} (R) - 2/\sqrt{m}$ and $\Phi'(z) = \frac{1}{\sqrt{2\pi}} e^{-z^2/2} \leq \frac{1}{\sqrt{2\pi}}$. 

For the second statement, we can work directly and write
\begin{align*}
R&(\cC_k) - R(\cC_{k+1}) \\
&= \frac{1}{2^{m+k}} \sum_{i=0}^r \left(\binom{m+k}{i}-\frac{1}{2}\binom{m+k+1}{i} \right) \\
&\overset{(a)}{=} \frac{1}{2^{m+k+1}} \binom{m+k}{r} \\
&\overset{(b)}{\leq} \frac{1}{2\sqrt{m+k}},
\end{align*}
where $(a)$ follows from $\binom{m+k+1}{i+1} = \binom{m+k}{i+1} + \binom{m+k}{i}$ via induction on $r$ and $(b)$ follows from using the factorial bound $\sqrt{2\pi n}\left(\frac{n}{e}\right)^n e^{\frac{1}{12n + 1}} < n! < \sqrt{2\pi n}\left(\frac{n}{e}\right)^n e^{\frac{1}{12n}}$ from \cite{Robbins-ams55} to verify that
\[\binom{m+k}{r} \leq \binom{m+k}{\lfloor (m+k)/2 \rfloor} \leq \frac{2^{m+k}}{\sqrt{m+k}}. \]
Using this, the stated result follows from
\begin{align*}
 R(\cC_0) - R(\cC_{k})
 &= \sum_{i=0}^{k-1} R(\cC_i) - R(\cC_{i+1})  \\
 &\leq \sum_{i=0}^{k-1} \frac{1}{2\sqrt{m+i}} \\
 & \leq \frac{k}{2\sqrt{m}}.  \qedhere
 \end{align*}
\end{proof}

\section{Subspace Codes and Sunflowers}
\label{app:sunflower}

Let $\mathcal{V} = \{V_0,V_1,\ldots,V_{M-1}\}$ be a collection of $t$-dimensional subspaces of $\mathbb{F}_2^{st}$ that are pairwise trivially intersecting (i.e., $V_i\cap V_j=\{0\}$ whenever $i\ne j$).
Such a collection forms a length-$st$ subspace code of size $M$ with constant dimension $t$ and distance $2t$~\cite{Trautmann-it13}.
Then, for $i\in [M]$, we can define $U_i = \mathbb{F}_2^{m-st} \times V_i$ to get a collection of $(m-(s-1)t)$-dimensional subspaces of $\mathbb{F}_2^m$ whose pairwise intersections all equal $\mathbb{F}_2^{m-st} \times \{0\}^{st}$.
From this, the fractional overlap of the subspaces equals
\[ |U_i \cap U_j|/|U_i|= 2^{m-st} / 2^{m-(s-1)t} = 2^{-t}. \]
Thus, a subspace sunflower containing $M$ subspaces can be easily constructed from a subspace code containing $M$ pairwise trivially intersecting subspaces of fixed dimension.

To construct an optimal subspace sunflower, we identify $\mathbb{F}_2^{st}$ with the extension field $F=\mathbb{F}_{2^{st}}$ and let $\alpha \in F$ be a primitive element.
Since $t\mid st$, the field $F$ contains the subfield $\mathbb F_{2^t}$. Define $M=(2^{st}-1)/(2^t-1)$ and $\beta=\alpha^M$, so $\mathbb F_{2^t}=\{0,1,\beta,\beta^2,\ldots,\beta^{2^t-2}\}$ is a $t$-dimensional $\mathbb F_2$-subspace of $F$. For $i\in[M]$, let $V_i=\alpha^i\mathbb F_{2^t}=\{0,\alpha^i,\alpha^i\beta,\ldots,\alpha^i\beta^{2^t-2}\}$. These $M$ subspaces are distinct, each has dimension $t$, and $V_i\cap V_j=\{0\}$ for $i\ne j$.
Thus, their subspace distance is $2t$.
Their nonzero elements partition $F^*$ and such a construction is called a $t$-spread.

This construction is well-known and the resulting code is called a cyclic orbit code~\cite{Trautmann-it13}.
It is an optimal spread code because the distinct subspaces intersect only in $\{0\}$ and their nonzero elements partition $\mathbb F_2^{st}\setminus\{0\}$.

\section{The EXIT Area Theorem}
\label{app:exit_area}

EXtrinsic Information Transfer (EXIT) charts were introduced by ten Brink in 1999 as a useful tool to understand the convergence of Turbo decoding for different component codes~\cite{tenBrink-elet99}.
His work led to the EXIT area theorem and this was put on rigorous mathematical footing by Ashikhmin, Kramer, and ten Brink in 2004~\cite{Ashikhmin-it04}.
The proof we give here follows the approach pioneered in~\cite{Measson-it08}.

Let $W$ be a BMS channel as described in 
Definition~\ref{def:bms} and define $W_t$ to be the cascade of $W$ followed by an erasure channel with erasure probability $t$.
For $W_t$, the output alphabet is $\cY' = \cY \cup \{*\}$ and the multiplicative noise $Z(t) \in \cY'$ equals the original multiplicative noise $Z$ with probability $1-t$ and $*$ with probability $t$.
For a channel input $X \in \cX$, the output of $W_t$ is $Y(t) = X Z(t) \in \cY'$ with the convention that $1\cdot*=*$ and $-1\cdot*=*$.

For multiple uses, each channel use has a different erasure probability and the erasures are independent.
We denote the erasure probability of the $i$-th channel use by $t_i$ and define $\bt=(t_0,\ldots,t_{N-1})$.
Let $\bX \in \cX^N$ be the channel input vector with pmf $P_{\bX} (\bx)$.
In this case, we define $\bZ (\bt) = (Z_0 (t_0),\ldots,Z_{N-1}(t_{N-1}))$ and write $\bY (\bt)  = \bX \odot \bZ (\bt)$ to indicate that $Y_i (t_i) = X_i Z_i (t_i)$ for $i\in [N]$.

To transmit a codeword over a BMS channel, each binary codeword $\bu \in \cC$ is mapped to a channel input sequence $\bx \in \{\pm 1\}^N$ via the binary phase-shift keying (BPSK) mapping defined by $x_i = (-1)^{u_i}$.
The resulting set of BPSK-modulated codeword sequences is denoted by  $\cC_x$.
In this work, we always assume that the automorphism group $\cG$ of the code $\cC$ is transitive and this property is automatically inherited by the distribution of $\bX$ because $P_{\bX} (\bx) \propto \ind_{\cC_x}(\bx) = \ind_{\cC_x}(\pi \bx)$ for all $\pi \in \cG$.

\begin{thm} \label{thm:exit_area}
If the automorphism group of $\cC$ is transitive and $\bX$ is uniform on $\cC_x$, then
\begin{align*}
\frac{1}{N} I(\bX ; \bY)
&= \int_0^1 I\big (X_0; Y_0 \mid Y_{\sim 0}(t) \big) \, dt. 
\end{align*}
\end{thm}

\begin{proof}
First, we note that
\begin{align*}
I(\bX ; \bY)
&= H(\bX) - H(\bX | \bY) \\
&= H(\bX | \bY(1)) - H(\bX | \bY(0)) \\
&= \int_0^1 \frac{d}{dt} H(\bX | \bY(t)) \, dt. 
\end{align*}
Since the erasure rate for each channel use is different (e.g., the erasure rate for $Y_i$ is $t_i$), we use the law of the total derivative to write
\begin{align*}
\frac{d}{dt} H(\bX | \bY(t)) = \sum_{i=0}^{N-1} \frac{d}{dt_i} H(\bX | \bY(\bt)) \bigg|_{t_0 = \cdots = t_{N-1} = t}.
\end{align*}
To simplify this expression, we use the chain rule of entropy and definition of an erasure channel to write
\begin{align*}
H&(\bX | \bY(\bt))
= H(X_i | \bY(\bt))
+H(X_{\sim i} | \bY(\bt),X_i)\\
&= \pr{Y_i \neq *} H(X_i | Y_{\sim i} (t_{\sim i}), Y_i) \tcb \tca \tc{\qquad} +
\pr{Y_i = *} H(X_i | Y_{\sim i} (t_{\sim i}))
 \tcb \tca \tc{\qquad \qquad} +H(X_{\sim i} | Y_{\sim i}(t_{\sim i}),X_i) \\
&= (1-t_i) H(X_i | Y_{\sim i} (t_{\sim i}), Y_i) \tcb \tca \tc{\qquad} +
t_i H(X_i | Y_{\sim i} (t_{\sim i}))
+H(X_{\sim i} | Y_{\sim i}(t_{\sim i}),X_i).
\end{align*}
From this, we see that
\begin{align*}
\frac{d}{dt_i} & H(\bX | \bY(\bt))
\\ &= H(X_i | Y_{\sim i} (t_{\sim i})) -  H(X_i | Y_{\sim i} (t_{\sim i}), Y_i) \\
&= I(X_i;Y_i | Y_{\sim i}(t_{\sim i})).
\end{align*}
From the transitive symmetry of the code and the full permutation symmetry of the channel uses, we see that this quantity does not depend on $i$ when $t_i = t$ for all $i \in [N]$.
Thus, we observe that
\begin{align*}
\frac{d}{dt} H(\bX | \bY(t))
&= \sum_{i=0}^{N-1} \frac{d}{dt_i} H(\bX | \bY(\bt))  \bigg|_{t_0 = \cdots = t_{N-1} = t} \\
&= N I\big (X_0; Y_0 \mid Y_{\sim 0}(t) \big).
\end{align*}
The proof is completed by substituting this into the integral and normalizing by $N$.
\end{proof}

\begin{thm} [{\cite[Lemma~13]{Reeves-isit23}}] \label{thm:exit_mmse_H}
Using this setup, we have
\[ 2\,\mathrm{BER}(X_0|Y_{\sim 0}) \leq \mmse(X_0|Y_{\sim 0}) \leq H(X_0 | Y_{\sim 0} ). \]
For a transitive code $\cC$ with rate $R(\cC)>0$, if $\bX$ is uniform on $\cC_x$, then $H(\bX)/N=R(\cC)>0$ and 
\[ H(X_0 | Y_{\sim 0} ) \leq 1 - (C-R(\cC)). \]
\end{thm}
\begin{proof}
For the first two inequalities, we note that each quantity is a simple function of the posterior probability $\pr{X_0=-1|Y_{\sim 0}=y_{\sim 0}}$.
For example, $\mathrm{BER}(X_0|Y_{\sim 0}=y_{\sim 0}) = b(\pr{X_0=-1|Y_{\sim 0}=y_{\sim 0}})$ for $b(x)=\min\{x,1-x\}$.
Similarly, we have $\mmse(X_0|Y_{\sim 0}=y_{\sim 0}) = m(\pr{X_0=-1|Y_{\sim 0}=y_{\sim 0}})$ for the function $m\colon [0,1]\to [0,1]$ defined by
\begin{align*} m(x) & =x (-1-(1-2x))^2 + (1-x) (1-(1-2x))^2 \\ &= 4x(1-x).
\end{align*}
Likewise, $H(X_0|Y_{\sim 0}=y_{\sim 0}) = h(\pr{X_0=-1|Y_{\sim 0}=y_{\sim 0}})$ where $h(x)=-x\log_2 (x) - (1-x) \log_2 (1-x)$ is the binary entropy function.
Since $2b(x) \leq m(x) \leq h(x)$ for $x\in [0,1]$ is easily verified, this implies that
\begin{align*} 2\underbrace{\ex{b(\pr{X_0=-1|Y_{\sim 0}})}}_{\mathrm{BER}(X_0|Y_{\sim 0})}
&\leq \underbrace{\ex{m(\pr{X_0=-1|Y_{\sim 0}})}}_{\mmse(X_0|Y_{\sim 0})} \\
&\leq \underbrace{\ex{h(\pr{X_0=-1|Y_{\sim 0}})}}_{H(X_0 | Y_{\sim 0} )}.
\end{align*}
For the last inequality in the theorem statement, we write
\begin{align*}
C &= I(X_0; Y_0) \\
&= H(Y_0) - H(Y_0|X_0) \\
&= H(Y_0) - H(Y_0|X_0,Y_{\sim 0} (t) ) \\
&= I(Y_0; X_0, Y_{\sim 0} (t) ) \\
&= I(Y_0; Y_{\sim 0}(t) ) + I(Y_0;X_0 | Y_{\sim 0} (t)) \\
&\leq I(X_0; Y_{\sim 0} (t) ) + I(Y_0;X_0 | Y_{\sim 0}(t) ) \\
&\leq I(X_0; Y_{\sim 0} ) + I(Y_0;X_0 | Y_{\sim 0}(t) ),
\end{align*}
where the first inequality holds because $Y_0 - X_0 - Y_{\sim 0} (t)$ is a Markov chain and the second inequality holds because $X_0 - Y_{\sim 0} - Y_{\sim 0}(t)$ is a Markov chain. 
Now, we can use the EXIT area theorem to write
\begin{align*}
\frac{1}{N} H(\bX)
&\geq \frac{1}{N} I(\bX;\bY) \\
&= \int_0^1 I\big (X_0; Y_0 \mid Y_{\sim 0}(t) \big) \, dt \\
&\geq \int_0^1 \big( C - I(X_0; Y_{\sim 0} ) \big) dt \\
&= C - I(X_0; Y_{\sim 0} ).
\end{align*}
Since $\cC$ is transitive and \(R(\cC)>0\), every \(X_i\) is uniform on $\{1,-1\}$ (i.e., no code bits are identically zero) and $H(X_0)=1$.
Thus, the above expression simplifies to $R(\cC) \geq C - 1 + H(X_0|Y_{\sim 0})$, which implies the desired conclusion.
\end{proof}

\section{Boolean Functional Analysis}
\label{app:bfa}

Here, we review Fourier analysis on the Hilbert space of functions from $\left\{ 0,1\right\} ^{n}$ to $\mathbb{R}$. Since the domain is a finite set (i.e., $\left|\left\{ 0,1\right\} ^{n}\right|=2^{n}$), this space of functions is a finite-dimensional vector space. On this space, we define the inner product 
\[
\left\langle f,g\right\rangle _{\mu}\triangleq\sum_{\bx\in\left\{ 0,1\right\}^{n}}\mu(\bx)f(\bx)g(\bx),
\]
where $\mu\colon\left\{ 0,1\right\} ^{n}\to\mathbb{R}_{\geq0}$ is a non-negative weight function that sums to 1.
For all $q\geq1$, we also define the norm
\[
\left\Vert f\right\Vert _{q}\triangleq\left(\sum_{\bx\in\left\{ 0,1\right\} ^{n}}\mu(\bx)|f(\bx)|^{q}\right)^{1/q}.
\]
Drawing the random variable $\bX\sim\mu$ makes the inner product equal to the expectation
\[
\left\langle f,g\right\rangle _{\mu}=\expt\left[f(\bX)g(\bX)\right].
\]
Likewise, the $q$-norm is given by $\left\Vert f\right\Vert _{q}=\left(\expt\left[|f(\bX)|^{q}\right]\right)^{1/q}$.

\subsection{The Fourier Transform}

For a fixed $\mu$, let the generalized Fourier transform of $f\colon\{0,1\}^n \to \mathbb{R}$ be defined by a set of orthonormal functions $\left\{ u_{S}(\bx)\right\}_{S\subseteq[n]}$  satisfying $u_{\emptyset} (\bx) = 1$ and 
$$\left\langle u_{S}, u_{S'}\right\rangle _{\mu} = \begin{cases} 1 & \text{if } S = S' \\ 0 & \text{if } S \neq S', \end{cases}$$
for all $S,S' \subseteq [n]$.
In this case, one can expand the function $f$ in this basis by computing inner products
\[
f(\bx)=\sum_{S\subseteq[n]}\left\langle f,u_{S}\right\rangle _{\mu}u_{S}(\bx)=\sum_{S\subseteq[n]}\hat{f}(S)u_{S}(\bx),
\]
where $\hat{f}(S)\triangleq\left\langle f,u_{S}\right\rangle _{\mu}$. Under these assumptions, we also have $\hat{f}(\emptyset)=\left\langle f,1\right\rangle _{\mu}=\expt\left[f(\bX)\right]$. \begin{lem}
[Parseval's Relation]\label{lem:parseval} For any orthonormal basis, $\left\{ u_{S}(\bx)\right\}_{S\subseteq[n]}$, the expansion $f(\bx)=\sum_{S\subseteq[n]}\hat{f}(S)u_{S}(\bx)$ implies that 
\[
\left\Vert f\right\Vert _{2}^{2}=\sum_{S\subseteq[n]}\hat{f}(S)^{2}.
\]
\end{lem}
\begin{proof}
Expanding $\left\Vert f\right\Vert _{2}^{2}$ and simplifying shows that
\begin{align*}
\left\Vert f\right\Vert _{2}^{2} & =\sum_{\bx\in\left\{ 0,1\right\} ^{n}}\mu(\bx)f(\bx)^{2}\\
 & =\left\langle f,f\right\rangle _{\mu}\\
 & =\left\langle \sum_{S\subseteq[n]}\hat{f}(S)u_{S}(\bx),\sum_{S'\subseteq[n]}\hat{f}(S')u_{S'}(\bx)\right\rangle _{\mu}\\
 & =\sum_{S\subseteq[n]}\sum_{S'\subseteq[n]}\hat{f}(S)\hat{f}(S')\left\langle u_{S},u_{S'}\right\rangle _{\mu}\\
 & =\sum_{S\subseteq[n]}\hat{f}(S)^{2},
\end{align*}
where orthonormality implies that $\left\langle u_{S},u_{S'}\right\rangle _{\mu}$ is 1 if $S=S'$ and 0 otherwise.
\end{proof}
It follows that the variance of $f$ satisfies
\begin{align}
\var{f(\bX)} & \triangleq\expt\left[f(\bX)^{2}\right]-\expt\left[f(\bX)\right]^{2}\nonumber \\
 & =\left\Vert f\right\Vert _{2}^{2}-\hat{f}(\emptyset)^{2}\nonumber \\
 & =\sum_{S\subseteq[n]:S\neq\emptyset}\hat{f}(S)^{2}.\label{eq:var_fx}
\end{align}

\subsection{Product Distributions}

Now, we specialize to the case where the inputs are drawn from a product distribution.
Historically, this idea was used in early works by Talagrand and Friedgut~\cite{Talagrand-ap94,Friedgut-comb98}.
Let $\mu (\bx)=\prod_{i=0}^{n-1}\mu_{i}(x_{i})$ be a product measure defined by $\mu_{i}\colon\left\{ 0,1\right\} \to[0,1]$ with $p_i\in (0,1)$, $\mu_{i}(0)=1-p_{i}$, and $\mu_{i}(1)=p_{i}$.
In this case, the analogous orthonormal basis is given by $\left\{ u_{S}(\bx)\right\} _{S\subseteq[n]}$ where $u_{\emptyset}(\bx)=1$, 
\begin{equation}
u_{S}(\bx)=\prod_{i\in S}r_{i}(x_{i}),\label{eq:product_basis}
\end{equation}
 and
\begin{align}
r_{i}(x_{i}) &=\sqrt{\frac{p_i}{1-p_i}} \left(-\frac{1-p_i}{p_i}\right)^{x_i} \tc{\nonumber} \tcb \tca =\begin{cases}
\sqrt{\frac{p_i}{1-p_{i}}} & \mbox{if }x_{i}=0\\
-\sqrt{\frac{1-p_{i}}{p_{i}}} & \mbox{if }x_{i}=1.
\end{cases}\label{eq:product_basis_r}
\end{align}

\begin{lem}
The functions $\left\{ u_{S}(\bx)\right\} _{S\subseteq[n]}$ defined by~(\ref{eq:product_basis}) and~(\ref{eq:product_basis_r}) satisfy 
\[
\left\langle u_{S},u_{S'}\right\rangle _{\mu}=\begin{cases}
1 & \text{if }S=S'\\
0 & \text{if }S\neq S'.
\end{cases}
\]
\end{lem}
\begin{proof}
First, we observe that $\mu_i$ can be written as
\begin{align}
\mu_i (x_i) = (1-p_i) \left(\frac{p_i}{1-p_i}\right)^{x_i}.
\end{align}
Let $\bX$ be drawn from the product distribution where $X_i \sim \mu_i$. Using this, we see that
\begin{align*}
&\expt \left[r_i(X_i) \right]
= \sum_{x_i \in \{0,1\}} \mu_i (x_i) r_i (x_i) \\
&\; = \!\! \sum_{x_i \in \{0,1\}} \!\! (1-p_i) \left(\frac{p_i}{1-p_i}\right)^{x_i} \!\!\! \sqrt{\frac{p_i}{1-p_i}} \left(-\frac{1-p_i}{p_i}\right)^{x_i} \\ &\;= \sum_{x_i \in \{0,1\}} (-1)^{x_i} \sqrt{p_i (1-p_i)} = 0,\\[2mm]
& \expt \left[r_i(X_i)^2 \right]
= \sum_{x_i \in \{0,1\}} \mu_i (x_i) r_i^2 (x_i) \\
&\;= \!\! \sum_{x_i \in \{0,1\}} \!\! (1-p_i) \left(\frac{p_i}{1-p_i}\right)^{x_i} \!\! \left(\frac{p_i}{1-p_i} \right) \! \left(-\frac{1-p_i}{p_i}\right)^{2 x_i} \\
&\;= p_i + (1-p_i) = 1.\\
\end{align*}

Recall that the symmetric difference between $S$ and $S'$ is defined by
\[ S \Delta S' \coloneqq (S \cup S') \backslash (S \cap S'). \]
Using the above results, we have
\begin{align*}
    \left\langle u_{S},u_{S'}\right\rangle _{\mu}
    &= \expt\left[u_{S}(\bX)u_{S'}(\bX)\right] \\
    &= \expt\left[\left(\prod_{i \in S \Delta S'} r_{i}(X_i) \right) \left(\prod_{i \in S \cap S'}r_{i}(X_i)^2 \right)\right] \\
    &= \left(\prod_{i \in S \Delta S'} \expt \left[r_{i}(X_i)\right]\right) \left( \prod_{i \in S \cap S'}\expt \left[r_{i}(X_i)^2 \right] \right) \\
    &= \begin{cases}
        1 & \text{if } S \Delta S' = \emptyset \\
        0 & \text{if } S \Delta S' \neq \emptyset, 
    \end{cases}
\end{align*}
where the expectation commutes with the product because the $X_i$ are independent.
\end{proof}

\subsection{A Proof of the Biased Small-Set Expansion Theorem}
\label{app:thm1025}

This result is a relatively straightforward application of the general hypercontractivity theorem in~\cite[Corollary 10.20]{ODonnell-2014} to a function $f\colon \{0,1\}^n \to \{0,1\}$.
It proves~\cite[Theorem 10.25]{ODonnell-2014} and is relegated to the appendix because its proof requires notation from~\cite{ODonnell-2014} that we do not use elsewhere in the paper.
 
\paragraph{Statement}

Let $\bX \in \{0,1\}^n$ have an i.i.d.\ distribution with $\Pr(X_0=1)=p\in(0,1)$.
For any $A \subseteq \{0,1\}^n$, let $f\colon\{0,1\}^n \to \{0,1\}$ be the indicator function of $A$ and define $\alpha = \ex{ f(\bX) }$.
Then, for $\lambda = \min\{p,1-p\}$ and $\rho = \frac{1}{q-1} \lambda^{1-2/q}$ with $q\geq 2$, we have
\[ \chi = \sum_{S\subseteq [n]} \hat{f}(S)^2 \rho^{|S|} \leq  \alpha^{2-2/q}. \]
If $\bX,\bY$ are Bernoulli($p$) with correlation $\rho$ and $q>2$, then $\Pr(\bX \in A, \bY \in A) = \chi$ and the name small-set expansion stems from the fact that the conditional probability of $\bY \in A$ given $\bX \in A$ vanishes as $\alpha \to 0$ for any $\rho<1$.

\begin{proof}
Here, we will use the standard noise operator $T_\rho$ defined in~\cite[Definition 8.27]{ODonnell-2014}.
By the general hypercontractivity result in~\cite[Corollary 10.20]{ODonnell-2014}, we have
\[ \| T_{\sqrt{\rho}} f \|_2 \leq \| f \|_{q'}, \]
where $q'=\frac{1}{1-1/q}$.
We note that we use the definition of $\rho$ from~\cite[Theorem 10.25]{ODonnell-2014} which differs from~\cite[Corollary 10.20]{ODonnell-2014} and this gives rise to the $\sqrt{\rho}$ above.
Using this definition, we note that
\[ \| T_{\sqrt{\rho}} f \|_2^2 = \langle f,T_\rho f \rangle = \sum_{S\subseteq [n]} \hat{f}(S)^2 \rho^{|S|}. \]
Thus, the stated result follows from computing
\[ \| f \|_{q'}^2 = \ex{|f(\bX)|^{q'}}^{2/q'} =  \alpha^{2/q'} = \alpha^{2-2/q}. \]
The connection to the name small-set expansion follows from
\[ \Pr(\bX \in A, \bY \in A) = \mathbb{E}[f(\bX)f(\bY)] = \langle f,T_\rho f \rangle. \qedhere \]
\end{proof}

\onecolumn

\section{Glossary of Notation} \label{app:gon}

\begin{itemize}

\item $\mathbb{F}_q$: Finite field with $q$ elements.
\item $\mathcal{C}$: A linear code, defined as a subspace of $\mathbb{F}_q^N$ (Definition~\ref{def:blc}).
\item $N$: Length of the code, representing the dimension of the vector space $\mathbb{F}_q^N$.
\item $R(\mathcal{C})$: Code rate, defined as $\dim(\mathcal{C})/N$ (Definition~\ref{def:blc}).
\item $\mathbb{S}_N$: Symmetric group of permutations on $N$ elements $\{0,1,\ldots,N-1\}$ (Section~\ref{sec:std_notation}).
\item $2^S$: Power set of $S$; for an index set $A$, we write $\cA = 2^A$ (Section~\ref{sec:std_notation}).
\item $\pi$: A permutation from the symmetric group $\mathbb{S}_N$ (Definition~\ref{def:code_equivalence}).
\item $\mathcal{G}$: Permutation automorphism group of the code $\mathcal{C}$ (Definition~\ref{def:code_auto}).
\item $\mathcal{C}|_A$: Projection of a code $\mathcal{C}$ onto a subset $A$ of coordinates (Definition~\ref{def:code_proj_nest}).
\item $\RM(r,m)$: Binary Reed--Muller code of order $r$ and length $N = 2^m$ (Definition~\ref{def:binary_rm}).
\item $\mathcal{F}(r,m)$: Set of $m$-variate polynomials with degree at most $r$ over $\mathbb{F}_2$ (Definition~\ref{def:binary_rm}).
\item $\theta_m$: Mapping from integer indices to binary expansions in $\mathbb{F}_2^m$ (Definition~\ref{def:binary_rm}).
\item $\text{BEC}(p)$: Binary erasure channel with erasure rate $p$ (Definition~\ref{def:bec}).
\item $\text{BSC}(p)$: Binary symmetric channel with error rate $p$ (Definition~\ref{def:bsc}).
\item $W$: A binary memoryless symmetric (BMS) channel (Definition~\ref{def:bms}).
\item $\mathcal{X}$, $\mathcal{Y}$: Input and output alphabets of a channel (Definitions~\ref{def:bec}, \ref{def:bsc}, and~\ref{def:bms}).
\item $X$, $Y$: Input and output random variables of a channel (Definition~\ref{def:bms}).
\item $Z$: Noise random variable in BMS channel decomposition $Y = XZ$ (Definition~\ref{def:bms}).
\item $f$: Decoding function mapping channel observations to estimates (Section~\ref{sec:dec_fun}).
\item $\beta$: The Bhattacharyya parameter of a BMS channel given by \eqref{eq:bhattacharyya}.
\item $A_w (\cC)$: The number of weight-$w$ codewords in the code $\cC$ (Section~\ref{sec:list_dec}).
\item $T_{\cC,\beta}(h)$: Truncated union bound for $\cC$ through Hamming weight $h$, given by \eqref{eq:competition_function}.
\item $P_e(\mathcal{C})$: Bit erasure probability for a code $\mathcal{C}$ on the BEC (Section~\ref{sec:bec}).
\item $\mathrm{Sym}(f)$: Symmetry group of a decoding function $f$ (Definition~\ref{def:symf}).
\item $g(y_{\sim 0})$: Boolean extrinsic decoding function indicating whether $X_0$ is not recoverable (Section~\ref{sec:bec}).
\item $A$, $B$, $C$, $D$: Subsets of bit indices for code projections (Section~\ref{sec:bec}).
\item $\rho$: Correlation parameter for random restrictions in Boolean function analysis (Section~\ref{sec:bec}).
\item $\Phi$: Cumulative distribution function of a standard Gaussian random variable (Lemma~\ref{lem:rate}).
\item $M(\mathcal{C})$: Extrinsic minimum mean-squared error (MMSE) of $X_0$ given $Y_{\sim 0}$ for a code $\mathcal{C}$ (Section~\ref{sec:bms}).
\item $P_b(\mathcal{C})$: bit-error probability for a code $\mathcal{C}$ on a BMS channel (Section~\ref{sec:bms}).
\item $\bm{X}, \bm{Y}, \bm{Z}$: Input, output, and noise sequences for multiple BMS channel uses (Section~\ref{sec:bms}).
\item $\cC_x$: Set of BPSK-modulated codeword sequences derived from $\cC$ (Section~\ref{sec:bms}).
\item $g(y_{\sim 0})$: Extrinsic conditional mean decoding function for $X_0$ (Equation~\eqref{eq:cond_mean_g}).

\item $\mu(\bx)$: Non-negative weight function $\mu(\bx) = p^{|\bx|} (1-p)^{n-|\bx|}$, where $|\bx| = \sum_{i=0}^{n-1} |x_i|$ (Section~\ref{sec:fourier}).
    
\item $u_S(\bx)$: Orthonormal basis indexed by $S \subseteq [n]$ of functions mapping $\{0,1\}^n \to \mathbb{R}$ (Section~\ref{sec:fourier}).
    
\item $\hat{f}(S)$: Fourier coefficient of $f$, defined as $\hat{f}(S) = \langle f, u_S \rangle_{\mu}$ (Section~\ref{sec:fourier}).
    
\item $f_A$: Restriction of $f$ to $A \subseteq [n]$ averaged over variables outside $A$ (Section~\ref{sec:fourier}).
    
\item $\mathrm{GL}(m,2)$: General linear group containing invertible linear transformations of $\mathbb{F}_2^m$ (Section~\ref{sec:aobf}).

\item $\Pi$: Random element of the symmetry group $\mathrm{Sym}(f)$, drawn according to the distribution $q$ (Lemma~\ref{lem:f_bool_restrict_var}).
    
\item $\dim(S)$: Dimension of the smallest subspace containing $S$ in $\mathbb{F}_2^m$ (Lemma~\ref{lem:gl_sym}).

\item $c(\lambda)$: Constant $1/(8\ln(1/\lambda))$, used in the  ``level-$k$'' inequality for Fourier coefficients (Lemma~\ref{lem:biased_level_k}).
     
\item $P_B(\cC)$: The block-error probability of code $\cC$ (Theorems~\ref{thm:fast_rm_err_bound_bec} and~\ref{thm:fast_rm_err_bound_bsc}).

\item $g_{\text{Maj}}$: The majority-vote decoding function for decoding codes with three looks (Proof of Lemma~\ref{lem:rm_level_k_bsc}).

\item $h(p)$: The binary entropy function $h(p) = -p \log_2(p) - (1-p) \log_2(1-p)$ (Theorem~\ref{thm:fast_rm_err_bound_bsc}).

\item $\cP=\cC_1\otimes\cC_2$: Tensor product of the linear codes $\cC_1$ and $\cC_2$ (Definition~\ref{defn:product_code}).

\item $\PRM(r_1,m_1,r_2,m_2)$: The RM product code $\RM(r_1,m_1)\otimes\RM(r_2,m_2)$ (Section~\ref{sec:product_rm}).

\item $d_1(S),d_2(S)$: Dimensions of the spans of the two coordinate projections of $S$ (Corollary~\ref{cor:product_gl_sym}).

\item $d_m$: The minimum distance of the code $\cC_m$ (Theorem~\ref{thm:bsc_to_bms}).

\item $B_m (p)$: The block-error probability for the code $\cC_m$ on a BSC$(p)$ (Theorem~\ref{thm:bsc_to_bms}).

\item $\alpha_m (p)$: Parameter characterizing block-error probability $B_m (p)$ given by \eqref{eq:alpha_m_p}.

\end{itemize}

\newpage

\bibliographystyle{IEEEtran}
\bibliography{WCLabrv,WCLnewbib,WCLbib}

\end{document}